\documentclass[twocolumn,prb,preprintnumbers,amsmath,amssymb]{aastex631} 
\usepackage{amsmath} 
\usepackage{amsfonts} 
\usepackage{comment}
\usepackage{graphicx}
\usepackage{dcolumn}
\usepackage{bm}
\usepackage{rotating}
\usepackage{xcolor,soul}
\sethlcolor{white}

\begin{document}
\title{Dust-Gas Coupling in Turbulence- and MHD Wind-Driven Protoplanetary Disks:\\ Implications for Rocky Planet Formation}
\author{Teng Ee Yap}
\author{Konstantin Batygin}
\affil{Division of Geological and Planetary Sciences, California Institute of Technology, Pasadena, CA 91125, USA.\\ Corresponding author: tyap@caltech.edu}

\begin{abstract}
\indent The degree of coupling between dust particles and their surrounding gas in protoplanetary disks is quantified by the dimensionless Stokes number. The Stokes number  ($St$) governs particle size and spatial distributions, in turn establishing the dominant mode of planetary accretion in different disk regions. In this paper, we model the characteristic $St$ of particles across time in disks evolving under both turbulent viscosity and magnetohydrodynamic (MHD) disk winds. In both turbulence- and wind-dominated disks, we find that collisional fragmentation is the limiting mechanism of particle growth. The water-ice sublimation line constitutes a critical transition point between dust settling, drift, and size regimes. For a fiducial value disk evolution parameter $\tilde{\alpha}\simeq10^{-3}$, silicate particles interior to the ice-line in our model are characterized by low $St$ ($\lesssim$ $10^{-2}$) and sizes in the sub-mm- to 1cm-scale. Icy particle/boulders beyond the ice-line are characterized by high $St$ ($\gtrsim 10^{-2}$) and sizes in the cm to dm size range. Hence, icy particles settle into a thin layer at the outer disk midplane and drift inward at velocities exceeding the gaseous accretionary flow due to substantial headwind drag. Silicate particles in the inner disk remain relatively well dispersed and are to a large extent advected inward with their surrounding gas. \\
\indent The $St$ dichotomy across the ice-line translates to distinct planet formation pathways between the inner and outer disk. While pebble accretion proceeds slowly for rocky embryos within the ice-line (across most of parameter space), it does so rapidly for volatile-rich embryos beyond it, allowing for the growth of giant planet cores before disk dissipation. Through simulations of rocky planet growth, we evaluate the competition between pebble accretion and classical pairwise collisions between planetesimals. We conclude that the dominance of pebble accretion can only be realized in disks that are driven by MHD winds, slow-evolving  ($\tilde{\alpha}\lesssim10^{-3.5}$), and devoid of pressure maxima that may concentrate solids and give rise of planetesimal rings in which classical growth is enhanced. Such disks are extremely quiescent, with Shakura-Sunyaev turbulence parameters $\alpha_{\nu}\lesssim 10^{-4}$. We conclude that for most of parameter space corresponding to values of $\alpha_{\nu}$ reflected in observations of protoplanetary disks ($\gtrsim 10^{-4}$), pairwise planetesimal collisions constitute the dominant pathway of rocky planet accretion. Our results are discussed in the context of super-Earth origins, and lend support to the emerging view that they formed in planetesimal rings. Moreover, these results argue against a significant contribution ($\gtrsim$ 10\%) of outer disk, carbonaceous, material to the proto-Earth in the form of pebbles, in agreement with chemical and isotopic investigations of Earth's accretion history. \\

\end{abstract}

\section{Introduction}
\indent  The evolution of the size and spatial distributions of dust particles in protoplanetary disks is intimately linked with the coupling of these particles to the \hl{H$_2$-He} gas in which they are embedded, and ultimately determines the large-scale architecture of planetary systems that emerge upon disk dissipation. Improvements in our understanding of the interplay between dust and gas in such disks thus translates to a more refined picture of the planet formation process, yielding insights into how our Solar System (SS) evolved to be a comparatively uncommon outcome of planet formation within the Galaxy. Indeed, the SS is devoid of the archetypal close-in rocky planets characteristic of many systems in the galactic exoplanetary census (\textit{e.g.,} Petigura et al., 2013; Mulders et al., 2018; Batygin \& Morbidelli, 2023), yet hosts gas/ice giants on long-period orbits, found only around $\sim15 \%$ of Sun-like stars (\textit{e.g.,} Rosenthal et al., 2022). \\
\begin{figure*} 
\centering
\scalebox{1.3}{\includegraphics{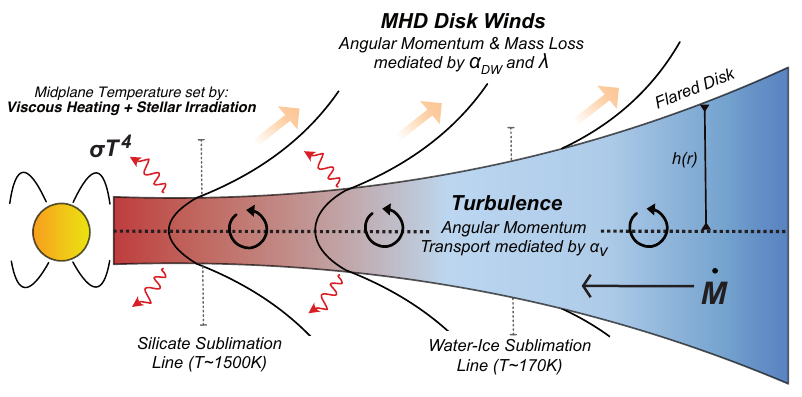}}
\caption{\footnotesize\textbf{Schematic of a protoplanetary disk evolving under both turbulence and MHD disk winds.} Turbulence manifests in an effective viscosity, and facilitates angular momentum transport. MHD disk winds arise from magnetic field lines that thread the disk, and facilitate angular momentum and mass loss. No solids exist interior to the silicate sublimation line, and the \hl{ice-line} demarcates the boundary between the inner and outer disk, rocky and icy dust particles. The midplane temperature of the flared disk is set by a balance between ``active" viscous heating and ``passive" stellar irradiation, and radiative loss. See Sections 2.1 \& 2.2.}
\label{fig:Figure 1}
\end{figure*}
\indent The degree of dust-gas coupling is quantified by the dimensionless ``Stokes number," defined as the ratio of the particle stopping time to the turnover time of the largest turbulent eddy in the disk, the latter customarily taken as the inverse of the Keplerian angular velocity (Cuzzi et al., 2001\hl{; Ormel \& Cuzzi, 2007; Sengupta et al., 2024}). As such, the Stokes number ($St$) represents the fraction of an orbit (given by the ratio of the stopping time to the orbital period) in radians required for a dust particle to equilibrate with its surrounding gas flow. The $St$ governs both the efficiency of planetesimal formation (\textit{i.e.,} the onset and growth rate of resonant drag instabilities; Youdin \& Goodman, 2005; Yang et al., 2017; Squire \& Hopkins, 2018) and the mode of planetary accretion favored in different regions of disks (\textit{e.g.,} Ormel, 2017). \\
\indent Recently, Batygin \& Morbidelli (2022) \hl{(BM22)} presented a self-consistent model for \hl{$St$} within a \hl{steady-state,} ``actively" heated viscous disk. In the context of this model, the change in dust composition from \hl{purely} silicate to \hl{icy} across the water-ice sublimation line \hl{(hereafter denoted simply as the ``ice-line")} manifests as a large ($>$ an order of magnitude) difference in $St$ between the inner and outer disk. This dichotomy in $St$ reflects the difference in fragmentation threshold velocities between silicate and ice grains (Blum \& Münch, 1993; Ormel \& Cuzzi, 2007;  Güttler et al., 2010), and underlies their key conclusion that pebble accretion (Ormel \& Klahr, 2010; Lambrechts \& Johansen, 2012) can only operate efficiently beyond the ice-line, in the outer disk. This suggests planets in the inner disk primarily grew via stochastic, pairwise collisions between planetesimals (\textit{e.g.,} Chambers \& Wetherill, 1998\hl{; Safronov, 1972}). While this conclusion is likely robust to leading order, the standard viscous disk theory on which the model is built invites layers of complexity that can yield additional insights into where and how planets form. \\
\indent Despite the historic appeal of turbulent viscosity as the driver of disk evolution, the extent to which disks are truly turbulent remains unknown. It is also unclear if mechanisms thought to source turbulence (\textit{e.g.,} the magnetorotational instability (MRI), Balbus \& Hawley, 1991; the vertical shear instability (VSI), Nelson et al. 2013) can do so with sufficient vigor to explain observed stellar mass accretion rates (\textit{e.g.,} Trapman et al., 2022). Magnetohydrodynamic (MHD) disk winds are garnering increasing attention as an alternative avenue for facilitating disk evolution. In essence, given that disks are sufficiently ionized at the surface, magnetic field lines threading the disk allow for the extraction of mass and angular momentum along disk-wind streamlines, driving accretion (Blandford \& Payne, 1982; Armitage, 2020). \\
\indent Tabone et al. (2022) introduced a phenomenological treatment of MHD disk winds, inspired by the functional $\alpha$-prescription for turbulent viscosity from Shakura \& Sunyaev (1973) and the self-similar solutions for disk surface density evolution from Lynden-Bell \& Pringle (1974).  Using their model, we revisit the exercise of modeling the characteristic $St$ of dust particles across protoplanetary disks, considering other processes beyond fragmentation that may set the characteristic $St$ in different regions of the disk, such as radial drift (Weidenschilling, 1977) and bouncing of colliding particles (Zsom et al., 2010; Windmark et al., 2012). Accordingly, our model illustrates how $St$ evolves both in time and in transition between turbulence-dominated and wind-dominated disks. \\
\indent The paper is structured as follows. In Section 2, we provide an overview of the disk model adopted and its parameters. We also define model constants that yield disk surface density and temperature profiles most consistent with heating from both viscous shear and stellar irradiation, as well as the initial disk mass prescribed. In Section 3, we derive characteristic $St$ profiles for two fiducial disks: one turbulence-dominated and the other wind-dominated. We show how these profiles translate to dust properties (\textit{i.e.,} vertical distribution, radial drift rate, physical size) and planetary accretion regimes across these disks in Section 4. \hl{We also assess the favored pathway of planetary accretion within and without the ice-line across parameter space, with a focus on the former, through consideration of mass-doubling timescales and simulations of rocky planet growth. Our results are discussed in the context of super-Earth origins and the Earth's accretion.} We consider the potential implications our assumptions have on \hl{$St$ across disks} in Section 5 before giving our concluding remarks in Section 6. 
\section{Disk Model}
\subsection{Overview}
\indent We begin by providing an overview of the formulation devised by Tabone et al. (2022) to describe the evolution of the disk surface density $\Sigma(r,t)$ under the influence of both turbulence and MHD disk winds (Fig. 1). We defer the reader to their work for an in-depth derivation of the equations to be introduced. Aside from the stellar mass accretion rate $\dot{M}$, a proxy for time (\textit{e.g.,} Hartmann et al., 1998), and the $\alpha$ turbulence parameter (henceforth denoted $\alpha_{\nu}$), the inclusion of disk winds necessitates the addition of two parameters to specify the evolution of $\Sigma(r,t)$. The first, denoted $\alpha_{DW}$, is analogous to $\alpha_{\nu}$ and quantifies the torque exerted on the disk by winds. The total torque exerted on the disk is encapsulated in $\tilde{\alpha}$, given by
\begin{equation}
\begin{split}
\tilde{\alpha} = \alpha_{\nu}+\alpha_{DW}. 
\end{split}
\end{equation}
The relative contribution of turbulence and disk winds to $\tilde{\alpha}$ is quantified by the ratio of $\alpha_{DW}$ to $\alpha_{\nu}$, denoted $\psi$:
\begin{equation}
\begin{split}
\psi = \alpha_{DW}/\alpha_{\nu}.
\end{split}
\end{equation}
\begin{figure*} 
\centering
\scalebox{0.775}{\includegraphics{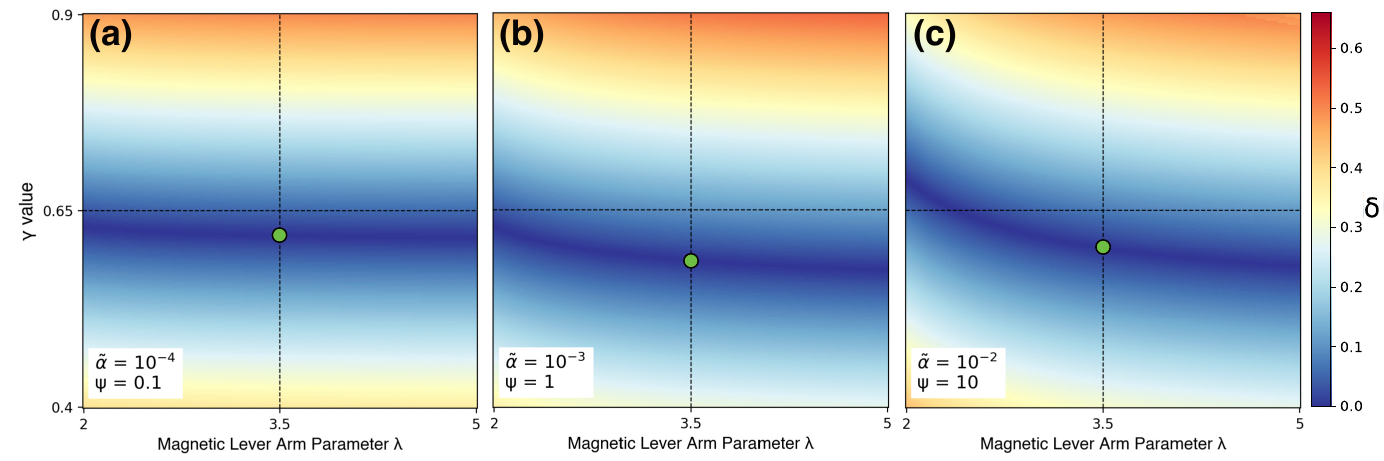}}
\caption{\footnotesize\textbf{Rasters of the deviation in the retrieved vs. assumed temperature power law index $\delta$ for (a) $\tilde{\alpha} = 10^{-4}$ and $\psi = 0.1$, (b) $\tilde{\alpha} = 10^{-3}$ and $\psi = 1.0$, and (c) $\tilde{\alpha} = 10^{-2}$ and $\psi = 10$ (\textit{i.e.,} diagonal across $\tilde{\alpha}-\psi$ parameter space).} It is clear that most of the variation in $\delta$ is attributed to $\gamma$. The green point along the center of all three plots, at $\lambda=3.5$ adopted as a constant in our disk model, represents the optimal values of $\gamma$, which changes with $\tilde{\alpha}$ and $\psi$. These optimal values yield $\delta\lesssim 0.1$. See Section 2.2.}
\label{fig:Figure 2}
\end{figure*}
Accordingly, disks characterized by high and low $\psi$ are wind- and turbulence-dominated, respectively. The second additional parameter is the magnetic lever arm parameter $\lambda$ (Blandford \& Payne, 1982), the angular momentum per unit mass extracted by disk-wind streamlines. The greater $\lambda$ is, the smaller the extracted mass required per unit time to explain the contribution to $\dot{M}$ from disk winds (\textit{i.e.,} to yield a given value of $\alpha_{DW}$). Thus, $\lambda$ ultimately sets the rate at which mass is extracted from the disk, given a specification for $\alpha_{DW}$ (or $\tilde{\alpha}$ and $\psi$). The mass loss rate due to disk winds $\dot{\Sigma}_{DW}$ is given by
\begin{equation}
\dot{\Sigma}_{DW} = \frac{3\alpha_{DW}c_{s}^{2}\Sigma}{4(\lambda-1)\Omega_{K}r^{2}} , 
\end{equation}
where $r$ is the radial/cylindrical distance from the star of mass $M$, $\Omega_{K}=\sqrt{GM/r^{3}}$ is the Keplerian angular velocity, and $c_{s}=\sqrt{k_{b}T/\mu}$ the isothermal sound speed, \hl{with} $k_{b}$ being the Boltzmann constant and $\mu$ the mean molecular mass of the disk gas ($\sim$2.4 proton masses). \hl{Note that the Keplerian \textit{orbital }velocity $v_K = r\Omega_K$, where $r$ is the radial distance from the star.}\\
\indent With \hl{$\alpha_{DW}$ and $\lambda$ defined}, the evolution of $\Sigma(r,t)$, given by mass and angular momentum continuity, is expressed as 
\begin{equation}
\begin{split}
\frac{\partial\Sigma}{\partial t} &= \frac{3}{r}\frac{\partial}{\partial r}\left[\frac{1}{r\Omega}\frac{\partial}{\partial r}(r^{2}\alpha_{\nu}\Sigma c_{s}^{2})\right]\\
&+ \frac{3}{2r}\frac{\partial}{\partial r}\left[\frac{\alpha_{DW}\Sigma c_{s}^{2}}{\Omega}\right] - \dot{\Sigma}_{DW}
\end{split}
\end{equation}
\indent Above, the first and second terms on the RHS of the equation represent viscous and wind-driven accretion, respectively. Assuming $\alpha_{\nu}$ and $\alpha_{DW}$ are constant in time and space, as well as a temperature profile scaling with $r$ as $T\sim r^{\gamma-3/2}$ (see Appendix C in Tabone et al., 2022), the ansatz to the above equation takes the form
\begin{equation}
\Sigma(r,t) = \Sigma_{c}(t)\left(\frac{r}{r_{c}(t)}\right)^{\xi-\gamma}exp\left[-(\frac{r}{r_{c}(t)})^{2-\gamma}\right] ,
\end{equation}
underpinned by the analytical solution from Lynden-Bell \& Pringle (1974) for the evolution of $\Sigma(r,t)$ in a viscous disk with constant $\alpha_{\nu}$. The variable $\xi$ is the so-called ``mass ejection index" describing the flattening of the steady-state $\Sigma$ profile (\textit{i.e.}, that without the exponential decay term, such that $\Sigma\sim r^{\xi-\gamma}; \xi>0$) due to the extraction of mass by disk winds, and is given by
\begin{equation}
\xi = \frac{1}{4}(\psi+1)\left(\sqrt{1+\frac{4\psi}{(\lambda-1)(\psi+1)^{2}}}-1\right) .
\end{equation}
Accordingly, $\gamma$ is the power law index of $\Sigma(r,t)$ in the limit of $\psi\ll1$ (\textit{i.e.,} $\xi\simeq 0$; no winds) and small $r$ (where the exponential term is insignificant). For $\gamma=1$ in the absence of disk winds, $\Sigma$ at small $r$ scales roughly as $r^{-1}$ and $T$ scales as $r^{-1/2}$ as in \hl{the} passively irradiated blackbody disk \hl{of Chiang \& Goldreich (1997)}.\\
\indent Additionally, $\Sigma_c(t)$ represents the characteristic surface density, and $r_c(t)$ the characteristic outer edge of the disk at which $\Sigma(r_c,t) = \Sigma_c(t)/e$. The former is given by
\begin{equation}
\Sigma_c(t) = \Sigma_c(0) \left(1+\frac{t}{(1+\psi)t_{acc,0}}\right)^{(5+2\xi+\psi)/[2(2-\gamma)]} ,
\end{equation}
and the latter by
\begin{equation}
r_c(t) = r_c(0) \left(1+\frac{t}{(1+\psi)t_{acc,0}}\right)^{1/(2-\gamma)} . 
\end{equation}
We identify $t=0$ with the end of the disk's infall stage, at which point the characteristic surface density $\Sigma_c(0)$ can be expressed with the initial disk mass $M_{disk}(0)$ and radial extent of the disk $r_c(0)$ as 
\begin{equation}
\Sigma_c(0) = \beta \frac{M_{disk}(0)}{2\pi r_c^{2}(0)} , 
\end{equation}
where we have introduced a factor of order unity $\beta$ used to ensure the integration of mass across the disk (\textit{i.e.,} $\int_{0.1 AU}^{\infty} 2\pi r\Sigma(r,0)dr$, where $0.1 AU$ is taken as the \hl{disk inner edge}) retrieves $M_{disk}(0)$. The initial accretion timescale $t_{acc,0}$ is defined as the time it takes to accrete a gas parcel across a distance of $r_c(0)/2$, keeping its radial velocity constant at its initial value $v_{r,gas,c}(0)$. It is expressed as
\begin{equation}
t_{acc,0} = \frac{r_{c}(0)}{2v_{r,gas,c}(0)} = \frac{r_c(0)v_{K,c}}{3(2-\gamma)^{2}c_{s,c}^{2}\tilde{\alpha}} , 
\end{equation}
where we have substituted $v_{r,gas,c}(0)$$=(3/2)(2-\gamma)^{2}\tilde{\alpha}c_{s,c}^{2}v_{K,c}^{-1}$ in the second equality. The variables $c_{s,c}$ and $v_{K,c}$ represent the sound speed and Keplerian orbital velocity at $r=r_c(0)$. \\
\indent To summarize, defining a $\Sigma(r,t)$ profile of a disk evolving under both turbulence and disk winds requires specification of eight parameters:  $\tilde{\alpha}$, $\psi$, $\lambda$, $\gamma$, $M_{disk}(0)$, $r_c(0)$, $\beta$, and $t$. In our model, we take $\tilde{\alpha}$, $\psi$, and $t$ as free parameters, letting $\tilde{\alpha}$ range between $10^{-4}$ and $10^{-2}$ (\textit{e.g.,} Andrews et al., 2009; Rosotti, 2023), and $\psi$ between 0.1 (turbulence-dominated) and 10 (wind-dominated). Note that $\alpha_{\nu}\sim10^{-2}$ corresponds to the maximum attainable value by MRI-generated turbulence (\textit{i.e.,} in the ideal MHD limit; Armitage, 2020), while $\alpha_{\nu}\sim10^{-4}$ approximates the minimum attainable value by VSI-generated turbulence (Nelson et al., 2013). Below, we discuss how we derive the values for $\lambda$ and $\gamma$ which yield $\Sigma(r,t)$ and $T(r,t)$ profiles most consistent with both viscous and irradiation heating for the range of $\tilde{\alpha}$ and $\psi$ explored. We set $M_{disk}(0)$ to $\simeq0.05M_{\odot}$ and $r_c(0)$ to $\simeq12AU$. These values roughly yield $\Sigma(r=1 AU)$ close to that of the ``minimum mass solar nebula" model of Hayashi (1981) (\textit{i.e.,} $\simeq 2\times 10^{4}$ kg/m$^{2}$) across an accretion timescale in both turbulence- and wind-dominated disks for an intermediate, fiducial value of $\tilde{\alpha}\simeq 10^{-3}$ (Fig. 4). \hl{For reference,} disks around Class-II young stellar objects like T-Tauri stars are inferred to have masses in the range $\sim3 \times 10^{-5}$ to $0.3M_{\odot}$ from measurements of dust emission (Boss \& Ciesla, 2014 and references therein; Ansdell et al., 2017; Miotello et al., 2022 ). We also \hl{found} that a $\beta$ value of $\simeq1.45$ yields $\Sigma$ profiles that reproduce $M_{disk}(0)$ to within $15\%$ for the optimal $\lambda$ and $\gamma$ found across $\tilde{\alpha}-\psi$ parameter space.
\begin{figure*} 
\centering
\scalebox{0.7}{\includegraphics{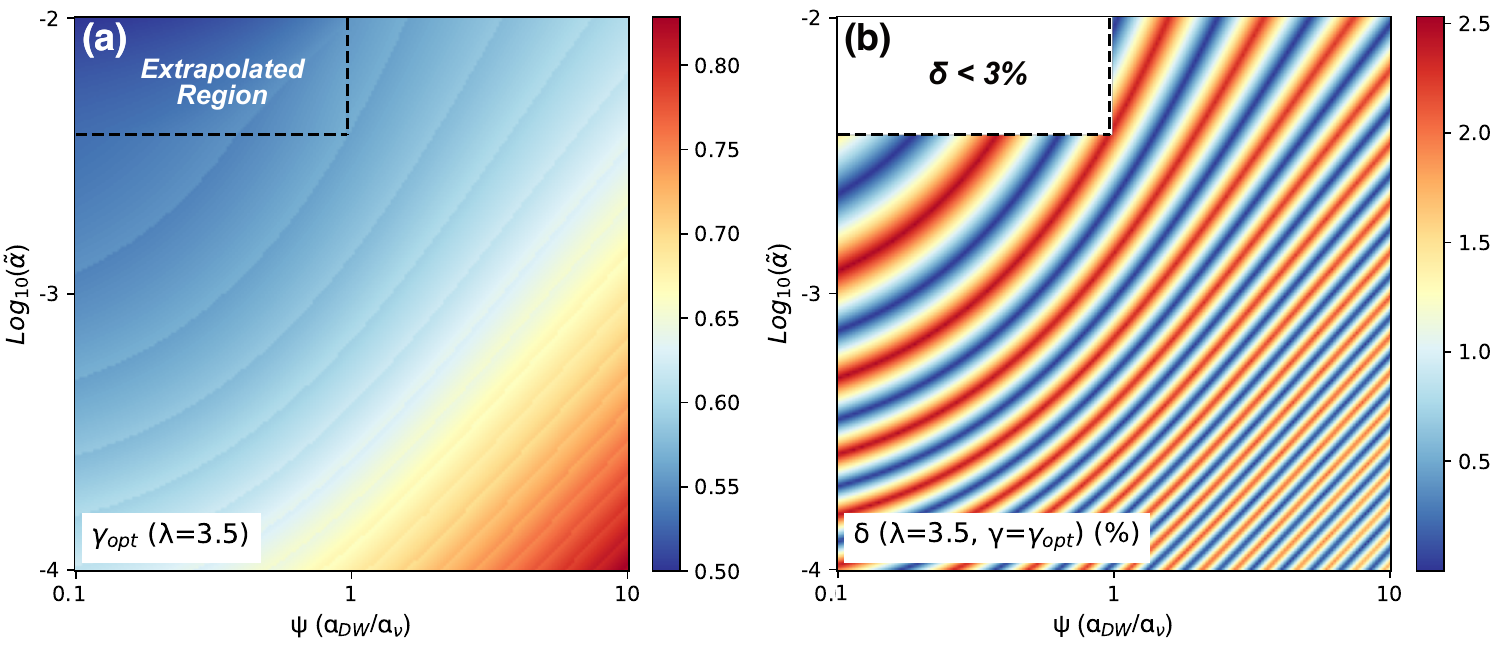}}
\caption{\footnotesize\textbf{Rasters of (a) $\gamma_{opt}$ and (b) the deviation (in \%) in the retrieved vs. assumed temperature power law index $\delta$ across $\tilde{\alpha}-\psi$ parameter space, assuming a constant $\lambda=3.5$.} In (a), values of $\gamma_{opt}$ in the top-left region of parameter space (\textit{i.e.,} high $\tilde{\alpha}$ and $\psi\lesssim1$) bounded by the dashed black lines are extrapolated (see text for discussion). Across all parameter space, $\gamma_{opt}$ ranges from $\sim0.50-0.83$, and $\delta$ values are $\lesssim 3\%$, indicating a high degree of self-consistency. See Section 2.2.}
\label{fig:Figure 3}
\end{figure*}
\subsection{Optimal Values for $\lambda$ and $\gamma$}
\indent To constrain the power law index of the disk's temperature profile (\textit{i.e.,} $\gamma-3/2$), we impose that the disk is optically thick and \hl{heated by both viscous shear between disk annuli and stellar irradiation.} The mid-plane temperature of such a disk $T(r,t)$ is established via equilibrium between heat generation and radiative loss, expressed as Armitage, 2020)
\begin{equation}
\sigma T^{4} \simeq \frac{3}{4}\tau F_{visc} + \sigma T_{irrad}^{4} ,
\end{equation}
where $F_{visc}$ is the heating rate per unit area from viscous shear, $T_{irrad}$ the interior temperature derived from models of passive radiative equilibrium disks (as opposed to blackbody disks), $\sigma$ the Stefan-Boltzmann constant, and $\tau$ the optical depth. The former is given by (Nakamoto \& Nagakawa, 1994)
\begin{equation}
F_{visc} = \frac{1}{2}\Sigma\nu(r\frac{\partial\Omega_{K}}{\partial r})^{2} = \frac{9}{8}\Sigma\alpha_{\nu}c_s^{2}\Omega_{K}, 
\end{equation}
where the turbulent viscosity $\nu = \alpha_{\nu}c_s h$ (Shakura \& Sunyaev, 1973), and the hydrostatic scale height of the disk (assumed to be vertically isothermal at all $r$) $h = c_{s}/\Omega_{K}$. The functional form of $T_{irrad}$ depends on the behavior of the geometrical aspect ratio $h/r$ with $r$, namely whether it remains virtually constant as in ``flat" disks, or increases as in ``flared" disks. The latter allows for greater interception of stellar irradiation in the outer disk, owing to its curvature away from the midplane. Analyses of spectral energy distributions from T-Tauri stars indicate that many disks are flared (Kenyon \& Hartmann, 1987), prompting us to adopt the expression for $T_{irrad}$ provided by Chiang \& Goldreich (1997) for flared radiative equilibrium disks, valid in the domain $0.4AU\lesssim r\lesssim 50AU$:
\begin{equation}
T_{irrad} = 150 (\frac{r}{1 AU})^{-3/7} K.
\end{equation}
This power-law fit is obtained for a T-Tauri star with effective temperature $T_{*}\simeq 4000 K$, mass $M_{*}\simeq 0.5M_{\odot}$, and radius $R\simeq 2.5R_{\odot}$. For reference, the sun has an effective $T_{\odot}\simeq 5800 K$.\\
\indent At temperatures below $\sim$1500K (\textit{i.e.,} the silicate sublimation line), dust constitutes the dominant source of opacity, and the optical depth $\tau$ takes the form (Bitsch et al., 2014)
\begin{equation}
\tau = \frac{1}{2}f_{d}\Sigma\kappa_{d} ,
\end{equation}
where $f_{d}\simeq 0.01$ is the dust-to-gas ratio and $\kappa_{d}\simeq 30$ m$^{2}$/kg the dust opacity. Bearing in mind the uncertainties and complications associated with the dependence of $\kappa_{d}$ on temperature, dust size distribution, and dust composition (\textit{e.g.,} Chambers, 2009), we refrain from adopting a specific model for variable $\kappa_{d}$. We note, nonetheless, that an analytic piecewise model for $\kappa_{d}$ has been implemented by Stepinski (1998), based on fits to opacities appropriate for dust size distributions centered below the tens of microns and of solar composition (Pollack et al., 1985). The applicability of these fits, given that grain growth of mm- to cm-sized grains via hit-and-stick collisions occurs rapidly, is likely unsubstantiated. \\
\indent Keeping in mind the relation $c_s = \sqrt{k_{b}T/\mu}$, Eqs. 11-14 yield a quartic equation in the mid-plane temperature with coefficients given by
\begin{equation}
\begin{split}
&p_{4} = 1;\ p_{3} = 0;\ p_{2} = 0\\
&p_{1} = -\frac{27f_{d}\kappa_{d}\alpha_{\nu}\Omega_{K}k_{b}\Sigma^{2}}{64\sigma\mu};\\ &p_{0} = -(7.3\times10^{27})\ r^{-12/7},
\end{split}
\end{equation}
where $p_{i}$ corresponds to the $i$th-order term in $T$. At small $r$ (\textit{i.e.,} \hl{$\lesssim r_c$}), the temperature profile $T(r,t)$ is dominated by viscous heating, and can be approximated as
\begin{equation}
T = \left(\frac{27f_{d}\kappa_{d}\alpha_{\nu}k_{b}\Omega_{K}\Sigma^{2}}{64\sigma\mu}\right)^{1/3}.
\end{equation}
Indeed, for intermediate values of $\tilde{\alpha}\simeq10^{-3}$ and $\psi\simeq 1$ ($\alpha_{\nu}=\alpha_{DW}$), $T(r=10 AU)$ as calculated from Eq. 16 deviates from that obtained by solving the quartic equation by only $\sim 12\%$. This relative deviation drops to $\sim 1\%$ at $5 AU$. At large $r$ (\textit{i.e.,} \hl{$\gtrsim r_c$}), stellar irradiation constitutes the dominant heat generation mechanism, in which case $T(r,t)$ can be approximated by Eq. 13 alone. For the same disk, $T(r=20 AU)$ as calculated with Eq. 13 deviates from the quartic root by $\sim 1\%$. \\
\indent Having specified $\tilde{\alpha}$ and $\psi$, we seek a pair of $\lambda$ and $\gamma$ that would yield a $\Sigma$ profile which, upon substitution into the quartic equation (Eq. 15), yields a $T$ profile with a power law index concordant with the value of $\gamma$ used (\textit{i.e.,} an index of $\gamma-3/2$; recall $T\sim r^{\gamma-3/2}$). As $\gamma$ enters directly into the power law index of $\Sigma$ (Eq. 5), while $\lambda$ does so through $\xi$ (and weakly; Eq. 6), the former is expected to be the more crucial parameter in this exercise. For a given $\tilde{\alpha}$ and $\psi$, we create a raster of $\Sigma$, and thus $T$, profiles using values of $\gamma$ ranging from 0.4 to 0.9 (corresponding to $T$ power law indices of $-0.6$ to $-1.1$), and $\lambda$ ranging from 2 to 5 (Tabone et al. 2022). The deviation of the power law index retrieved via a fit to one of such $T$ profiles from its respective, assumed value of $\gamma - 3/2$ constitute a measure of self-consistency, which we denote $\delta(\lambda, \gamma)$ for the sake of clarity. The values of $\delta(\lambda, \gamma)$ are compared, with the minimum designating the optimal pair of  $\lambda$ and $\gamma$ for the choice of $\tilde{\alpha}$ and $\psi$. The results of this procedure for three cases spanning the range explored for $\tilde{\alpha}$ and $\psi$ are shown in Fig. 2. \\
\indent As expected, most of the variation in $\delta$ (given $\tilde{\alpha}$ and $\psi$) is attributed to $\gamma$, motivating us to adopt an intermediate value of $\lambda=3.5$ regardless of $\tilde{\alpha}$ and $\psi$ for simplicity. While it appears there is minimal variation in the optimal $\gamma$ (henceforth denoted $\gamma_{opt}$) between Figs. 2a, 2b, \& 2c ($\lesssim0.05$), $\gamma_{opt}$ can vary appreciably between turbulence-dominated disks (low $\psi$) of high $\tilde{\alpha}$ and wind-dominated disks (high $\psi$) of low $\tilde{\alpha}$. This can be seen in Fig. 3a, which showcases $\gamma_{opt}$ across parameter space, with each pair of $\tilde{\alpha}$ and $\psi$ (each pixel, if you will) a product of the computational exercise described above\textemdash simplified, of course, by having $\lambda$ constant at $3.5$. Values of $\gamma_{opt}$ range from $\simeq0.50$ to $0.83$, corresponding to $T$ power law indices of $\simeq-0.67$ to $-1.00$. In the top-left region of Fig. 3a, bounded by the black dashed lines, $\gamma_{opt}$ values are extrapolated because our algorithm yielded erratic and spurious results therein. This likely stems from inconsistencies in solving the quartic equation for $T$. In Fig. 3b, we depict $\delta$ values (expressed in \%) \hl{obtained using respective $\gamma_{opt}$ values from Fig. 3a, all of which lie $\lesssim3\%$.} \\
\indent Before moving on, we note that excluding the bottom right corner of Fig. 3a where $\alpha_{\nu}=\tilde{\alpha}/(\psi+1)\sim 10^{-5}$, $\gamma_{opt}$ ranges from $\simeq 0.50$ to $0.61$, corresponding to $T$ indices of $\simeq -0.89$ to $-1.00$. This is similar to the $T$ index of $-0.9$ for the steady-state, ``actively" heated viscous disk of \hl{BM22}, for which the power law index of $\Sigma$ is $\simeq -0.6$. Recalling that the steady-state (not exponential) term in Eq. 5 scales as $r^{\xi-\gamma}$, it is clear that for $\psi\ll1$, the present model naturally reduces to the viscous disk model of \hl{BM22} for a wide range of $\tilde{\alpha}$, as the power law index of $\Sigma$ then becomes $(\xi-\gamma)\simeq-\gamma\sim -0.6$. This smooth transition in the turbulent regime despite the inclusion of passive irradiation, along with the retrieval of $\gamma_{opt}\sim0.6$ for high $\psi$ and $\tilde{\alpha}$ hints at (i) the significance of the disk at small $r$ (\textit{i.e.,} where the steady-state term in $\Sigma$ and viscous heating dominate) in establishing the power law indices of $\Sigma$ and $T$, and (ii) reaffirms the notion that the form of $\Sigma(r,t)$ is primarily controlled by $\gamma$ (the ``mass ejection index" $\xi$ is small for a wide range of $\psi$ so long as $\lambda\gtrsim 2$; see Fig. 3 in Tabone et al., 2022). It appears only for $\alpha_{\nu}\lesssim 10^{-4}$ (see Eq. 16) does irradiation begin to influence the form of $T(r,t)$ \hl{(higher $\gamma$ corresponds to flatter $T$ profile; Fig. 3a)}.\\

\begin{figure*} 
\centering
\scalebox{0.675}{\includegraphics{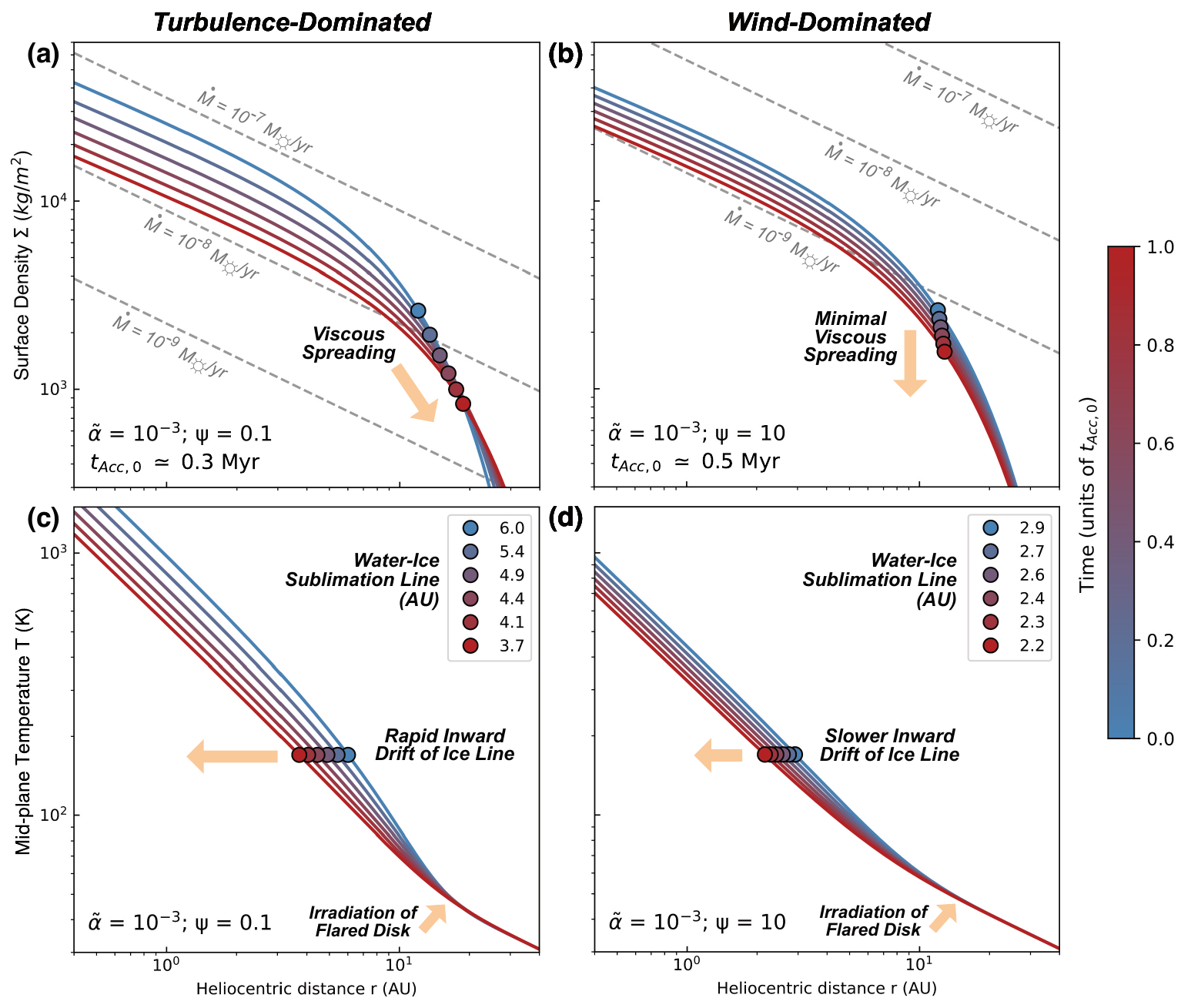}}
\caption{\footnotesize\textbf{Profiles of the surface density $\Sigma(r)$ and mid-plane temperature $T(r)$ for the turbulence-dominated (a,c) and wind-dominated (b,d) fiducial disks as a function of time.} Time steps (moving from blue to red) are in units of the initial accretion timescale as calculated with Eq. 10. The turbulence-dominated disk is characterized by viscous spreading and the faster inward drift of the \hl{ice-line} relative to the wind-dominated disk, in which viscous spreading is virtually absent. The flared disk profile facilitates interception of stellar irradiation which prevents $T(r,t)$ from plummeting with $\Sigma(r,t)$. See Section 2.3.}
\label{fig:Figure 4}
\end{figure*}

\subsection{Fiducial $\Sigma$ and $T$ Profiles}
\indent With all parameters defined, we depict the $\Sigma(r,t)$ and $T(r,t)$ profiles for two fiducial cases corresponding to (i) a turbulence-dominated disk ($\tilde{\alpha}\simeq 10^{-3}$, $\psi\simeq$ 0.1) (Figs. 4a, c) and (ii) a MHD wind-dominated disk ($\tilde{\alpha}\simeq 10^{-3}$, $\psi\simeq$ 10) (Figs. 4b, d). The obtained $\gamma_{opt}$ for these disks are $\simeq 0.55$ and $\simeq 0.69$, respectively, corresponding to $T$ power law indices of $\simeq0.95$ and $\simeq 0.81$. The gray dashed lines in Fig. 4a and 4b represent the $\Sigma$ profiles from the model of \hl{BM22} for different values of $\dot{M}$. Turbulence facilitates angular momentum \textit{transport} between disk materials, while disk winds facilitate mass and angular momentum \textit{removal} from the disk. Consequently, viscous spreading accompanies stellar accretion in the presence of turbulence, and is virtually absent is a wind-dominated disk. The extent of viscous spreading is controlled by Eq. 8, where $\psi$ governs the evolution rate of $r_{c}(t)$ with every step of $t_{acc,0}$. The $\Sigma$ profiles in the turbulence-dominated disk match those of the steady-state viscous disk for $\dot{M}$ between $10^{-8}$ to $10^{-7}M_{\odot}$/yr, deviating from them past $r\simeq 10 AU$ due to the physical cutoff of the disk implemented via the exponential decay term in Eq. 5. In the wind-dominated disk, flattening of the $\Sigma$ profile at small $r$ due to the increase in the mass ejection index $\xi$ is resisted by higher $\gamma_{opt}$. The increase in $\Sigma$ for a given $\dot{M}$ \hl{(from BM22; driven only by $\alpha_{\nu}$)}, as well as the overall decrease in temperatures in the viscously-heated, small-$r$ portion of the disk (by a factor of $\sim2$ for $t=0$) reflects the decrease in $\alpha_{\nu}$ by close to an order of magnitude. This in turn yields ice-lines at shorter distances from the star and an increase in $t_{Acc,0}$. The slower evolution as evidenced by the closer proximity of $\Sigma$ and $T$ profiles across time is attributed to high $\psi$, and can be understood by inspection of Eq. 7. This, and to a lesser extent the higher $t_{Acc,0}$, culminates in an inward drift rate of the ice-line that is much slower ($\sim1.4$ AU/Myr) than that in the turbulence-dominated disk ($\sim7.7$ AU/Myr). \\
\indent The inclusion of stellar irradiation ensures $T(r,t)$ beyond $r_{c}(t)$ does not plummet with $\Sigma(r,t)$, \hl{as} would be the case if only viscous heating was present (Eq. 16). As higher temperatures correspond to greater turbulence (Eq. 17) and particle drift velocities (Eq. 18) in the Shakura-Sunyaev prescription, irradiation indirectly stalls particle growth in the outer disk (see Section 3.5). 
\section{The Stokes Number}
\indent The Stokes number is intrinsically related to particle size, and its characteristic value at any given region in the disk is defined through the competition between several barriers to particle growth therein (Birnstiel et al., 2012). Each of these barriers predicts a maximal $St$, and thus size, attainable by particles undergoing hit-and-stick collisions, and the characteristic $St$ is given by the minimum of these maximal values. In other words, particle growth starting at the micron-scale is halted by the first barrier it encounters, and the mechanism enforcing that barrier is location-dependent. Here, we provide an overview of the said barriers, and construct the characteristic $St(r,t)$ profiles for our fiducial disks. We emphasize that the characteristic particles at a given location in the disk are those that hold most of the solid mass, not necessarily those that are most numerous. The persistence of optically-thick disks over millions of years suggests the prolonged preeminence of micron-sized dust (Williams \& Cieza, 2011), in turn hinting at the \hl{prevalence} of particle collisions/fragmentation in disks.

\subsection{Turbulent Fragmentation}
\indent Underlying our work here is the assumption that hit-and-stick collisional growth operates at a sufficiently high efficiency such that most particles have $St$ close to the maximal attainable value. Numerical simulations have shown that this is indeed the case when $St$ is limited by the most generic of barriers: turbulent fragmentation (Birnstiel et al., 2011). The maximal $St$ in this regime, denoted $St_{tf}$, is expressed though the equivalence between the average relative velocity of similar-sized dust particles in turbulent motion $\Delta v_t = \sqrt{3\alpha_{\nu}St}c_s$ (valid for $St\lesssim1$) and the the fragmentation threshold velocity of those particles $v_f$ (Ormel \& Cuzzi, 2007): 
\begin{equation}
St_{tf} = \frac{1}{3}\frac{v_{f}^{2}}{\alpha_{\nu}c_s^{2}}
\end{equation}
\indent Laboratory experiments suggest that the onset of silicate dust fragmentation occurs at $v_{f,rock}\simeq$ 1 m/s (Blum \& Münch, 1993; Güttler et al., 2010), and we adopt this value for particles inward of the \hl{ice-line}. Icy particles beyond the ice-line, \hl{constituting ice-rock aggregates}, are characterized by surface energies greater than their silicate counterparts, and thus expected to be ``sticker," fragmenting at higher collisional velocities (Gundlach et al., 2011; Gundlach \& Blum, 2015). We adopt $v_{f,ice}\simeq5$ m/s for these particles, a factor of two lower than that used by \hl{BM22}. This is because in our model,  $v_{f,ice}\simeq10$ m/s yields icy \textit{boulders} up to the dm- to m-scale in the turbulence- and wind-dominated fiducial disks, respectively, at odds with astronomical observations suggesting dust layers composed of abundant cm-sized particles \hl{(\textit{e.g.,} Testi et al., 2003; Ricci et al., 2010)}. The viscous disk model of \hl{BM22} retrieved cm-sized icy particles because it employed a steady-state $\Sigma$ profile, that is, one without an exponential decay term that would otherwise decrease $\Sigma$ and $T$ (solely from viscous heating; Eq. 16), \hl{decrease} $\Delta v_{t}$, and thus increase $St_{tf}$ and particle sizes beyond the cm-scale. We note that $v_{f,ice}\simeq5$ m/s is substantially lower than the $\simeq15$ to $20$ m/s obtained assuming cm-sized particles composed of ``weak ice" from the hydrodynamic collision simulations of Leinhardt \& Stewart (2009). This discrepancy is insignificant to the key conclusions of our work, which pertain to the inner disk \hl{(see Section 4.3)}. Moreover, the assumption that ice fragments at higher velocities than silicates has been called into question by recent laboratory experiments at temperatures appropriate for most of the outer protoplanetary disk ($\lesssim150K)$, which suggests icy particles therein are comparable to their silicate counterparts in tensile strength (Gundlach et al., 2018; Musiolik \& Wurm, 2019; Pinilla et al., 2021). Our choice of a lower $v_{f,ice}$ thus reflects a step in the right direction. It also bears mention that $v_{f,rock}$ is expected to increase under high ($\gtrsim1000K$) temperatures (Pillich et al., 2011), and $v_{f}$ in general depends on porosity, composition (itself related to temperature), and size (\textit{e.g.,} Benz \& Asphaug, 1999; Leinhardt \& Stewart, 2009; Beitz et al., 2011). We omit these complications here for simplicity, but briefly discuss them in the context of particle sizes obtained in Section 4.1.2. We consider the potential implications of the temperature dependence more deeply in Section 5.

\subsection{Radial Drift}
\indent Drift towards the Sun, or host star, poses another barrier to particle growth. The \hl{H$_2$-He} gas composing the bulk of protoplanetary disks is partially supported against gravity by a \hl{radial} pressure gradient, and thus orbits the central star at sub-Keplerian azimuthal velocities. The larger a particle grows \hl{(for $St\lesssim 1$)}, the more susceptible it is to drag from headwind induced by the sub-Keplerian gas flow around it (Weidenschilling, 1977). The increase in drag (\textit{i.e.,} angular momentum loss per unit time) accelerates the particle's drift inward. Thus, a characteristic particle size (and $St$) for a local dust population may be defined through a balance between hit-and-stick collisional growth and radial drift. Consideration of the radial and azimuthal equations of motion for an orbiting particle undergoing drag forces leads to the well-known expression for its radial drift velocity, given by (Nagakawa et al., 1986)
\begin{equation}
v_{r} = \frac{St^{-1}v_{r,gas} + |\epsilon|(c_s^{2}/v_{K})}{St+St^{-1}},
\end{equation}
where $|\epsilon|$ is the power law index of the disk's midplane pressure profile $P=\rho_{g} c_{s}^{2}\sim r^{\epsilon}$, $\rho_{g} = \Sigma/(\sqrt{2\pi}h)$ being the gas density. The two terms on the RHS represent the contribution to $v_{r}$ from advection with the accretionary gas flow, and the sub-Keplerian headwind, respectively. Recall that $v_{r,gas} = (3/2)(2-\gamma)^{2}\tilde{\alpha}c_{s}^{2}v_{K}^{-1}$, and note that for (aerodynamically) small particles that are tightly coupled to the surrounding gas ($St<<1$), $v_{r}\simeq v_{r,gas}$ meaning these particles are virtually advected towards the central star with the gas. At very large sizes ($St>>1$), keeping in mind acceleration from drag scales with the inverse of particle \hl{size/radius} $s$ ($\sim 1/s$), particles/boulders decouple from the gas with $v_{r}\simeq 0$. In between these two extremes, $v_{r}$ rises to a maximal value of $\simeq (1/2)|\epsilon|c_{s}^{2}v_{K}^{-1}$, which occurs at $St\simeq 1$. In the turbulence- and wind-dominated disks at $t=0$ and $r=1$ AU, this evaluates to $\simeq 140$ m/s and $\simeq 60$ m/s, respectively. Physically, the peak in $v_r$ describes particles that are sufficiently coupled to the gas to travel at a sub-Keplerian velocity, and yet have enough mass such that the centrifugal force acting on them fails to counteract gravity. This peak constitutes the so-called ``meter-size barrier" (Weidenschilling \& Cuzzi, 1993; Blum \& Wurm, 2008), which depletes the local disk region of $St\simeq 1$ particles rapidly relative to the growth timescale ($\tau_{grow}\sim s/\dot{s}$) of those particles. This is made clear with an example. Assuming particles are in the Epstein drag regime, applicable for sizes $s\lesssim 9/4$ times the mean free path of gas molecules $\lambda_{mfp}$ (Armitage, 2020; see Section 4.1.2 below), the growth timescale takes the simple form $\tau_{grow} = 1/\Omega_{K}f_{d}$, independent of $s$ (Birnstiel et al., 2012). The ``meter-size" drift timescale $\tau_{drift}$, assuming a ``locality" in the disk has a radial (bin) width of $\simeq 1$ AU, is simply the ratio of $1$ AU to the peak $v_{r}$. Evaluation of $\tau_{drift}/\tau_{grow}$ at $r\simeq 20$ AU yields $\sim0.1$ for both the turbulence- and wind-dominated disks, meaning that $St\simeq 1$ particles therein are doubling their mass about ten times slower than the time it takes for them to drift across a hundred million kilometers. For simplicity, we set the maximal $St$ in the drift-limited regime, denoted $St_{rd}$, to unity.

\subsection{Fragmentation from Relative Drift Velocities}
\indent In disks with low turbulence, fragmentation of colliding particles may be facilitated by their relative drift velocities (Birnstiel et al., 2012). The maximal $St$ in this alternate fragmentation-limited regime, denoted $St_{df}$, is expressed through an equivalence between some characteristic relative drift velocity between similar-sized particles $\Delta v_{r}$ and the fragmentation threshold velocity $v_{f}$. The radial drift velocity of a particle in the disk is given by Eq. 18. Assuming the smaller of the colliding particles has $St_{2} = $$NSt_{1}$, where $N<1$, their relative drift (or collision) velocity is
\begin{equation}
\begin{split}
\Delta v_{r} = \frac{c_s^{2}}{v_{K}}&\left(\frac{3\tilde\alpha(2-\gamma)^{2}/2 + |\epsilon|St_{1}}{St_{1}^{2}+1}\right.\\&\ \ \ \ \ \ \ \ \left.- \frac{3\tilde\alpha(2-\gamma)^{2}/2 + |\epsilon|NSt_{1}}{N^{2}St_{1}^{2}+1} \right).
\end{split}
\end{equation}
Setting $\Delta v_{r} = v_{f}$ and rearranging for $St_{1}$ (\textit{i.e.,} $St_{df}$), Eq. 26 yields a quartic equation with coefficients given by
\begin{equation}
\begin{split}
p_{4} = \frac{v_{f}v_{K}N^{2}}{c_s^{2}} &;\ p_{3} = |\epsilon|(N-N^{2})\\
p_{2} = \frac{v_{f}v_{K}(N^{2}+1)}{c_s^{2}} &+ \frac{3\tilde{\alpha}(2-\gamma)^{2}(1-N^{2})}{2}\\
p_{1} = |\epsilon|(N-1)&;\ p_{0} = \frac{v_{f}v_{K}}{c_s^{2}}
\end{split}
\end{equation}
where $p_{i}$ corresponds to the $i$th-order term in $St_{df}$. We set N = 0.5 (Birnstiel et al., 2012), and obtain the solution to $St_{df}$ at each location in the disk numerically as done for $T(r,t)$ (Eq. 5). As in the case of turbulent fragmentation, we adopt $v_{f,rock}\simeq 1$ m/s and $v_{f,ice}\simeq 5$ m/s for particles inward and outward of the ice-line, respectively. To ensure the quartic equation at each location in our fiducial disks is solved, we substituted the $St_{df}(r,t)$ obtained back into Eq. 19 and checked that $\Delta v_{r}(r,t) \simeq v_{f}(r,t)$. 
\begin{figure*} 
\centering
\scalebox{0.7}{\includegraphics{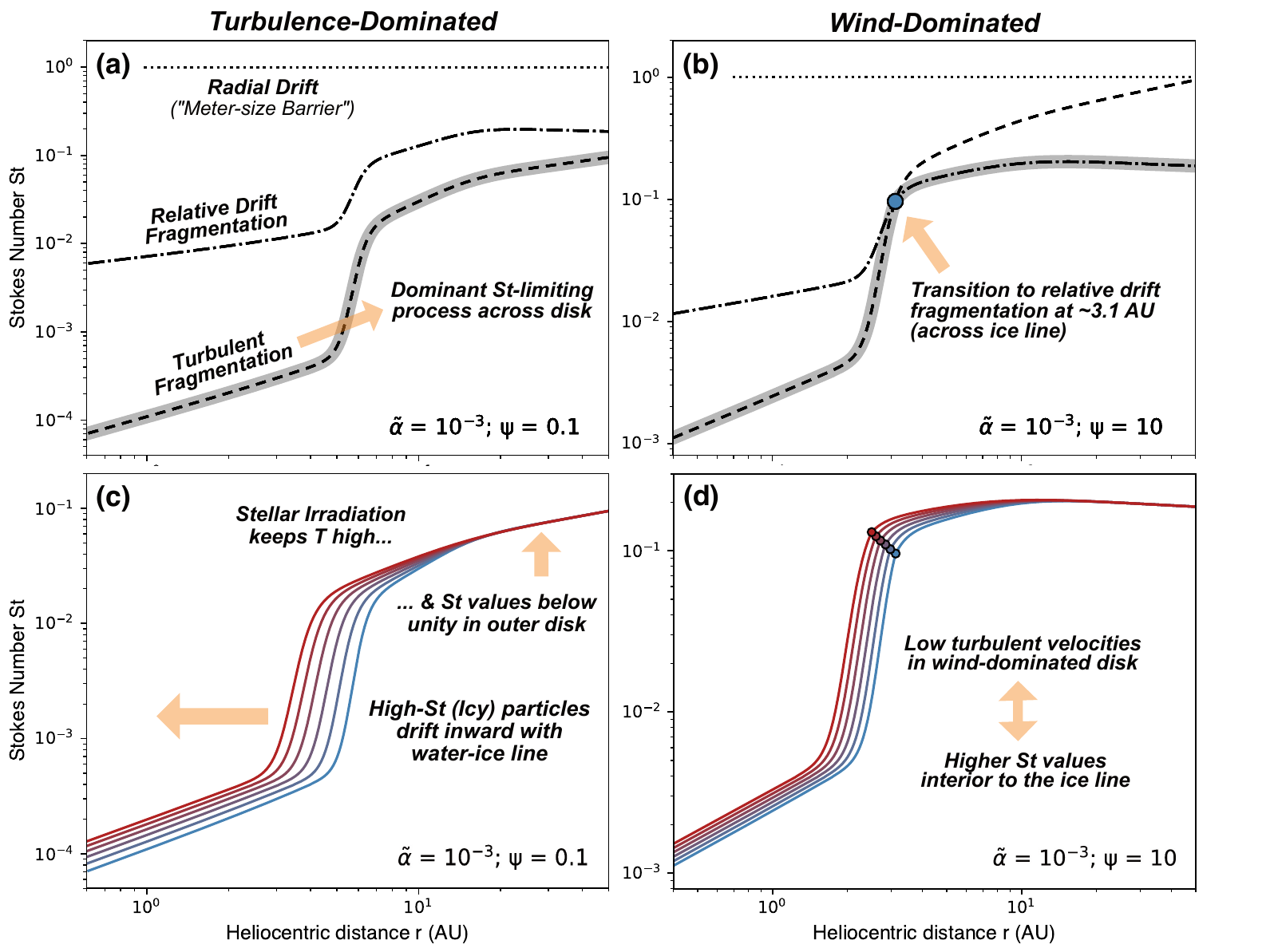}}
\caption{\footnotesize\textbf{Profiles of the maximal Stokes number predicted by competing barriers to particle growth at $t = 0$ (a \& b) and the characteristic Stokes number over time (c \& d) for the two fiducial disks.} In (a) and (b), the grey band defines the minimum of the three maximal $St$ profiles (\textit{i.e.,} the characteristic $St$ profile). As in Fig. 4, time steps in units of $t_{Acc,0}$ are represented in blue to red. Turbulent fragmentation is the dominant growth-limiting mechanism in the turbulence-dominated disk and inner wind-dominated disk. In the latter, fragmentation from relative drift velocities limits growth beyond $\sim 3.1$ AU. The ice-line constitutes a major transition point in $St$, owing to the higher fragmentation threshold velocity of ice ($\simeq 5$ m/s) assumed. In (c), the flared disk profile keeps temperatures high in the outer disk, prohibiting a rapid rise in $St_{tf}$ (Eq. 17). In (d), low temperatures, and hence turbulent velocities, result in higher $St$ values in the inner disk. Points on $St$ profiles represent the transition from $St_{tf}$ to $St_{df}$ at each time step. In both disks, high-$St$ particles drift inward with the cooling of the disk. See Section 3.5.1.}
\label{fig:Figure 5}
\end{figure*}
\subsection{Bouncing Collisions}
\indent A host of laboratory experiments have demonstrated that a possible outcome of particle collisions, in addition to sticking and fragmentation, is bouncing (Blum \& Münch, 1993; Güttler et al., 2010).  Dust collision models suggests that bouncing can halt growth before particle velocities are sufficiently high for fragmentation to be relevant (Zsom et al., 2010). Despite experimental support for such a ``bouncing barrier,"  its significance to dust evolution in protoplanetary disks remains nebulous. Numerical simulations indicate that bouncing may only occur during collisions involving compact aggregates with relatively high coordination numbers (\textit{i.e.,} the number of neighboring contacts per particles in the aggregate), unrepresentative of dust particles characterized by low bulk densities ($\lesssim100 $ kg/m$^{3}$) as generally expected in disks (Wada et al., 2011). It is important to note, however, that the bulk densities of such particles remains poorly constrained. It is possible, for instance, that the majority of particles were sub-mm- to mm-sized chondrules, or at least chondrule-like in their densities ($\rho_{\circ}\gtrsim 3000$ kg/m$^{2}$; \textit{e.g.,} Friedrich et al., 2015 and references therein) such that bouncing was significant. Such a proposition is supported by the abundance of chondrules in primitive meteorites from both the inner and outer SS. The caveat here is that chondrules may not be as ubiquitous in the SS as the meteorite record suggests. In particular, it has been suggested that most of the material in the SS is CI-(Carbonaceous Ivuna)like, with densities and tensile strengths too low to survive passage through Earth's atmosphere (Sears, 1998; Yap \& Tissot, 2023). \hl{This view is supported by the similarity between CI chondrites and asteroid Ryugu samples returned by Hayabusa2 (Yokoyama et al., 2022), as well as \textit{in-situ} thermal inertia measurements of asteroid Bennu's surface, indicating abundant high-porosity boulders (Rozitis et al., 2020).} Even with bouncing in effect, it is unclear how easily it may be overcome. Windmark et al. (2012) postulated that cm-sized particles may catalyze growth beyond the mm/sub-mm-scale to which particles are confined by bouncing collisions, and recently, Steinpilz et al. (2020) argued that the bouncing barrier is insignificant in consideration of triboelectric charging of colliding particles. \\
\indent Given the uncertainty surrounding the bouncing barrier as well as its postulated precedence over fragmentation and radial drift, we do not include it in our primary analysis, but assess its implications for our results separately and only in the turbulence-dominated disk. Here, we derive the maximal $St$ attainable in the bouncing-limited regime (denoted $St_{b}$) by equating the turbulent velocity dispersion of dust particles $\Delta v_{t}=\sqrt{3\alpha_{\nu}St}c_{s}$ to the mass-dependent bouncing threshold velocity $\Delta v_{b}$ deduced by Weidling et al. (2012) from collision experiments. With the target mass $\gtrsim$ the mass of the projectile $m_{proj}$, $\Delta v_{b}$ (in m/s) is given by
\begin{equation}
\Delta v_{b} = 0.01 \left(\frac{m_{proj}}{m_{b}}\right)^{-5/18} , 
\end{equation}
where the normalizing constant is $m_{b}$ is calibrated at $\simeq3.3\times 10^{-6}$ kg. We assume that particles limited by bouncing are confined to the Epstein drag regime and confirm this assumption \textit{a posteriori}. In this regime, the relationship between $St$ and $s$ is given by
\begin{equation}
St(s)_{Eps} = \sqrt{\frac{\pi}{8}}\frac{s\rho_{\circ}\Omega_{K}}{\rho_{g}c_{s}}.
\end{equation}
Here, $\rho_{\circ}$ is the material density of the particle, for which we assumed typical values for compact rock and water-ice aggregates ($\rho_{\circ,rock}\simeq 3300$ kg/m$^{3}$; $\rho_{\circ,ice}\simeq 1000$ kg/m$^{3}$) given the aforementioned uncertainties in this parameter. Note that $\rho_{\circ}$ does not enter into the expressions derived for $St_{tf}$ (Eq. 17) nor $St_{df}$ (Eq. 20). Assuming spherical particles with radii $s$, $St_{b}$ takes the form
\begin{equation}
St_{b} \simeq 0.01\left(\frac{m_{b}\rho_{\circ}^{2}\Omega_{K}^{3}}{\rho_{g}^{3}\alpha_{\nu}^{9/5}c_s^{33/5}}\right)^{5/24}.
\end{equation}

\begin{figure} 
\scalebox{0.525}{\includegraphics{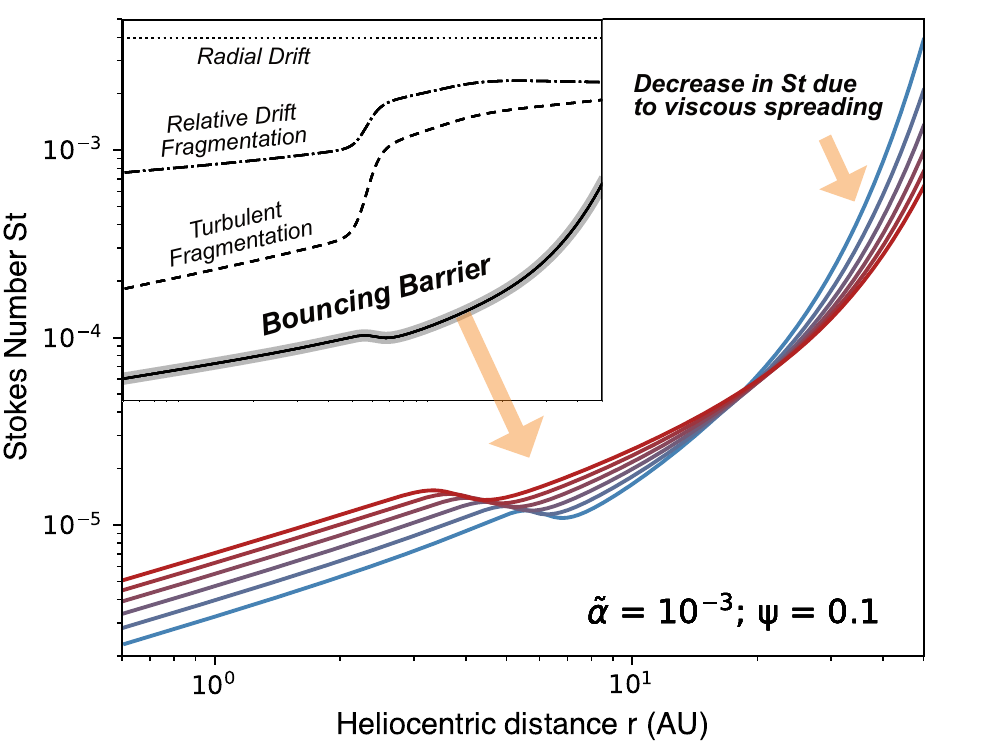}}
\caption{{\footnotesize\textbf{Profiles of the characteristic Stokes number in the bouncing-limited turbulent disk over time.} The bouncing barrier limits particles to much lower $St$ than turbulent fragmentation. Here, the transition across the ice-line reflects the change in particle material density, as opposed to fragmentation threshold velocities as in Fig. 5. Viscous spreading increases $\rho_{g}$ in the outer disk over time (Eq. 23), leading to lower $St$. See Section 3.5.2.}}
\label{fig:Figure 6}
\end{figure}

\subsection{Stokes Number Profiles}
\subsubsection{Without Bouncing}
\indent The $St(\tilde{\alpha},\psi,r)$ profiles constructed for our fiducial cases are shown in Fig. 5. As \hl{BM22} have shown, $St$ rises dramatically across the \hl{ice-line} owing to the higher fragmentation velocities of \hl{icy relative to silicate particles}. In the turbulence-dominated case, it is clear that turbulent fragmentation sets the characteristic $St$ throughout the disk (Fig. 5a). While particles interior to the ice-line are confined to $St< 10^{-3}$, those beyond it possess $St$ values $\gtrsim10^{-2}$, indicating strong dust-gas coupling in the former, and weak coupling in the latter. Stellar irradiation of the flared outer disk prohibits temperatures from plummeting with $\Sigma$ and hence $St_{tf}$ from rising dramatically past $r_{c}(t)$. The lack of a decrease in $St$ over time for $r\gtrsim r_{c}(t)$ (Fig. 5c), as would be expected from viscous spreading (which increases $\Sigma$ and $T$ in the outer disk, translating to to faster turbulent velocities in the Shakura-Sunyaev prescription), reaffirms the dominance of irradiation over viscous heating therein.\\
\indent The maximal $St$ predicted by the two fragmentation barriers increase as turbulence gives way to disk winds (Fig. 5b, d). In the wind-dominated disk, particles possess characteristic $St$ values about an order of magnitude greater than those in the turbulence-dominated disk. Silicate particles in the inner disk are confined to $St<10^{-2}$, while icy particles in the outer disk possess $St$ $\sim0.1$. Although turbulent fragmentation remains the $St$-limiting process for the former, fragmentation due to relative drift velocities dominates beyond $\sim3.1$ AU and limits the growth of the latter. In both fiducial disks, high-$St$ icy particles drift inwards over time with the ice-line, owing to the cooling of the disk \hl{accompanying} the decrease in $\Sigma$ (Eq. 7). \\

\subsubsection{With Bouncing}
\indent The $St(\tilde{\alpha},\psi,r)$ profile for the turbulence-dominated disk with the inclusion of the bouncing barrier is shown in Fig. 6. As expected, bouncing sets the characteristic $St$ across the entire disk, confining particles up to $r_{c}\sim 20$ AU below $St\simeq10^{-4}$ and limiting growth far before turbulent fragmentation takes effect. The slight depression in $St$ across the sublimation line reflects the change in particle density, and viscous spreading leads to lower $St$ over time in the outer disk, owing to the increase in $\Sigma$ and thus $\rho_{g}$ (Eq. 23). The parametrization of the bouncing threshold velocity given in Eq. 21 is based on experiments on millimeter-sized aggregates composed of micron-sized poly-disperse SiO$_{2}$ particles. \hl{Assuming icy particles are `stickier" than silicates, it is likely that $\Delta v_{b,ice}$  $\gtrsim$ $\Delta v_{b,rock}$. This would result in a dichotomy in $St_{b}$ akin to that for $St_{tf}$ and $St_{df}$ in our two fiducial disks.}

\section{Results \& Discussion}
\subsection{Dust Properties}
\subsubsection{Dust Scale Height \& Drift Velocities}
\indent The $St(\tilde{\alpha},\psi,r)$ profiles derived above permit determination of the vertical distribution, radial drift rate, and physical size of the characteristic dust particles across their disks. Here, we compare how these three features vary between the turbulence- and wind-dominated disks absent of the bouncing barrier, before extending our analysis to the bouncing-limited, turbulent disk. \\
\indent At the macroscopic level, dust in protoplanetary disks can be treated as a trace fluid separate from the bulk \hl{H$_2$-He} gas, modulo the Schmidt number ($S_{c}\simeq 1)$ and regions where pressure bumps may lead to $\rho_{\circ}\gtrsim\rho_{g}$. By virtue of a balance between turbulent diffusion and gravitational settling, dust particles establish a solid sub-disk with a $St$-dependent scale height $h_{\circ}$($\lesssim h$) given by (Dubrulle et al., 1995)
\begin{equation}
{h_{\circ}} = \frac{h}{\sqrt{1+St/\alpha_{\nu}}}.
\end{equation}
As mentioned above, most of the dust mass is assumed to be held in particles with $St$ close to the maximal value allowed by the growth barriers described. However, the profile of $h_{0}$ as calculated using the characteristic $St$ values from above is not that which dominates the observational appearance of disks. The optical thickness of a disk is primarily attributed to micron-sized dust, which, tightly coupled to the gas as they are  (with $St\ll 1$), have $h_{\circ}\simeq h$. Fig. 7 depicts the profiles of the normalized scale height $h_{\circ}/h$ for our two fiducial cases. Plotted as well are profiles of the normalized radial drift rate $|v_{r}/v_{K}|$ as calculated using Eq. 18. \\
\indent In both the turbulence- and wind-dominated disks, the water ice sublimation line constitutes a critical transition point in the degree of dust settling and particle drift regimes. Generally, silicate dust particles in the inner disk are well coupled to their surrounding gas, and form a relatively dispersed and well-mixed layer ($h_{\circ}\gtrsim 0.1 h$) about the disk midplane. Their inward drift is predominantly facilitated by advection with the accretionary gas flow, and so proceeds at rather slow velocities ($< 10^{-4} v_{K}$). In the turbulence-dominated disk (Fig. 7a), strong dust-gas coupling ($St<10^{-3}$) in the inner disk manifests as a virtual equivalence between $h_{0}$ and $h$, and between $v_{r}$ and $v_{r,gas}$. The increase in disk winds, and hence $St$, leads to greater settling ($h_{0}\lesssim 0.5 h$) and faster (by a factor of a few) inward drift relative to the gas (Fig. 7b).\\
\indent Icy particles in the outer disk, being larger and loosely coupled to the gas, settle into a thin layer ($h_{0}\lesssim0.1h$) at the mid-plane. Their inward drift is driven primarily by headwind drag, and proceeds at velocities that are substantially greater fractions ($>10^{-4} v_{K}$). As with silicate particles, icy particles in the wind-dominated disk settle more readily, and drift faster, than those in the turbulence-dominated disk.\\
\indent As $St_{df}$ tapers off at $\sim 0.1$ in both fiducial disks (Figs. 5a \& 5b), only past the point where there is no solution to $St_{df}$ can $St$ approach unity (\textit{i.e.,} $St_{rd}$) in the outer disk. While this point may be too far in the outskirts of the disk to be meaningful, it is bound to exist since $c_s^{2}/v_{K}$ falls with $r$ (Eq. 19), unless $v_{f}$ falls equally fast or faster (say due to the aforementioned temperature dependence). Where $St_{rd}$ becomes the dominant growth limiting process, $v_{r}$ will take on its peak value of $\sim (1/2)|\epsilon|c_s^{2}/v_{K}$.\\
\begin{figure*}
\centering
\scalebox{0.72}{\includegraphics{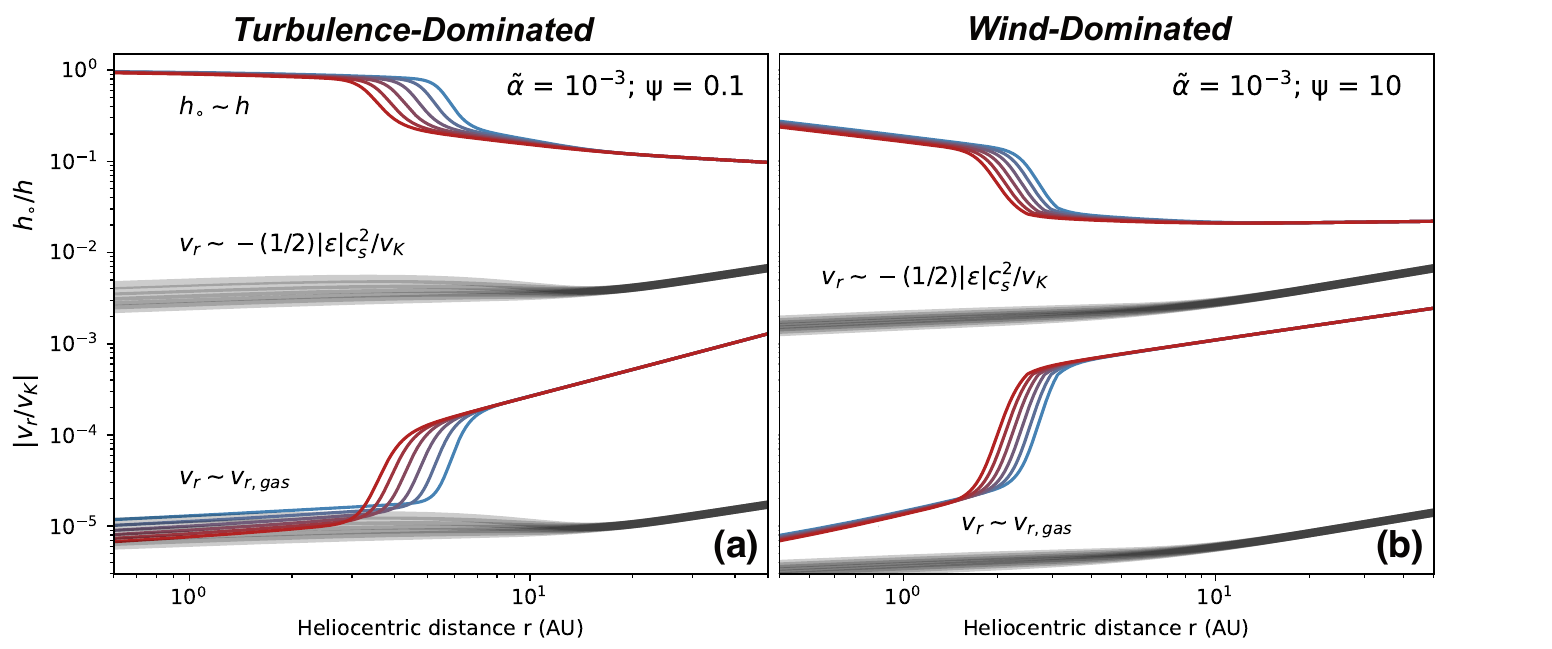}}
\caption{\footnotesize\textbf{Profiles of the normalized scale height and radial drift velocity of the characteristic dust particles in the (a) turbulence- and (b) wind-dominated disks. } Grey bands represent the upper and lower bounds to the drift velocity for all six time steps, the former corresponding to $St\simeq1$. The degree of dust settling as well as the headwind contribution to the dust drift velocity change dramatically across the ice-line. In general, silicate particles in the inner disk are vertically dispersed and drift inward with at velocities close to that of the surrounding gas. Icy particles concentrate in a thin sub-disk at the mid-plane, and drift inwards faster than the surrounding gas. See Section 4.1.1.}
\label{fig:Figure 7}
\end{figure*}
\indent Figs. 8a \& 8b shows the profiles of $h_{0}/h$ and $|v_{r}/v_{K}|$ for the bouncing-limited turbulent disk. As expected from the small \hl{$St$/$s$} (Figs. 6, 8c) to which the bouncing barrier relegates particles across the disk, the scale height and drift velocity of dust is essentially identical to those of the gas. Only beyond the outer edge at $\sim r_c$ do particles begin to settle and drift faster than their surrounding gas. Even then, $h_{\circ}\gtrsim0.5 h$ and $v_{r}$ is only a factor of a few greater than $v_{r,gas}$.
\subsubsection{Dust Particle Sizes}
\indent The relationship between $St$ and particle \hl{size/radius} $s$ is dependent on the drag law in effect.  In the Epstein drag regime, valid for $s\lesssim 9/4\lambda_{mfp}$, $St(s)$ is given by Eq. 22. In the Stokes drag regime, for which $s\gtrsim9/4\lambda_{mfp}$, $St(s)$ is instead given by (Armitage, 2020)
\
\begin{equation}
St(s)_{Sts} = \frac{8}{3}\frac{s\rho_{\circ}\Omega_{K}}{C_{D}\rho_{g}v_{rel}},
\end{equation}
where $v_{rel} = v_{K}[1-(1-|\epsilon|c_{s}^{2}/v_{K}^{2})^{1/2}]$  is the relative velocity between dust and gas (\textit{i.e.,} the headwind velocity), and $C_{D}$ the drag coefficient dependent on the fluid Reynolds number $Re$, given by
\begin{equation}
Re = \frac{\sqrt{2\pi}sv_{rel}}{c_{s}\lambda_{mfp}} = \frac{\sqrt{2\pi}sv_{rel}n\sigma_{mol}}{c_{s}}.
\end{equation}
In the second equality, we have replaced $\lambda_{mfp}$ with $(n\sigma_{mol})^{-1}$, $n = \rho_{g}/\mu$ being the number density of gas molecules, and $\sigma_{mol}\sim 2\times 10^{-19}$ m$^{2}$ the cross-section for ($H_{2}$) molecular collisions (Armitage, 2020). For $Re\lesssim1$ (Stokes Regime I), $C_{D}\simeq 24Re^{-1}$, while for $1\lesssim Re\lesssim 800$ (Stokes Regime II), $C_{D}\simeq 24 Re^{-0.6}$. For $Re\gtrsim 800$ (Stokes Regime III), $C_{D}\simeq 0.44$ (Whipple 1972; Weidenschilling 1977). In constructing $s(r,t)$ profiles for our two fiducial disks, depicted in Figs. 9a \& 9b, care was taken to ensure self-consistency in the drag law used at each location $r$. The $s(r,t)$ profiles for the bouncing-limited turbulent disk is depicted in Fig. 8c. \\
\indent In the turbulence-dominated disk, all particles are confined to the Epstein drag regime. While the inner disk is characterized by sub-mm-scale silicate particles, the outer disk hosts icy particles millimeters to centimeters in size. Particles in the wind-dominated disk are universally larger by $\sim$two orders of magnitude, with $\sim$ one-cm particles and cm-to-decimeter-scale particle/boulders in inner and outer disk, respectively. Here, with the exception of the region around $r\sim 2.5$ AU to $\sim5$ AU, wherein particles are in the Stokes regime, all other particles are in the Epstein regime. Particles around $r\sim3.5$ AU transition from Stokes Regime II to I over an initial accretion timescale $t_{Acc,0}$ of $\sim 0.5$ Myr (Fig. 4b). Particles in the bouncing-limited turbulent case are, of course, deep within the Epstein regime, with sizes in the tens of microns. Although not shown, bouncing-limited icy particles in the wind-dominated disk reach sizes of around a hundred microns.\\
\indent The large, decimeter-sized icy boulders predicted right past the ice-line in the wind-dominated disk may be an artifact of assigning a fixed fragmentation threshold velocity $v_{f}\simeq 5$ m/s to all icy particles beyond the ice-line, regardless of locally variable disk conditions. Stated differently, they reflect negligence of the effects temperature and particle size have on $v_{f}$. We note, nonetheless, that experimental and theoretical efforts to understand particle growth in disks have shown that dust grains can feasibly grow up to decimeters in size by hit-and-stick collisions (Blum \& Wurm, 2008).\\
\indent With respect to particle size, for colliding particles $\lesssim 100$ meters in size, the threshold for catastrophic disruption $Q_{D}\sim v_{f}^{2}\sim$ $St_{tf}/St_{df} = f(s)$  decreases with $s$ within the so-called ``strength dominated" regime (Benz \& Asphaug, 1999; Leinhardt \& Stewart, 2009; Armitage, 2020). That is, the larger a particle grows, the less resistant it is to fragmentation (\textit{i.e.,} lower $v_f$). The most self-consistent way to evaluate $s$ at each location in the disk (when growth is fragmentation-limited) is to seek out a value of $v_{f}\sim \sqrt{Q_{D}}$ there which, along with local disk conditions, yields a value of $s$ concordant with the $Q_{D}$ used. The implementation of this procedure, of course, assumes a reliable relationship between $Q_{D}$ and $s$ has been established. Given (i) the uncertainties that still remain concerning the nature and behavior of colliding dust particles in protoplanetary disks and (ii) the observation that $Q_{D}$ seems to change by only a factor of a few with a change in $s$ over two orders of magnitude (Leinhardt \& Stewart, 2009), we had omitted the size dependence of $v_{f}$. More crucial to $v_{f}$ and thus the dubious decimeter boulders we predict for the wind-dominated disk is the temperature dependence. If $v_{f,ice}$ evolves quickly to $\sim v_{f,rock}$ past the ice-line, most of the icy particles at $r\lesssim 20$ AU  would be mm- and cm-sized in the turbulence- and wind-dominated disks, respectively. While a dichotomy in $St$ would be absent \hl{(apart from a ``bump" across the ice-line; see Section 5)}, a \hl{mild} dichotomy in $s$ would still persist, reflecting a change in dust material density $\rho_{\circ}$. In this case, however, particle sizes fall to below a millimeter rather rapidly, between $\sim20$ to $30$ AU \hl{given our choice of $r_c(0)$}. \\
\begin{figure*} 
\centering
\scalebox{0.63}{\includegraphics{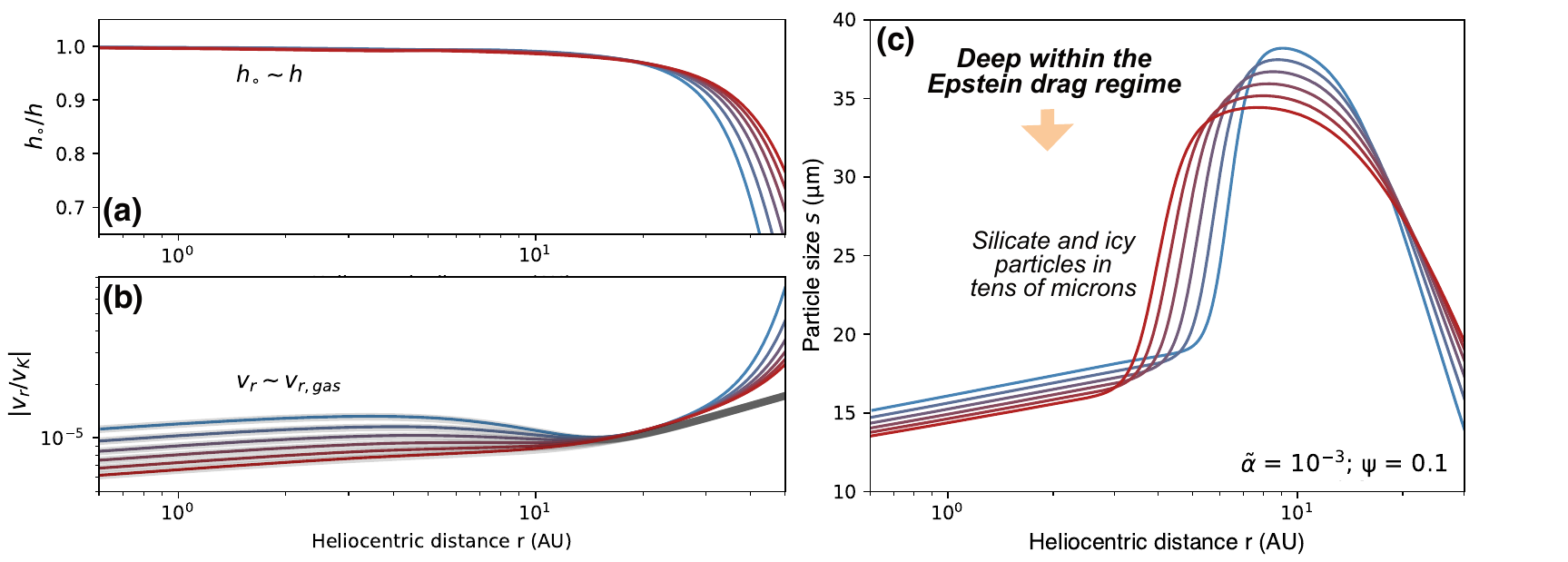}}
\caption{\footnotesize\textbf{Profiles of the (a) normalized dust scale height, (b) normalized dust radial drift velocity, and (c) characteristic particle size in the bouncing-limited turbulent disk.} Interior to the outer edge of the disk at $r=r_{c}$, the dust particles tens of microns in size are completely coupled to the gas. The scale height of the dust sub-disk is equivalent to the hydrostatic scale height of the gas, and dust particles drift inward with the accretionary gas flow. These particles sit deep within the Epstein drag regime. See Sections 4.1.1 \& 4.1.2.}
\label{fig:Figure 8}
\end{figure*}
\indent Turning our attention to silicate particles in the inner disk, we note that the sizes obtained in the turbulence-dominated disk (upper hundreds of microns) closely match those of \hl{BM22} for a viscous disk with $\alpha_{\nu}\simeq 10^{-3}$ and $\dot{M}\simeq 10^{-8} M_{\odot}$/yr.  In their model, silicate particles are just short of a millimeter in size, a factor of order unity higher than our predictions. This slight discrepancy reflects the lower $\Sigma$ values in their model for the choice of $\dot{M}$ (Fig. 5a), which results in lower $T$ and thus higher $St_{tf}$ (Eq. 17).\\
\indent The growth of silicate particles is limited by turbulent fragmentation, which is expected to operate at higher $St$ (\textit{i.e.,} larger particle sizes) than the bouncing barrier. However, the collision outcome model of Windmark et al. (2012) predicts that for similar-sized colliding particles, perfect bouncing (\textit{i.e.,} no sticking whatsoever) is achieved for particles hundreds of microns to a millimeter in size, with fragmentation setting in beyond the mm-scale. This seems to suggests the sizes we have derived are underestimated. The value of $v_{f,rock}\simeq 1$ m/s is based on collision experiments performed on millimeter-sized grains composed of $\sim 0.5$ micron particles ZrSiO$_{4}$ in free-fall and vacuum (Blum \& Münch, 1993), and is possibly unrepresentative of particles in actual disks. Due to sintering effects at high ($T\sim1000K$) temperatures in the inner disk (Poppe, 2003), \hl{accompanied by dehydration and possible phase changes (Pillich et al., 2021, 2023),} silicate particles therein may have $v_{f,rock}>1$ m/s. Moreover, $v_{f,rock}$ could increase if progressive compaction of silicate particles (\textit{i.e.,} evolution in their material density) occurs (Blum \& Wurm, 2008), and does so on a meaningful timescale, say, less than the radial drift timescale. We revisit the discussion of the $v_f$ temperature dependence in Section 5. Before moving on, note that regardless of whether particles are in the Epstein (Eq. 22) or Stokes (Eqs. 25 \& 26) regimes, higher $\rho_{0}$ lead to smaller $s$. Thus, if $\rho_{\circ,rock}<3300$ kg/m$^{3}$ as assumed, silicate particles would be larger than presently predicted. \\
\indent The particle sizes predicted in our bouncing-limited turbulent disk, in the tens of microns, also appear inconsistent with the expectation that perfect bouncing occurs for particles exceeding a hundred microns. Nonetheless, there is little cause for concern due to two main reasons. First, our results are obtained for a disk with $\tilde{\alpha}\simeq 10^{-3}$ and $\psi\simeq 0.1$. As can be deduced from Eq. 23 and the observation that particles sizes increase $\psi$ (due to the decrease in temperatures; Fig. 9), a decrease in $\tilde{\alpha}$ and/or a transition to a more wind-dominated disk will lead to higher bouncing-limited particle sizes. Second, in the model of Windmark et al. (2012), similar-sized colliding particles between ten to a hundred microns are not strictly within the sticking regime. Experiments suggests there is no sharp boundary between sticking and bouncing regimes, but a so-called ``sticking-to-bouncing" transition in parameter space, where collision outcomes are modeled with a probability distribution (Weidling et al., 2012). This indicates at least that collision outcomes and the parameters that control them remain \hl{uncertain}. We note that the tens of microns we have obtained are consistent with the fragmentation-limited particle sizes from the fiducial turbulence-dominated disk, in the upper hundreds of microns.\\
\indent Inspection of Fig. 9 suggests the closest match to inferences in particle sizes from astronomical observations may be found for $\psi\sim 1$, in between the two extreme cases we have considered. For $\psi\simeq 1$ and $\tilde{\alpha}\simeq 10^{-3}$, silicate and icy particles up to $r\simeq 40$ AU are confined between a millimeter and decimeter.

\subsubsection{A Note on Planetesimal Formation}
\indent \hl{The low characteristic $St$ of silicate particles within the ice-line ($<10^{-2}$) do not appear conducive to efficient planetesimal formation via streaming (or more generally, resonant drag) instabilities, which develop readily at higher $St$ (Youdin \& Goodman, 2005; Squire \& Hopkins, 2018). Our model, however, does not consider the potential for radial concentration of dust by pressure bumps (\textit{e.g.,} associated with the silicate sublimation line; Morbidelli et al., 2022), and other substructures that would hasten the growth rate of instabilities, which depend on both $St$ and the dust-to-gas ratio (metallicity) $f_d$. An increase in $f_d$ beckons an increase in $St$ as turbulence becomes less capable, of sufficiently energetic, to combat the settling of dust towards the midplane (see Section 4 in Batygin \& Morbidelli, 2020). Settled dust particles grow to larger sizes/$St$ owing to a smaller turbulent velocity dispersion $\Delta v_t$. In Section 4.3.3 below, we discuss how the buildup of dust at the silicate sublimation line can result in planetesimal ``rings" in which rocky planets may be efficiently accreted (\textit{e.g.,} Batygin \& Morbidelli, 2023).}

\begin{figure*} 
\centering
\scalebox{0.675}{\includegraphics{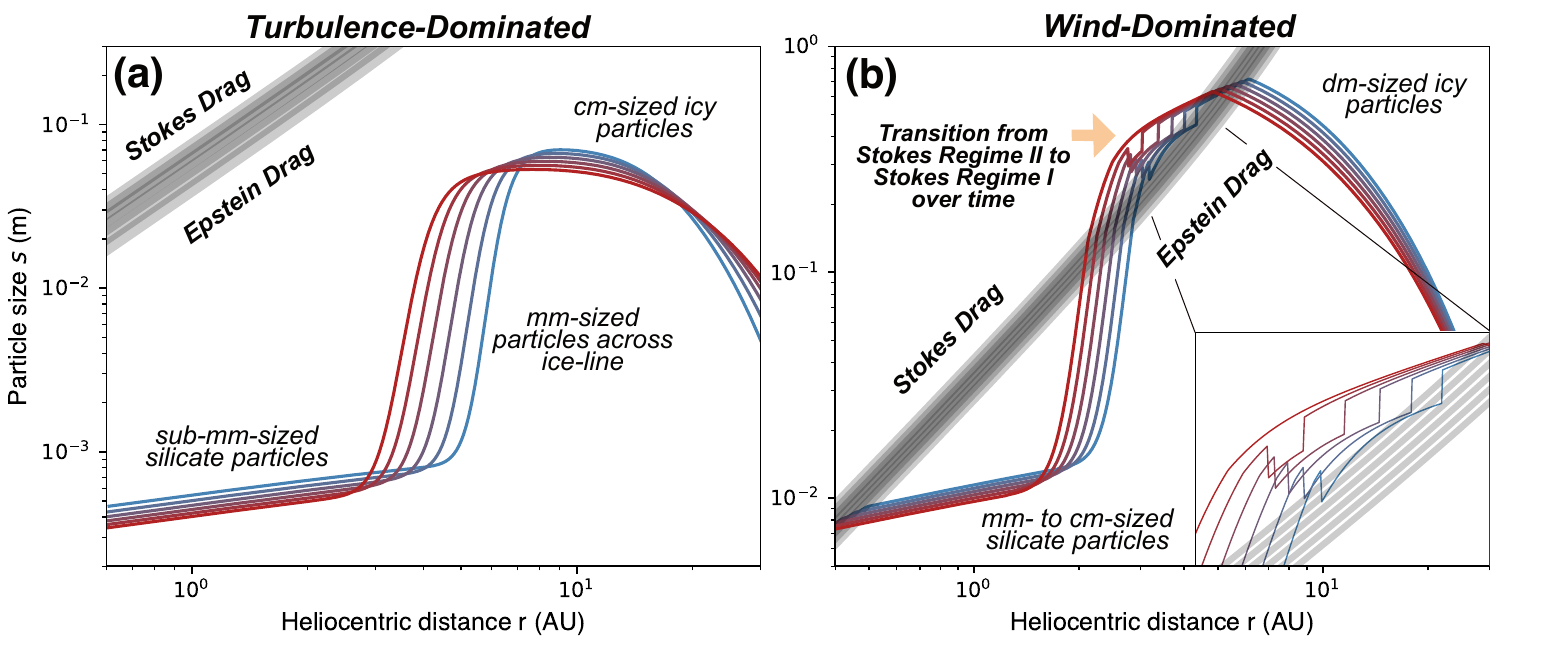}}
\caption{\footnotesize\textbf{Profiles of the characteristic particle size for the (a) turbulence- and (b) wind-dominated disks.} Grey bands represent the boundary between the Epstein and Stokes drag regimes across the six time steps, defined by $s=(9/4)\lambda_{mfp}$. In (a), silicate and icy particles across the disk are in the Epstein drag regime. Sizes of the former are confined to the sub-mm scale, while those of the latter range from millimeters to centimeters in size. In (b), only particles between $r\sim2.5$ AU to $\sim5$ AU enter the Stokes drag regime. A small portion of this region begins in Stokes Regime II, but evolves into Stokes Regime I within the initial accretion timescale. All other particles fall within the Epstein regime. Silicate particles in the inner disk are millimeters to a centimeter in size, while icy particles/boulders in the outer disk are in the cm- to dm-scales. See Section 4.1.2.}
\label{fig:Figure 9}
\end{figure*}
\subsection{2D vs. 3D Pebble Accretion}
\indent Pebble accretion (PA) is the process by which the gravity of a planetary body detaches particles with $10^{-4}\lesssim St\lesssim 1$ from the background flow of gas, inducing considerable drag which causes them to spiral into the body. In the ensuing analysis and discussion, we use the term ``planetary embryo" to denote both km- to 100-km-sized bodies formed via the gravitational collapse of \hl{pebble clouds} (\textit{i.e.,} planetesimals) and larger, 1000-km-sized bodies which have outgrown the planetesimal stage (\textit{i.e.,} protoplanets). A common misconception about PA, perhaps stemming from its success in resolving the so-called ``timescale conflict" in gas/ice-giant core accretion (Lambrechts \& Johansen, 2012), is that it is always efficient. After all, the collisional cross sections characteristic of PA are generally perceived to exceed those in the classical, Safronov-type model for planetary growth involving pairwise planetesimal collisions (Safronov, 1969; Chambers et al., 1998). In reality, the mass accretion rate $\dot{m}_{p}$ of an planetary embryo undergoing PA is strongly dependent on the vertical distribution of the surrounding pebbles, namely whether they are concentrated in a thin layer, or diffusely distributed about the disk mid-plane (Ormel, 2017). In the former case, PA proceeds efficiently in the 2D regime, where (in the headwind/Bondi sub-regime; see below) 
\begin{equation}
\dot{m}_{p,2D,Bondi} \simeq f_{d}\Sigma\sqrt{8Gm_pStv_{rel}/\Omega_{K}}, 
\end{equation}
while in the latter case, PA proceeds more slowly in the 3D regime, where 
\begin{equation}
\dot{m}_{p,3D} \simeq 6\pi f_{d}\Sigma R_{H}^{3}St\Omega_{K}/\sqrt{2\pi}h_{\circ}, 
\end{equation} 
and the planetary Hill radius $R_{H} = r(10St m_{p}/3M_{*})^{1/3}$, $M_{*}$ being the mass of the central star (Ormel, 2017\hl{; Fig. 10}). Recall from above that $v_{rel} = v_{K}[1-(1-|\epsilon|c_{s}^{2}/v_{K}^{2})^{1/2}]$. The critical planetary-to-stellar mass ratio for the transition between the two regimes is obtained by setting $\dot{m}_{p,2D,Bondi} = \dot{m}_{p,3D}$, which yields 
\begin{equation}
\left(\frac{m_{p}}{M_{*}}\right)_{2D-3D} \simeq \frac{4\alpha_{\nu}|\epsilon|h^{4}}{\pi r^{4} St(St+\alpha_{\nu})}.
\end{equation}
Above this critical ratio, PA operates in the 2D regime. \\
\indent 2D pebble accretion is further divided into two sub-regimes, depending on the relevant length scale over which pebble capture ensues. In the headwind or Bondi regime, (Eq. 28) this length scale is the Bondi radius $R_{B} = Gm_{p}c_s\sqrt{StM_{*}/m_p|\epsilon|}/v_{K}^{3}$, whereas in the (generally) more efficient shear or Hill regime, it is the Hill radius. For the latter,  $\dot{m}_{p,2D}$ is given by (Ormel, 2017)
\begin{equation}
\dot{m}_{p,2D,Hill} \simeq 2f_{d}\Sigma R_{H}^{2}\Omega_{K}St^{2/3}.
\end{equation} 
The critical mass ratio demarcating the boundary between these two sub-regimes is obtained by setting $R_{B} = R_{H}$, yielding:
\begin{equation}
\left(\frac{m_p}{M_*}\right)_{B-H} \simeq \frac{100|\epsilon|^{3}c_s^{6}}{9Stv_{K}^{6}}. 
\end{equation}
As the planetary embryo grows, its Bondi radius will eventually surpass its Hill radius, at which point PA transitions from the headwind to shear regime (\textit{e.g.,} Ida et al., 2016; Ormel, 2017). \hl{Moving forward, it is important to keep in mind that the PA rates introduced} serve only as approximations, and overestimates for that matter. Indeed, they assume a maximal \hl{pebble} impact parameter, obtained by equating the pebble settling time (\textit{i.e.,} the time it takes for a pebble to settle onto the embryo from gravity and drag) and the pebble-embryo encounter time (\textit{i.e.,} the characteristic time over which the embryo exerts the greatest gravitational tug on the pebble). \\
\indent Pebble accretion does not \hl{proceed} indefinitely. It is self-limiting, and terminates when the Hill radius of the accreting protoplanet becomes comparable to the (local) hydrostatic scale height of the disk ($R_{H}\sim h$) (Lambrechts et al., 2014). At this stage, the protoplanet is sufficiently large to perturb the disk structure, carving out a gap of high pressure that halts the inward drift of pebbles. This ``pebble isolation mass" takes the form:
\begin{equation}
\left(\frac{m_p}{M_*}\right)_{Iso}\simeq \frac{1}{2}\left(\frac{h}{r}\right)^{3}
\end{equation}
For regions of a disk with $(m_p/M_*)_{Iso}<(m_p/M_*)_{2D-3D}$, PA is confined to the 3D regime for all planetary masses. \\
\begin{figure*} 
\centering
\scalebox{1.65}{\includegraphics{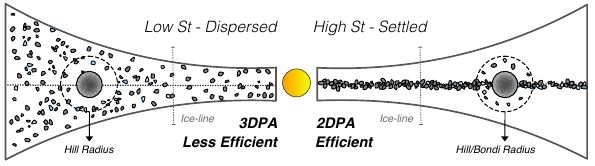}}
\caption{\hl{\footnotesize\textbf{Illustration of pebble accretion in the (relatively) inefficient 3D (left) and efficient 2D (right) regimes, relevant for low and high pebble $St$, respectively.} The relevant pebble capture length scale in the former is the Hill radius, while that in the latter is either the Bondi or Hill radius. See Section 4.2.}}
\label{fig:Figure 10}
\end{figure*}
\indent Profiles of these transitory mass scales for our two fiducial disks are shown in \hl{Fig. 11}, where $(m_p/M_*)_{2D-3D}$ is represented by solid lines, $(m_p/M_*)_{B-H}$ in faded lines, and $(m_p/M_*)_{Iso}$ in dotted lines. Here and in Section 4.3, a Sun-like star (\textit{i.e.,} $M_*\simeq 2\times10^{30}$ kg) is assumed. The results for the bouncing-limited turbulent case are not shown, as they are readily inferred from the conclusions drawn for \hl{silicate} particles in the turbulence-dominated disk (\hl{Fig. 11a}). A salient feature of Fig. 10 is that, regardless of whether disk evolution is driven by turbulence or MHD winds, planetary embryos interior to the ice-line are prohibited from partaking in 2DPA as they would reach the pebble isolation mass ($m_p/M_*\sim 10^{-5}$ to $10^{-4}$) before the transition point between the 3D an 2D regimes ($m_p/M_*>10^{-4}$).  The situation changes dramatically across the ice-line, beyond which sufficiently massive bodies can undergo Bondi and Hill 2DPA. \\
\indent Across the ice-line in the wind-dominated disk (\hl{Fig. 11b}), Earth-sized embryos beyond $\simeq3$ AU (\textit{i.e.,} the ice-line) are ensured 2DPA from time zero, and for $r\lesssim 5$ AU, transition from the Bondi to Hill regime in $\sim$ a cycle of $t_{Acc,0}$ ($\simeq$0.5 Myr; Fig. 4b). \hl{Mars-sized embryos beyond $\simeq3$ AU can partake in Bondi 2DPA from time zero, while Moon-sized embryos do so after $\sim$ half a cycle of $t_{Acc,0}$.} Embryos of mass $m_p\sim 10^{-5} M_*$ are firmly situated in the Hill regime past $\simeq 4 $ AU from time zero. Conditions for 2DPA are far more stringent in the turbulence-dominated disk (\hl{Fig. 11a}). Only embryos larger than Earth with $m_{p}\gtrsim10^{-5} M_{*}$ can partake in 2DPA within the first accretion timescale ($\simeq$0.3 Myr; Fig. 4a), and even so, only at $r\gtrsim 10$ AU and in the Bondi regime. The result that embryos within the ice-line are confined to 3DPA remains true even after many ($\gtrsim5$) cycles of $t_{Acc,0}$. \\
\indent As the disk cools and the ice-line migrates inwards over time, $(m_p/M_*)_{2D-3D}$ at any location in the disk decreases owing to the increase in $St$. However, the rate of ice-line migration far exceeds that at which $(m_p/M_*)_{2D-3D}$ falls towards $(m_p/M_*)_{Iso}$. In the turbulence-dominated disk, Earth- and Mars-sized embryos (at the present-day locations of those planets in the SS) never enter the 2D regime, even after the ice-line passes 1 AU at $\gtrsim 10$ $t_{Acc,0}$ \hl{(the efficiency of 3DPA increases across the ice-line nonetheless; see below).} In the wind-dominated disk, lower $(m_p/M_*)_{2D-3D}$  values enable those same embryos to partake in 2DPA after $\gtrsim 3$ cycles of $t_{Acc,0}$. This, however, does not constitute 2DPA within the ice-line\textemdash larger, more settled icy particles have simply moved into the feeding zone of those planets (see \hl{Section 4.3.1}). \\
\begin{figure*} 
\centering
\scalebox{0.75}{\includegraphics{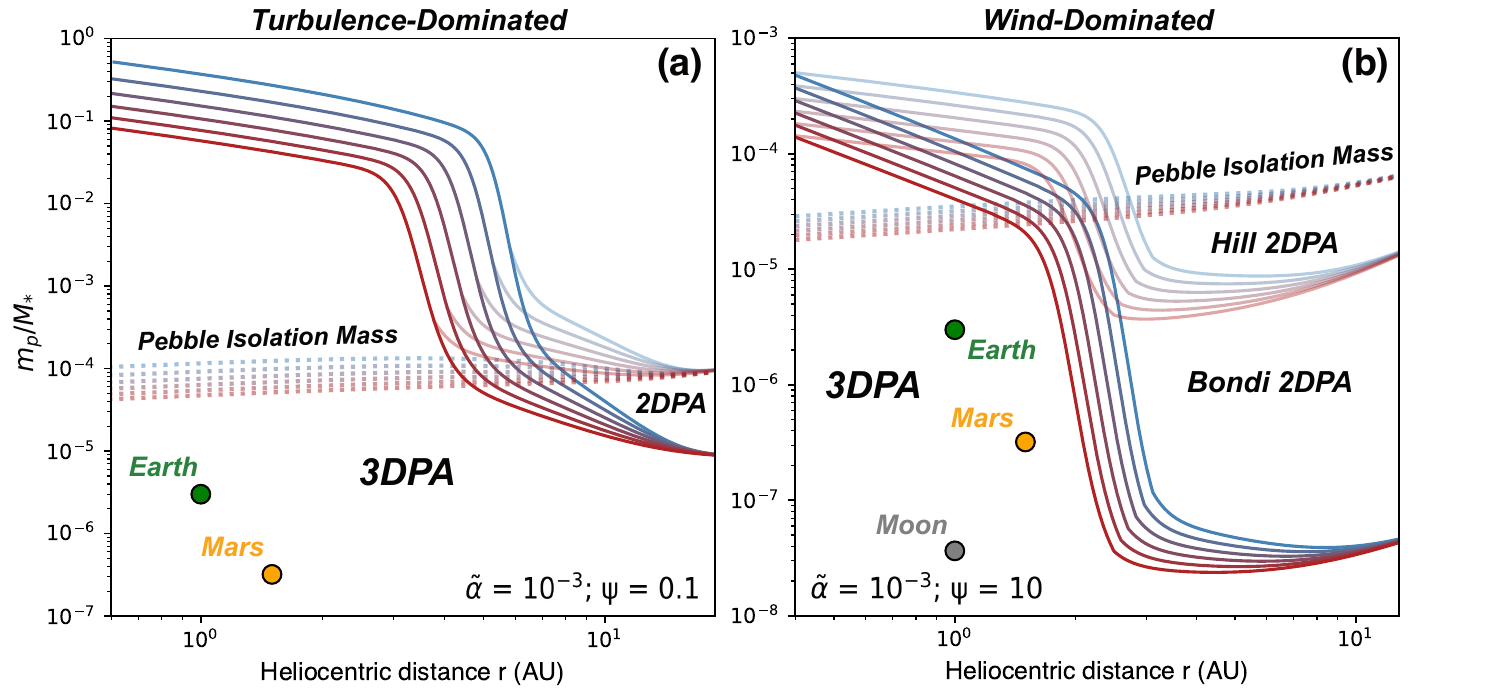}}
\caption{\footnotesize\textbf{Profiles of the critical planetary-stellar mass ratio for transitioning between the 2D to 3D pebble accretion regime (solid lines) and between 2D Bondi and Hill pebble accretion (faded lines), as well as the pebble isolation mass (dotted lines), over time for the (a) turbulence- and (b) wind-dominated disks. } In both scenarios, planetary embryos interior to the ice-line are prohibited from partaking in 2DPA. Embryos beyond the ice-line are granted 2DPA given masses that surpass a threshold value which decreases in both time, and with increased contribution from disk winds. See Section 4.2.}
\label{fig:Figure 11}
\end{figure*}
\indent Considering the similarity in $\Sigma(r,t)$ between our fiducial disks and the steady-state viscous disk of \hl{BM22}(Figs. 4a \& 4b) for the inner disk (\textit{i.e.,} small \textit{r}), in which viscous heating dominates (see Section 2.2), it comes at no surprise that our key results here corroborate theirs. Namely, rocky embryos within the ice-line are confined to 3DPA due to considerable vertical dispersion of silicate particles, the growth of which is limited by turbulent fragmentation. Given the inefficiency of 3DPA relative to 2DPA, \hl{BM22} interpreted this as indicating the accretion of such embryos is primarily driven by mutual pairwise collisions. In Section 4.3.2 below, we show that this is true only in disks that are not driven by MHD winds ($\psi$ $\gtrsim 1$) and characterized by low $\tilde{\alpha}$ ($\lesssim 10^{-3.5}$). That is, \hl{disks that are not quiescent, with $\alpha_{\nu}\lesssim 10^{-4}$.} The lack of 2DPA alone is an insufficient criterion for the dominance of classical growth. \\
\indent Bearing in mind the fact that particles across the entire bouncing-limited turbulent disk are in tens of microns (Fig. 8c), it is clear that 2DPA (and virtually PA in general) cannot be realized therein. The existence of gas and ice giants in our SS indicate that, at least exterior to the ice-line, the bouncing barrier was insignificant if not wholly absent. This follows from the aforementioned “timescale conflict” associated with gas/ice giant planet formation: pairwise collisions, at large distances from the host star, are incapable of facilitating the rapid core growth necessary for the accretion of gaseous envelopes prior to disk dissipation (\textit{e.g.,} Rafikov, 2011) \hl{(see Fig. 13b)}. Cm-sized (or larger) pebbles must have been present in the outer disk to fuel the growth of, say, proto-Jupiter. Related to this point, our results in \hl{Fig. 11} \hl{hint at an outer SS disk with $\alpha_{\nu} = \tilde{\alpha}/(\psi+1)$ between $\sim10^{-4}$ and $10^{-3}$}, such that giant planet cores in the outer disk could have partaken in 2DPA fairly early in SS history. \hl{This range is in broad agreement with $\alpha_{\nu}$ values inferred from telescopic [{i.e.,} Atacama Large Millimeter/sub-millimeter Array (ALMA)] measurements of molecular line broadening and disk $h_0/h$ ratios (Rosotti, 2023 \& references therein). Nonetheless, it should be taken with a grain of salt. Pebble accretion in the 2D regime is not necessarily required for rapid growth in the outer disk, where high $St$ and low $h_0$ serve to enhance the 3DPA rate (Eq. 28). Indeed, for a disk with intermediate $\tilde{\alpha}\simeq 10^{-3}$ and $\psi\simeq 1$, the mass-doubling timescale for a Mars-sized planetary embryo just beyond the ice-line (\textit{i.e.,} close to Jupiter's inferred formation region; Kruijer et al., 2017a) is on the order of $\sim$ few times $0.01$ Myr despite undergoing 3DPA. This indicates the few Myr lifespan of the SS disk is ample time for giant planet core accretion, especially considering PA rates increase with embryo size (see Section 4.3.2) and the potential for accelerated growth by elevated solid densities induced by principal (\textit{e.g.,} water-ice, CO) sublimation lines (see Section 4.1.3 \& discussion on planetesimal rings in Section 4.3.3).}

\subsection{Rocky Planet Accretion}
\indent In this section, we explore the competition between pairwise collisions and PA in rocky planet formation across parameter space. Our results culminate in conditions \hl{(\textit{i.e.,} $\tilde{\alpha}$, $\psi$)} under which each growth pathway is dominant, and are discussed in the context of super-Earth origins and Earth’s accretion history. The motivating questions for each subsection are: Is there a region of parameter space in which rocky embryos can partake in 2DPA (Section 4.3.1)? How do the mass-doubling timescales from pairwise collisions and PA compare for rocky embryos (Section 4.3.2)? How do the contributions from the two competing processes to the final planet mass compare, and how do they change under different envisioned models for rocky planet formation (Section 4.3.3)? Finally, what do our results imply for the origin of super-Earths (Section 4.4), and how do they fare with isotopic constraints on the building blocks of our Earth from meteoritics (Section 4.5)? 

\subsubsection{Possibility of 2DPA in the Inner Disk?}
\indent We have demonstrated, albeit under simplifying assumptions in place (\textit{e.g.,} no temperature dependence on $v_{f}$ , constant $\alpha$ parameters; see Section 5), that 2DPA is not attainable inward of the ice-line for $\tilde{\alpha}\simeq 10^{-3}$. Nonetheless, there may be region of $\tilde{\alpha}-\psi$ parameter space in which sufficiently massive rocky embryos can partake in 2DPA. The shrunken gap between $(m_p/M_*)_{2D-3D}$ and $(m_p/M_*)_{Iso}$ in the inner disk accompanying the transition from \hl{turbulence- to wind dominance (Fig. 11)} suggests this region lies at high $\psi$. Moreover, the region likely resides towards low $\tilde{\alpha}$ as this translates to lower relative turbulent velocities (\textit{i.e.,} $\Delta v_{t}$), higher characteristic $St$, and thus lower values of $(m_p/M_*)_{2D-3D}$. Here, we systematically seek out pairs of $\tilde{\alpha}$ and $\psi$ that yield $(m_p/M_*)_{2D-3D}<(m_p/M_*)_{Iso}$ at $r=1$ AU for different times following disk infall. \\
\indent At each time step (\textit{i.e.,} 0, 1, and 3 Myr), we generate a raster of $(m_p/M_*)_{2D-3D}$ at 1 AU for all values of $\tilde{\alpha}$ and $\psi$ explored. This is compared to a similar raster of $(m_p/M_*)_{Iso}$, and the pairs (pixels) that yield $(m_p/M_*)_{2D-3D}<(m_p/M_*)_{Iso}$ are determined. It is important to differentiate the potential to partake in 2DPA within the ice-line from that arising from ice-line migration past 1 AU. This is especially relevant for disks that evolve relatively rapidly due to \hl{large} $\tilde{\alpha}$ (\textit{i.e.,} $\gtrsim 10^{-2.5}$) and thus low $t_{Acc,0}$ (scaling as $1/\tilde{\alpha}c_{s,c}^{2}$; Eq. 10). As a related point, it is meaningful to define a proxy for disk dissipation arising from the E/FUV photoevaporation of disk gas (\textit{e.g.,} Störzer \& Hollenbach, 1999; Ercolano \& Pascucci, 2017; Concha-Ramírez et al., 2019). Here, we take the time at which the stellar mass accretion rate $\dot{M} = 2\pi r \Sigma v_{r,gas}\simeq 10^{-10}M_{\odot}$/yr at 1 AU as heralding the end of the disk’s life. \\
\indent Results from this exercise are shown in \hl{Fig. 12}. At \textit{t = 0} (\hl{Fig. 12a}) and for high $\tilde{\alpha}$ and low $\psi$  (top left corner), no solids can condense at and within 1 AU, which lies interior to the silicate sublimation line at \textit{$T=1500K$}.  For most of parameter space, rocky embryos are confined to 3DPA by the pebble isolation mass. Only in the bottom right (\textit{i.e.,} low $\tilde{\alpha}$ and high $\psi$ as predicted), beyond the solid black curve, can 2DPA be granted for a sufficiently high mass [$(m_p/M_*)_{2D-3D}\gtrsim 10^{-6}-10^{-5}$]. The region in which 2DPA is accessible increases in size with increasing $m_p/M_*$ (the dashed red curve for Earth sits above that for Mars), reflecting more $\tilde{\alpha}-\psi$ pairs for which $(m_p/M_*)_{2D-3D}<m_p/M_*$. After 1 Myr (\hl{Fig. 12b}), the silicate sublimation line has migrated within 1 AU for all parameter space, and a region wherein $(m_p/M_*)_{2D-3D}<(m_p/M_*)_{Iso}$ has developed in the top right (\textit{i.e.,} high $\tilde{\alpha}$ and high $\psi$). This results from ice-line migration past 1 AU, and as such, does not represent the possibility of 2DPA \textit{within} the ice-line. In fact, for all disks with $\tilde{\alpha}\gtrsim10^{-2.5}$, the ice-line lies interior to 1 AU by this time. The reason most embryos there are still confined to 3DPA can be understood by comparing \hl{Figs. 11a \& 11b}. Greater turbulence (\textit{i.e.,} lower $\psi$; higher $\alpha_{\nu}$ for a given $\tilde{\alpha}$) results in higher temperatures across the disk due to viscous heating (see Section 2.2), in turn leading to lower $St$ and higher $(m_p/M_*)_{2D-3D}$. For high $\tilde{\alpha}$ and moderately low $\psi$ $(\lesssim 5)$, the entire $(m_p/M_*)_{2D-3D}$ profile lies above $(m_p/M_*)_{Iso}$. The rapidity of ice-line migration for these disks arises not only from their \hl{large $\tilde{\alpha}$}, but stellar irradiation of the flared disk, which keeps $T$ (or $c_s$) higher than it would be at $r_{c,0}$ (where $t_{Acc,0}$ is defined\hl{; see Eq. 10}) in its absence \hl{(Fig. 4c, d)}. After 3 Myr (\hl{Fig. 12c}), disks at the upper right corner of parameter space have dissipated, and the region in which embryos at 1 AU are prohibited from 2DPA has shrunken considerably, owing mostly to ice-line migration. In quiescent disks ($\alpha_{\nu}\lesssim$ few times $10^{-4}$) at the bottom right, the ice-line still sits beyond 1 AU. By now, this region where $(m_p/M_*)_{2D-3D}<(m_p/M_*)_{Iso}$ \textit{interior to the ice-line} has expanded only slightly due to \hl{large} $t_{Acc,0}$. \\
\indent We briefly note that $t=0$ in our model is not anchored to SS time zero, conventionally defined by the condensation of Calcium-Aluminum-rich Inclusions (CAIs). Moreover, the definition of $t_{Acc,0}$ is rather arbitrary, such that its values should not be given much weight [especially for high $\tilde{\alpha}$, where $t_{Acc,0}$ is extremely low (on the order of $\sim 0.01$ Myr)]. It mainly serves to show how the morphology of parameter space evolves with time. Clearly, so long as embryos in even moderately turbulent disks sit within the ice-line (\textit{i.e.,} accrete ``dry"), they are confined to 3DPA. 
\begin{figure} 
\scalebox{0.64}{\includegraphics{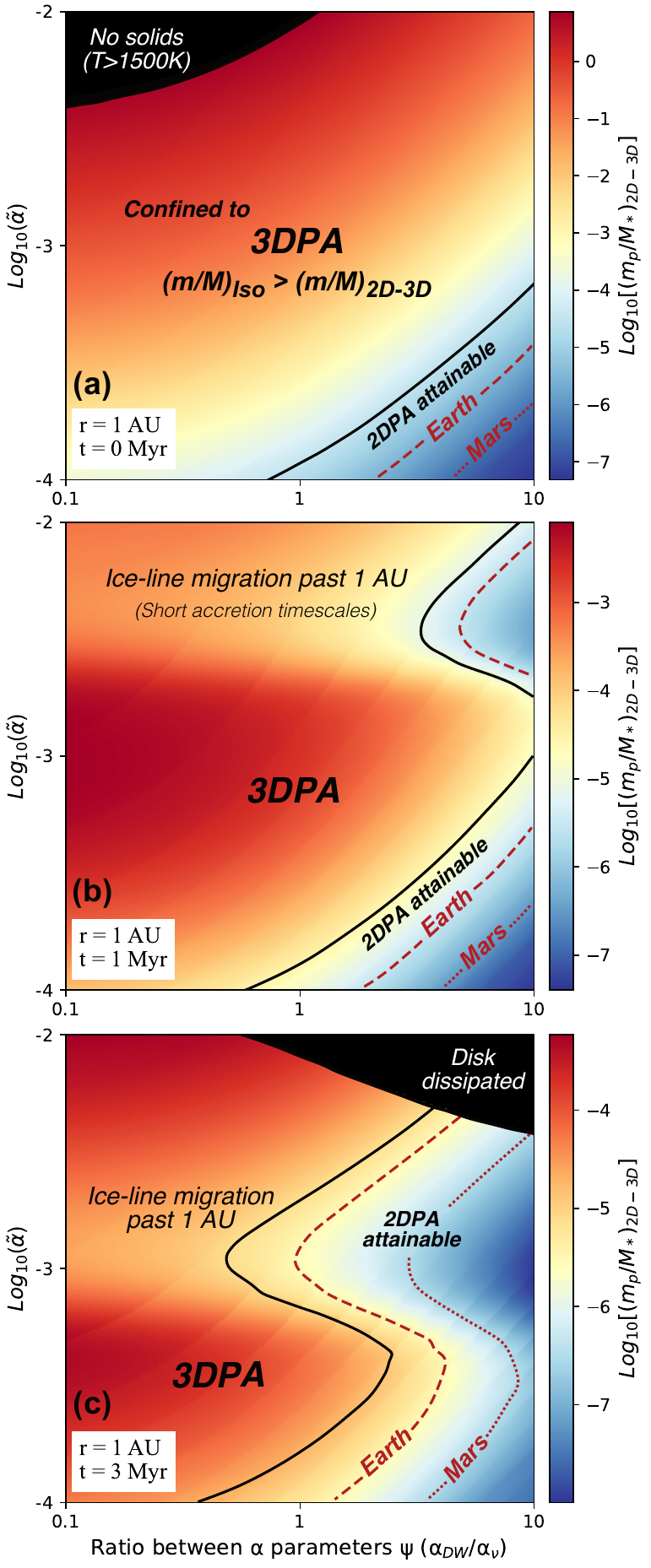}}
\caption{\textbf{\footnotesize Rasters of the critical planetary-to-stellar mass ratio for transitioning from 3DPA to 2DPA at $r=1$ AU and after (a) 0 Myr, (b) 1 Myr, and (c) 3 Myr of disk evolution.} For most of parameter space, planetary embryos in the inner disk are confined to 3DPA by the pebble isolation mass. Only in wind-dominated disks with low $\tilde{\alpha}$ can sufficiently massive embryos \textit{interior to the ice-line} partake in 2DPA. The red dashed and dotted lines denote the Earth-to-Sun and Mars-to-Sun mass ratios, respectively. See Section 4.3.1.}
\label{fig:Figure 12}
\end{figure}
\subsubsection{Embryo Mass-Doubling Timescales}
\indent Our investigation of parameter space indicates that for slow-evolving (\textit{i.e.,} low $\tilde{\alpha}$) disks driven by MHD winds (\textit{i.e.,} high $\psi$), \hl{rocky embryos} can accrete pebbles in the efficient 2D regime. While such embryos are relegated to 3DPA for most \hl{$\tilde{\alpha}-\psi$ pairs}, it would be unseemly to assume classical growth by pairwise collisions thus constitutes their primary mode of accretion. Instead, a quantitative statement about the favored pathway for rocky planet formation can be made via consideration of the planetary mass-doubling timescale $\tau_{p} \simeq m_p/\dot{m_p}$. The accretion rates $\dot{m_p}$ for the different regimes of PA are provided above by Eqs. 27, 28, \& 30, while that for pairwise collisional growth, in account of gravitational (Safronov) focusing, is given by (Lissauer, 1993; Kokubo \& Ida, 2000)
\begin{equation}
\dot{m}_{pw} \simeq f_{d}\Sigma\pi R_{p}^{2}\Omega_{K}(1+\Theta)/\sqrt{2\pi}.
\end{equation}
Here, we assume a Mars-sized embryo with $m_p\simeq 6.4\times 10^{23}$ kg, and $R_{p}\simeq 3.4\times10^{6}$ m. The Safronov number $\Theta = (v_{esc}/\sigma_{v})^{2}$ quantifies the degree of gravitational focusing and is generally taken to be of order \hl{a few}. The embryo escape velocity $v_{esc} = \sqrt{2Gm_{p}/R_{p}}$ and $\sigma_{v}$ is the local velocity dispersion of embryos participating in mutual collisions.\\
\indent A raster of $\tau_{p}$ can be generated for PA assuming the appropriate regime (dependent on \textit{r} and \textit{t}) for each pair of $\tilde{\alpha}$ and $\psi$. We denote values in this raster $\tau_{p,PA}$ to differentiate them from those in a similar raster generated assuming pairwise collisional growth, denoted $\tau_{p,PW}$. In our calculation of $\tau_{p,PA}$ and $\tau_{p,PW}$, we assume the total solid mass $f_{d}\Sigma$ in the feeding zone ($r\simeq1 AU$) is distributed evenly between pebbles and embryos. That is, the fraction of solid mass held in pebbles $f_{Peb} = 1-f_{Emb} \simeq 0.5$ (see below). We use the ratio $\tau_{p,PA}/\tau_{p,PW}$ as a metric for evaluating the competition between PA and classical growth, neglecting for now the dependence of $\tau_p$ on time (see Section 4.3.3). Assuming $\Theta\simeq 5$, our results for $t=0$ and two locations $r=1$ AU and $r = 20$ AU, corresponding to within and without the ice-line, are shown in \hl{Fig. 13}.\\
\indent Interior to the ice-line (\hl{Fig. 13a}), rocky Mars-sized embryos in disks characterized by relatively low $\tilde{\alpha}$ ($\lesssim 10^{-3.5}$; below the black dashed line) have $\tau_{p,PA}/\tau_{p,PW}<1$, indicating PA constitutes their primary mode of accretion. At higher $\tilde{\alpha}$, PA also dominates growth in disks driven by MHD winds ($\psi\gtrsim$ a few). A key observation is that for most of the pebble-dominated parameter space, PA operates in the 3D regime, suggesting \textit{the inaccessibility of 2PDA does not necessarily entail the dominance of classical growth. } At $\psi\simeq 1$ and $\tilde{\alpha}\simeq 10^{-3.5}$ (keeping in mind our choice of $f_{Peb}\simeq 0.5$ and $\Theta\simeq 5$) 3DPA is $\sim$10 times more efficient than growth by pairwise collisions. In accordance with \hl{Fig. 12}, only in wind-dominated ($\psi\simeq 10$) disks of low $\tilde{\alpha}$ $(\simeq10^{-4})$ are the rocky Mars-sized embryos accreting pebbles in the 2D regime. For a turbulence-dominated disk with $\tilde{\alpha}\simeq 10^{-3}$, such as that of \hl{BM22}, pairwise collisions remain $\sim5$ times more efficient than PA. \\
\indent Beyond the ice-line (\hl{Fig. 13b}), $\tau_{p,PA}/\tau_{p,PW}<1$ across all of parameter space, indicating embryos therein grow primarily from pebbles. Notably, PA is $\sim$10,000 times more efficient than pairwise collisions. The 2DPA region has expanded owing to greater settling of icy particles, though embryos in turbulence-driven disks at relatively high $\tilde{\alpha}$ are still confined to 3DPA \hl{(see Section 4.3.1)}. For all disks wherein Mars-sized embryos at \textit{r }= 20AU (and 1 AU) partake in 2DPA, they do so in the Bondi regime. As can be deduced from inspection of \hl{Fig. 11}, only embryos more massive than Earth in relatively wind-dominated disks can enter the Hill regime. \\
\indent In both \hl{Figs. 13a \& 13b}, the lack of a gradient in the 2DPA field reflects the effect of the square root in Eq. 27, which dampens the variations in the characteristic $St$ with $\tilde{\alpha}$ and $\psi$. Consideration of $\tau_{p,PA}$ in going from the bottom right to top left of parameter space (\textit{i.e.,} deeper into the 3DPA regime) makes the inefficiency of 3D- relative to 2DPA apparent. Within the ice-line (Fig. 13a), for instance, $\tau_{p,PA}$ is roughly consistent at $\simeq$0.003 Myr within the 2D regime, and remains so across the 2D-3D boundary. Once in the 3D regime, however, $\tau_{p,PA}$ increases rapidly, reaching $\gtrsim$ 5 Myr at intermediate values $\tilde{\alpha}\sim 10^{-3}$ and $\psi\sim1$, and $> 10$ Myr farther up.\\
\indent At this stage, it is worthwhile to consider how our results for the inner disk in \hl{Fig. 13a} change with $f_{Peb}$ and $\Theta$. The scenario explored so far, with $f_{Peb}\simeq0.5$, is likely “pebble optimistic.” While the reservoir of pebbles in the planetary feeding zone may be constantly replenished by their inward drift from farther disk regions, the emergence of embryos from the collapse of \hl{concentrated pebble clouds} followed by their subsequent growth necessitates a decline in $f_{Peb}$ \hl{(and increase in $f_d$)}. This process is inferred to be rapid and extensive in the SS. Indeed, Hf-W chronology of iron meteorites coupled with thermal modeling of planetesimals internally heated by \textsuperscript{26}Al decay indicate that planetesimals $\gtrsim$ 100 km in size accreted within $\simeq$0.5 to 1 Myr following \hl{CAI condensation (preceding time zero in our model)}, both within and without the ice-line (Kruijer et al., 2014, 2017a; Hunt et al., 2018; Spitzer et al., 2021). 
Rapid ($\lesssim 1$ Myr) planetesimal growth in the inner disk is further supported by dust coagulation models (\textit{e.g., }Izidoro et al., 2021a; Morbidelli et al., 2022). By the time Mars-sized embryos are present and ubiquitous in the disk, $f_{Peb}$ has likely fallen more closely to $\simeq$0.1 or 0.2, if not lower. As for the Safronov number $\Theta$, we note that even if $v_{esc}$ exceeds $\sigma_{v}$ by a factor of a few due to say, gas drag, collisional damping, and dynamical friction, $\Theta$ increases to $\sim 10$. \\
\indent As $f_{Peb}$ decreases and $\Theta$ increases, the pebble-dominated region of parameter shrinks towards the bottom right (high $\psi$ and low $\tilde{\alpha}$). For $f_{Peb}\simeq0.5$ and $\Theta\simeq5$ (\hl{Fig. 13a}), $\sim 42\%$  of parameter space (excluding the region where $T\gtrsim 1500K$) is dominated by classical growth. This fraction increases to $\sim66\%$ for $f_{Peb}\sim$ 0.1, and to $\sim72\%$ with an additional increase in $\Theta$ to $\sim10$. Accounting for the fact that the equations for $\dot{m}_{p,2D}$ and $\dot{m}_{p,3D}$ are overestimates (see Section 4.2), this suggests the growth of Mars-sized embryos to $\sim$twice their mass proceeds primarily via pairwise collisions across parameter space. Time, however, has not been considered. \\
\begin{figure*} 
\centering
\scalebox{0.7}{\includegraphics{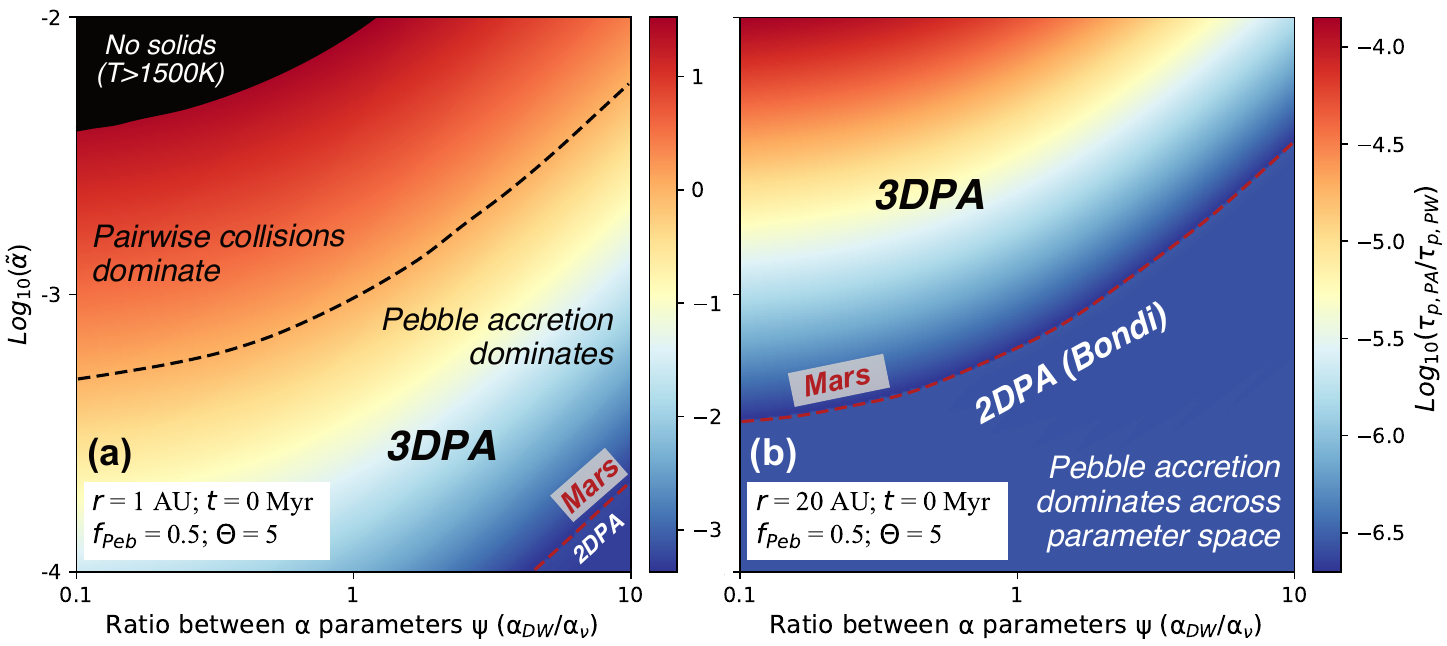}}
\caption{\footnotesize\textbf{Rasters of  $Log_{10}(\tau_{p,PA}/\tau_{p,PW})$ at $t=0$ for a Mars-sized embryo at (a) $r = 1$ AU and (b) $r=20$ AU.} In (a), values corresponding to disks with low $\tilde{\alpha}$ and to a lesser extent high $\psi$ are $< 1$, indicating the dominance of pebble accretion in the mass-doubling growth of the embryo. In (b), values are $<1$ across all parameter space. See Section 4.3.2.}
\label{fig:Figure 13}
\end{figure*}
\indent It is important to realize that as accretion progresses, \hl{while the disk is still present (see below)}, PA grows in its dominance. This can be understood as a result of both the increase of the embryo mass $m_p$ and more crucially the $St$ of particles with time. Regarding to the former, $\dot{m}_{p,3D}$ (Eq. 28) scales as $\sim R_{Hill}^{3}\sim m_p$ while $\dot{m}_{pw}$ (Eq. 33) scales as $\sim R_p^{2}\sim m_p^{2/3}$. As $m_p$ grows, then, the ratio $\dot{m}_{p,3D}/\dot{m}_{pw}$ does so too as $\sim m_p^{1/3}$\textit{. } Note that in the less applicable 2D regime, $\dot{m}_{p,2D,Bondi}/\dot{m}_{pw}\sim m_p^{-1/2}$ (Eq. 27) and $\dot{m}_{p,2D,Hill}/\dot{m}_{pw}\sim m_p^{0}$ (Eq. 30). As time progresses, pebble \textit{St }increases owing to the decrease in $\Sigma$ and thus $T$. This leads to an increase in $\dot{m}_{p,3D}$ due to both (i) the increase in the PA impact parameter (or equivalently, the decrease in the pebble settling time; see Section 4.2) (Ormel, 2017) and (ii) the decrease in the scale height of the solid sub-disk $h_{0}\sim St^{-1/2}$ (note that $h_{0}$ also decreases with $h\sim\sqrt{T}$ ). For rocky embryos at $r \simeq 1$ AU (\hl{Fig. 13a}), the pebble-dominated region (\textit{i.e.,} where $\tau_{p,PA}/\tau_{p,PW}<1$) grows rapidly with time, and slightly with mass. The rapid rise in $St$ is largely due to the short initial accretion timescale $t_{Acc,0}$ across parameter space, owing to stellar irradiation at $r_{c,0}$, the parametrization for which is independent of $\tilde{\alpha}$ and $\psi$ (Eq. 13). Short $t_{Acc,0}$ also results in a rapid decline in $\Sigma$, which \hl{slows accretion}.  Thus, for embryos much larger than Mars, while $\tau_{p,PA}/\tau_{p,PW}$ (evaluated at $t=0$) may be $>1$ for most $\tilde{\alpha}-\psi$ pairs, this dominance of pairwise collisions across parameter space is not necessarily reflected in their contribution to the final planetary mass. This motivates our work in Section 4.3.3, wherein the contributions from pebbles and embryos throughout the full accretion history of a planet are compared. We emphasize that time zero in our model is physically undefined and may well succeed the onset of planetary accretion, at which point higher $\Sigma$ and $T$ yielded lower $St$, translating to a greater dominance of classical growth. Moreover, $t_{Acc,0}$ values are \hl{not necessarily meaningful (see end of Section 4.3.1)}, prompting us to consider accretion in steady-state disks in Section 4.3.3 below. \\
\indent To summarize, PA can constitute the dominant pathway of \hl{rocky} planetary accretion \textit{regardless of whether it operates in the 2D or 3D regime}. Pairwise collisions between embryos dominate the initial growth of Mars-sized embryos for most of parameter space, given low $f_{Peb}$ and accounting for the fact that $\dot{m}_{p,2D}$ and $\dot{m}_{p,3D}$ are upper limits. This dominance may not apply across the full planetary accretion process given the rapid decline in $\Sigma$ (and thus $T$) with time due to short disk evolution timescales governed by $t_{Acc,0}$. The rapid decline in $\Sigma$ increases the time required to accrete a given mass, and as more time passes, the rise in pebble $St$ and decline in the disk hydrostatic scale height $h$ increase the efficiency of PA relative to pairwise collisions. The physical cutoff to this process is disk dissipation, \textit{beyond which pairwise collisions are the sole contributors to planetary growth}. The post-CAI timing of SS disk dissipation is constrained to $\sim4$ Myr from paleomagnetic analyses of angrites and CO chondrites (Wang et al., 2017; Borlina et al., 2022) as well as extinct radionuclide systematics of CAIs and chondrules (Dauphas \& Chaussidon, 2011). This is concordant with disk dissipation timescales inferred around T-Tauri stars (few to $\sim$10 Myr; \textit{e.g.,} Podosek \& Cassen, 1994; Briceño et al., 2001). \\
\indent As a prelude to Section 4.3.3, we note that the dominance of PA in the growth of a $\sim$ 100 km-sized embryo to an Earth- or super-Earth-sized planet necessitates (i) most of the growth to be completed within $\sim$ few Myr, and (ii) the dominance of PA across virtually the entire growth period. 

\subsubsection{Building a Super-Earth:\\ Pebble vs. Planetesimal Contributions}
\indent Time is clearly a crucial parameter in evaluating the competition between classical growth and PA. In our model, it determines the density of solids (and hence the rapidity of planetary accretion) as well as the pebble \textit{St} (and hence the efficiency of PA) at any location in the disk. Furthermore, it enables the account of disk dissipation, and thus the cessation of PA, in the planetary growth \hl{process}. To include the impact of time in our analysis, we simulate the growth of a $\sim500$ km-sized embryo towards $\sim$$8M_{\oplus}$ as inferred for most “super-Earths” (Wu, 2019) (see Section 4.4), truncating growth at the pebble isolation mass which decreases over time (\textit{i.e.,} through the hydrostatic scale height $h$; see Eq. 32). In doing so, we keep track of the contribution of pebbles and other embryos to the final planetary mass, denoting them by $M_{PA}$ and $M_{PW}$, respectively. We use time steps of $\Delta t = 0.05$ Myr, adding the contributions $\dot{m}_{p,PA}(t)\Delta t$ and $\dot{m}_{p,pw}(t)\Delta t$ to the embryo mass at each step. The characteristic size of planetesimals formed via streaming instabilities is estimated to be on the order of 100 km from both theory and numerical simulations (Youdin \& Goodman, 2005; Schäfer et al., 2017; Klahr \& Schreiber, 2020; Gerbig \& Li, 2023). Thus, a starting size of 500 km is within reason. \hl{We emphasize that our ensuing analyses, while capable of providing insight into the key factors influencing planetary growth, are fundamentally approximate and do not constitute a replacement for N-body simulations.}\\
\indent Given that our embryo is a ``so-called" first-generation planetesimal,  $t=0$ here is presumably close to (say,  $\lesssim 1$ Myr) the timing of CAI formation. While it is instructive to simulate planetary growth in an evolving disk, the uncertainty in $t_{Acc,0}$ values (\hl{setting} the clock for disk evolution; see Eqs. 7 \& 8), beckon a consideration of accretion assuming steady state disks as well. Similar to the exercise in Section 4.3.2, we generate a raster of $M_{PA}/M_{PW}$ for all pairs of $\tilde{\alpha}$ and $\psi$ at $r = 1$ AU. We explore how $M_{PA}/M_{PW}$ vary across parameter space, assuming both time evolution and steady state. Note that while the pebble \textit{St} remains constant in a steady state disk, accretion rates from both pairwise collisions and PA still account for the increasing planetary mass $m_p$. We first consider, as done in Section 4.3.2, disks with smooth solid densities. That is, disks without substructures of concentrated solids (\textit{e.g.,} dust traps) induced by, say, pressure maxima. In these “smooth” disks, we set  $f_{Peb} = 0.1$, as expected to be more representative of the full accretion history compared to the conservative value of 0.5 used in Section 4.3.2. The Safronov number $\Theta$ is kept at 5. The dust/solid-to-gas ratio $f_{d}$ “seen” by the growing embryo (\textit{i.e.,} excluding its own mass) at 1 AU is assumed to remain $\sim$ constant throughout accretion, being replenished from the inward drift of solids. Note that the solid density $f_d\Sigma$ applies only before disk dissipation. Following the photoevaporation of disk gas, while $f_d\Sigma$ falls to zero, embryos can persist and proceed to partake in pairwise collisions for tens of Myr. Moreover, $f_{Emb}f_d\Sigma$ underestimates the solid mass held in embryos over time. As embryos proceed to form by the collapse of dust clouds and mutual collisions, $f_{Emb}$ will approach unity even for a constant pebble abundance, and $f_d$ will naturally grow even in the absence of disk substructures. \\
\indent The raster described for disks evolving in time and in steady state are shown in \hl{Figs. 14a \& 14b}, respectively. Recall from Section 4.3.2 that disk dissipation timescales are on the order of a few Myr following planetesimal formation. That said, $\tilde{\alpha}-\psi$ pairs which yield planetary accretion times exceeding 10 Myr (to be conservative) are discarded. In the time-evolving case (Fig. 14a), these disks severely overestimate the pebble contribution to the final mass, owing to the increase in $St$ with time \hl{(see Section 4.3.2)}. Within the discarded region, planetary masses at 10 Myr range from $\sim$ \hl{few times} $10^{-4}$ to $2$ $M_{\oplus}$ from high $\tilde{\alpha}$ (and to a lesser extent low $\psi$) to low $\tilde{\alpha}$ with most below $1$ $M_{\oplus}$. \hl{This slow rate of accretion reflects the rapid fall in $\Sigma$ over time (due to short $t_{Acc,0}$), leading to small $\dot{m}_{pw}$ (Eq. 33) and $\dot{m}_{p,3D}$ (Eq. 28).} Corresponding ratios of $M_{PA}/M_{PW}$ \textit{evaluated at $t=10$ Myr} range from $\sim 0.1$ to $120$. The pebble isolation mass across the entire discarded region is greater than $2$ $M_{\oplus}$. Thus, for most disks wherein the accretion time exceeds disk dissipation, growth is far from completion, and with high $\tilde{\alpha}$, dominated by classical growth thus far. This suggests that if rocky \hl{Mars-sized} and larger planets reside in such disks, they grew primarily by pairwise collisions persisting after disk dissipation. For all $\tilde{\alpha}-\psi$ pairs for which accretion takes place in $<10$ Myr, PA dominates planetary growth. These disks are, as expected, those characterized by relatively low $\tilde{\alpha}$ and driven by MHD winds. The final planetary masses reached in these disks are their respective pebble isolation masses at $1$ AU, ranging from $\sim 2$ to $7$ $M_{\oplus}$ with increasing $\psi$. While the initial temperature at $1$ AU is higher with greater turbulence, translating to a higher pebble isolation mass, $t_{Acc,0}$ and the (3D) PA accretion rate decrease with turbulence. This means that the pebble isolation mass falls \hl{faster} towards a slower growing planet in turbulence-driven disks, explaining the smaller final planetary masses in them relative to wind-driven disks. As our work in Section 4.3.1 illustrates (\hl{Fig. 12}), the region of parameter space within which 2DPA is accessible expands towards low $\psi$ with time, owing mostly to the drift of the ice-line past $1$ AU. This phenomenon is exhibited in \hl{Fig. 14a} by the higher of the two regions wherein 2DPA took place. \\
\indent The leading order observations from \hl{Fig. 14a} are reprised in the steady state case (\hl{Fig. 14b}). Here, the inward drift of the ice-line is prohibited such that only embryos in wind-dominated disks at low $\tilde{\alpha}$ can partake in 2DPA (\hl{Fig. 12}). Without the aid of time evolution (or more specifically, the accompanying increase in $St$), more disks are characterized by accretion times exceeding 10 Myr. Across the PA dominated region at low $\tilde{\alpha}$ and high $\psi$, $M_{PA}/M_{PW}$ values are comparable to those in the time-evolving case, owing to relatively large $t_{Acc,0}$. In other words, these disks evolve slowly so the removal of time evolution hardly impacts embryo growth within them. The final planetary masses in these disks range from $\sim 6$ to $8$ $M_{\oplus}$. In the discarded region, planetary masses at 10 Myr \hl{increase} from high $\tilde{\alpha}$ and low $\psi$ towards the boundary with the non-discarded region, going from $\sim 10^{-3}$ to $8$ $M_{\oplus}$. Corresponding ratios of $M_{PA}/M_{PW}$ at 10 Myr range from $\sim 10^{-3}$ to $10$. As in the time-evolving case, it is safe to infer that rocky \hl{Mars-sized} and larger embryos in such disks grow primarily through pairwise collisions.\\
\begin{figure*} 
\centering
\scalebox{0.67}{\includegraphics{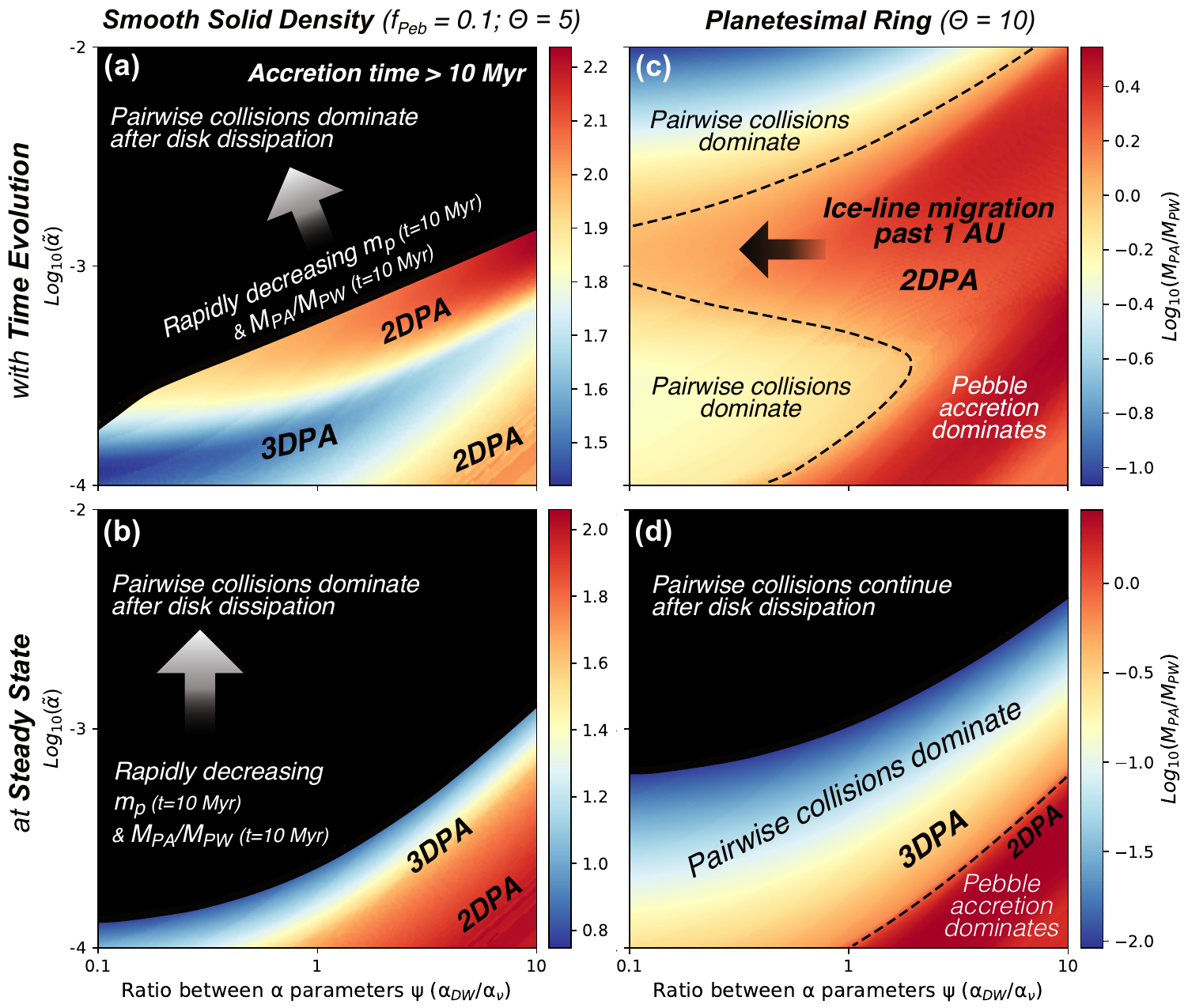}}
\caption{\footnotesize\textbf{Rasters of $Log_{10}(M_{PA}/M_{PW})$ for the growth of a 500 km-sized rocky embryo towards 8 $M_{\oplus}$ at $r=1$ AU, assuming accretion in (a) "smooth", time-evolving disks, (b) "smooth", steady state disks, (c) a planetesimal ring within time-evolving disks, and (d) a planetesimal ring within steady state disks.} In disks with "smooth" solid densities (a \& b), PA dominates growth for high $\psi$ and low $\tilde{\alpha}$ (\textit{i.e.,} bottom right of parameter space). Pairwise planetesimal collisions dominate growth at lower $\psi$ and higher $\tilde{\alpha}$, as PA ceases with disk dissipation. In planetesimal rings (c \& d), it appears a substantial region of parameter space is still PA-dominated. However, $M_{PA}/M_{PW}$ in this region is at most $\sim3$. Given the overestimation of PA rates, the omission of ``pebble filtering" effects, etc. (see text for full discussion), pairwise collisions likely dominate growth across virtually all parameter space. See Section 4.3.3.}
\label{fig:Figure 14}
\end{figure*}
\indent The paradigm of planet formation in disks with “smooth” solid densities have recently been challenged by models suggesting rapid embryo growth in planetesimal rings arising at pressure maxima in protoplanetary disks, notably associated with principal (\textit{i.e.,} silicate and water) sublimation lines (Cai et al., 2022; Izidoro et al., 2022; Morbidelli et al., 2022; Batygin \& Morbidelli, 2023). These models are buttressed by observations of ringed structures in T-Tauri disks by ALMA (\textit{e.g.,} Andrews et al., 2016; Dullemond et al., 2018). \hl{As an alternative to accretion in ``smooth" disks, we} simulate the growth of the $500$ km-sized embryo to super-Earth mass in this emerging framework, once again with and without time evolution. In our model, the planetesimal ring that develops around $1$ AU constitutes an annulus with a radial width of $\sim0.3$ AU hosting $\sim 2.5$ $M_{\oplus}$. This translates to a solid density (held in embryos) of $\sim500$ kg/m\textsuperscript{3} as assumed by Batygin \& Morbidelli (2023). Within such a ring, inelastic collisions coupled with aerodynamic drag dampens the embryo velocity dispersion, enhancing gravitational focusing. This motivates us to adopt a slightly higher Safronov number of $\Theta = 10$. The pebble flux into the ring is taken to be $\dot{M}_{peb}\sim100$ $M_{\oplus}$/Myr (Bitsch et al., 2019; Lenz et al., 2019), \hl{such that the solid (pebble) density $\Sigma_{peb}\simeq \dot{M}_{peb}/2\pi r v_r$, with $r = 1AU$ and $v_r$ evaluated there accordingly.} Results from our simulations are shown in \hl{Figs. 14c \& 14d}.\\ 
\indent In the time-evolving case (\hl{Figs. 14c}), growth is completed within 10 Myr across parameter space. The dominance of PA (\textit{i.e.,} $M_{PA}/M_{PW}>1$) in all but turbulence-driven disks at high and moderately low $\tilde{\alpha}$ (\textit{i.e., }$> 10^{-2.5}$ and $<10^{-3}$) is attributed to the efficient 2D regime, the widespread accessibility to which results from rapid ice-line migration past $1$ AU for high $\tilde{\alpha}$ ($\gtrsim10^{-3}$) (\hl{Fig. 12}). Recall from Section 4.3.1 that for high $\tilde{\alpha}$, the ice-line sits within $1$ AU after only $1$ Myr has passed. Nonetheless, for low $\psi$ (\textit{i.e.,} substantial turbulence), the transitory stellar-planet mass ratio between the 3D- and 2DPA (\textit{i.e., } $(m_p/M_*)_{2D-3D}$) sits above the pebble isolation mass ratio $(m_p/M_*)_{Iso}$. It is due to this confinement in 3DPA that enables the domination of pairwise collisions. Growth to the terminal mass of  $8 M_{\oplus}$ can still occur in $< 10$ Myr in these disks owing to both enhanced collisional growth in the planetesimal ring and the sharp increase in $St$ accompanying the introduction of icy pebbles past the ice-line. Keeping in mind inferences on the volatile-poor nature of super-Earths (\textit{e.g.,} Owen \& Wu, 2017; Gupta \& Schlichting, 2019; see Section 4.4), it is unclear if the substantial accretion of icy \hl{particles} in high $\tilde{\alpha}$ disks as suggested here (and in Fig. 4a) is meaningful, motivating consideration of the steady state case \hl{in which ice-line migration is prohibited}. Before moving on, we note that while PA prevails over classical growth across a broad swath of parameter space, it does so by only a factor of \hl{a few}. Considering the overestimated PA accretion rates and probable reductions in the pebble flux over time due to say, “pebble filtering” by planets forming at farther radii (\textit{e.g.,} Izidoro et al., 2021a), the significance of PA is dubious. It is likely that in most disks ($\tilde{\alpha}-\psi$ pairs), rocky embryos in the planetesimal ring grew primarily via pairwise collisions. Only in wind-dominated disks could the contribution of pebbles and other embryos to growth be comparable. \\
\indent Interestingly, the final planetary masses (\textit{i.e.,} pebble isolation masses) in the planetesimal ring model decrease with increasing $\psi$, a trend opposite of that in our model of a “smooth” disk solid density (see discussion above for \hl{Fig. 14a}). This reflects the faster growth rates in planetesimal rings, which enable embryos in turbulence-driven disks to exceed pebble isolation masses in wind-driven disks before pebble isolation masses in their own disks truncate their growth. \\
\indent In the steady state case (\hl{Fig. 14d}), the prohibition of ice-line migration changes the morphology of parameter space dramatically. Here, only rocky embryos in wind-dominated disks at low $\tilde{\alpha}$ can access 2DPA, and grow primarily via PA. As in the time-evolving case, the cumulative contribution from pebbles surpasses that from collisions with other embryos by only a small factor ($<3$). Given the aforementioned advantages conferred on PA in our model, it is safe to conclude that pairwise collisions contribute a comparable if not greater amount of mass to planetary growth, constituting the primary mode of accretion even in such quiescent disks. Across the rest of parameter space, rocky embryos in planetesimal rings owe virtually all their growth to pairwise collisions. It is worth noting that in for high $\tilde{\alpha}$ and low $\psi$ (\textit{i.e.,} the discarded region), the planetary mass at $10$ Myr far exceeds $1 M_{\oplus}$, ranging from $\sim 6.7 M_{\oplus}$ to just short of $8 M_{\oplus}$. Thus, rapid classical growth is facilitated across all parameter space by planetesimal rings (see Section 4.4). \\
\indent In reality, the evolution of most disks likely falls somewhere between the steady state and (rapidly) time-evolving cases explored. The $10$ Myr cutoff to PA is likely an overestimate, given that disks are inferred to dissipate only a few Myr following planetesimal formation. The application of our model to the SS, for instance, requires implementation of a dissipation time $< 4$ Myr for consistency with paleomagnetic meteorite analyses (see Section 4.3.2). Dissipation timescales $< 10$ Myr would lend support to the dominance of classical growth, as less of the final planetary mass can be attributed to PA. Additional support for classical growth, to summarize key points from above, include (i) the overestimation of PA rates, (ii) the negligence of “pebble filtering” effects from planet formation beyond 1 AU, which would decrease the pebble flux at 1 AU over time, (iii) the fact that the \hl{density of solids stored in embryos} ought to increase with time \hl{(not decrease as $f_{Emb}f_d\Sigma$ does in our model, since $f_{Emb}$ and $f_d$ are kept constant)}, and finally, (iv) the prevalence of disk substructures such as concentric rings and gaps in a host of spatially resolved observations of protoplanetary disks, \hl{suggestive of} dust trapping and planet formation in planetesimal rings. \hl{Even in the absence of rings, (iii) would accelerate accretion by pairwise collisions, which is clearly underestimated in high $\tilde{\alpha}$ disks of Figs. 14a \& 14b}. All things considered, it appears the dominance of PA in rocky planet formation can only be realized in disks (i) driven by MHD winds (\textit{i.e.,} $\psi\gtrsim 1$), (ii) undergoing low torques (\textit{i.e., }$\tilde{\alpha}\lesssim 10^{-3.5}$ ; slow-evolving), and (iii) absent of pressure maxima that evolve rapidly into planetesimal rings within which classical growth is strongly enhanced. \hl{Criteria (i) and (ii) imply disks with $\alpha_{\nu}\lesssim 10^{-4}$, significantly lower than the range inferred for protoplanetary disks by astronomical observations (Rosotti, 2023), which probe $\alpha_{\nu}$ far beyond the ice-line ($\alpha_{\nu}$ is likely higher closer in to the star; see Section 5). This further supports the notion that rocky planets grow primarily by pairwise planetesimal collisions.}

\subsection{Origin of Rocky Super-Earths\\ in Planetesimal Rings}
\indent Exoplanet surveys from the \textit{Kepler} space telescope have revealed the ubiquity of low mass (several $M_{\oplus}$) planets orbiting close (<< 1 AU) to their host stars (Petigura et al., 2013; Mulders et al., 2018), and that such planets exhibit a so-called “radius valley” in their population between $\sim 1.5$ to $2 R_{\oplus}$ (Fulton et al., 2017).  It is conventionally accepted that the smaller “super-Earths” ($R\lesssim1.5 R_{\oplus}$) originated as the larger “sub-Neptunes” ($R\gtrsim2 R_{\oplus}$), having lost their primordial H\textsubscript{2} atmospheres (Bean et al., 2021). The provenance of these planets in their protoplanetary disks remains enigmatic, largely reflecting the uncertainty in their bulk compositions. The two prevailing models for how sub-Neptunes evolve into super-Earths—photoevaporation (\textit{e.g.,} Owen et al., 2013; Owen \& Wu, 2017) and core-powered mass loss (Gupta \& Schlichting, 2019)—constrain sub-Neptune core compositions to be volatile-poor. Along with predictions that super-Earths originating in the outer disk will be enriched in ices (Izidoro et al., 2021b), these constraints point to their accretion within the ice-line. This picture of “dry” accretion has been challenged by increasingly detailed analyses of \textit{Kepler} planet size distributions, suggesting a subset of super-Earths may in fact be water-rich (\textit{e.g.,} Teske et al., 2018). Adding to the complexity, initially water-rich super-Earths ($\lesssim 3 M_{\oplus}$) can lose virtually all their water ice to photoevaporation close to their host stars ($\lesssim 0.03$ AU) (Kurosaki et al., 2014). Moreover, accretion of rocky pebbles during the inward migration of icy cores can result in final, \hl{largely} volatile-poor compositions (Bitsch et al., 2019).\\
\indent Although it is unclear whether super-Earths (\textit{i.e.,} sub-Neptune cores) formed interior to the ice-line (\textit{e.g.,} Chatterjee \& Tan, 2013; Ida \& Lin, 2010) or across a broad swath of the disk (\textit{e.g.,} Coleman \& Nelson, 2016), there is general consensus that they did not form \textit{in-situ} (\textit{i.e.,} at their present-day locations) (Inamdar \& Schlichting, 2015; Ogihara et al., 2015). This sets a timescale constraint for rocky super-Earth accretion in the inner disk, as growth must be sufficiently rapid as to allow for disk-driven (Type-I) inward migration prior to disk dissipation. While such rapidity can be achieved by volatile-rich cores undergoing PA in the outer disk, it cannot, for most of parameter space, be realized by rocky cores in the 3DPA-confined inner disk assuming a “smooth” solid density (see Section 4.3.3). Indeed, only in relatively slow-evolving and wind-dominated “smooth” disks (\hl{Figs. 14a \& 14b}) may embryos grow sufficiently fast to drift inwards before disk dissipation (recall that disk dissipation timescales are a few Myr, not the conservative $10$ Myr assumed). In both the time-evolving and steady state cases (Figs. 14a, b), only 500 km-sized rocky embryos in disks with $\tilde{\alpha}\lesssim 10^{-3}$ and $\psi\gtrsim 1$ reach the size of Mars in $< 4$ Myr. While PA in such wind-dominated disks can facilitate rapid growth, we stress that this assumes pebbles are in constant supply. Furthermore, as discussed in Section 4.5, the terrestrial isotopic composition does not support a considerable contribution of pebbles sourced from the outer SS to the Earth’s accretion. \\
\indent Planetesimal rings offer a more compelling scenario for rapid planetary growth. In the time-evolving  (\hl{Fig. 14c}) and steady state (\hl{Fig. 14d}) cases, rocky embryos grow to Mars size in $<1.5$ Myr and $<2.5$ Myr across parameter space, respectively. In the model for super-Earth formation recently posited by Batygin \& Morbidelli (2023), accretion by mutual planetesimal collisions unfolds in a planetesimal ring centered on the silicate sublimation line. The ring arises from rocky grains that condense from silicate vapor diffusing outward across the said line early in the disk’s life, which participate in gravito-hydrodynamic instabilities to form planetesimals. While the ring may instead owe its origin to the inward drift and pile-up of rocky pebbles at pressure bumps (\textit{e.g., }Flock et al., 2019), the exact mechanism by which it forms is hardly as important as the benefits it reaps. Not only \hl{is} the planetesimal ring advantageous from a timescale perspective, it lends itself nicely to explaining the galactic abundance and intra-system uniformity of super-Earths. Regarding the latter point, growth in the ring proceeds until planetary masses are sufficiently large for removal by Type-I migration. In this way, the terminal masses of super-Earths “manufactured” in a planetesimal ring, within a given disk, are regulated. In the framework of Batygin \& Morbidelli (2023), the dearth of super-Earths in our SS is simply a consequence of excessive turbulence, which prevented the construction of a ring sufficiently massive to induce growth in time for disk-driven inward migration.\\
\indent To conclude, the emerging paradigm of planetesimal rings is supported by both the\hl{astronomically observed ubiquity} of disk cavities and dust traps, and its ability to explain the statistics of a substantial exoplanet population. If super-Earths formed ``dry", planetesimal rings offer an avenue for accelerated growth in the inner disk, allowing for disk-driven migration close to their host stars before disk dissipation.

\subsection{Earth's Accretion from Planetesimal Collisions}
\indent In Section 4.3.3, we concluded that PA can dominate rocky planet formation only in \hl{quiescent} disks characterized by “smooth” solid densities. The prevalence of classical growth suggested by this result, from a statistical and observational standpoint, has two key implications for Earth’s accretion. First, it argues against the rapid accretion of the proto-Earth (\textit{i.e., }pre-Moon-forming Giant Impact; MGI) by pebbles, as suggested previously by several authors (\textit{e.g.,} Johansen et al., 2015; Schiller et al., 2020). It does not, however, preclude the possibility that the proto-Earth accreted rapidly by pairwise planetesimal collisions before the dissipation of the SS disk in a few Myr. As illustrated in \hl{Figs. 14c \& 14d}, and discussed above, classical growth can be greatly accelerated in planetesimal rings, and dominates rocky embryo growth across a broad span of parameter space. If the proto-Earth accreted within such a ring, growth must still have been slow enough such that the SS disk dissipated before substantial inward migration occurred. Interestingly, the relatively low abundance of \textsuperscript{182}W in the present-day mantle indicates a series of large impacts (\textit{i.e.,} planetesimal collisions) necessarily accompanies rapid early accretion. This reflects the fact that an early (\textit{i.e.,} pre-dissipation) proto-Earth mantle would be enriched in \textsuperscript{182}W today from the decay of \textsuperscript{182}Hf, lest large impacts facilitate extensive core-mantle equilibration (Olson \& Sharp, 2023). \\
\indent A significant contribution to Earth’s accretion from pairwise collisions is indirectly supported by the MGI, which may be considered the final major step in forming Earth. The timing inferred for the MGI ($\gtrsim 50$ Myr after CAIs) from Hf-W isotope systematics of lunar materials (Barboni et al., 2017; Kruijer et al., 2017b; Thiemens et al., 2021) postdates disk dissipation by tens of Myr, and is consistent with a prolonged accretion of the proto-Earth.\\
\indent Another implication of our work is that the proto-Earth did not incorporate a substantial fraction of outer SS (\textit{i.e.,} carbonaceous; CC) material in the form of pebbles. This is in agreement with analyses of nucleosynthetic isotope anomalies in multi-element space (\textit{e.g.,} Burkhardt et al., 2021) that suggests the Earth accreted $\lesssim 10\%$ of CC material by mass, assuming enstatite chondrites (which bear the closest isotopic resemblance to the Earth; \textit{e.g.,} Dauphas, 2017), constitute the Earth’s non-carbonaceous (NC) component. Chemical considerations also support an inner disk, predominantly NC origin of the Earth (\textit{e.g.,} Sossi et al., 2022; Tissot et al., 2022; Liu et al., 2023). Notably, our results contradict the model for Earth’s accretion recently proposed by Onyett et al. (2023) on the basis of Si isotope anomalies. This model considers only Si anomalies, and assumes the primary building blocks of Earth are derived from a mixture of NC achondrite material and CI (carbonaceous Ivuna) material. As a consequence, it suggests that $\gtrsim 25 \%$ of the proto-Earth mass is attributed to CI-like pebbles from the outer SS. The assumption of NC achondrite-CI mixing is founded upon an inferred temporal evolution in the isotopic composition of inner disk materials from the introduction of CI-like pebbles, itself based on the achondrite-chondrite dichotomy observed in Si anomalies. This dichotomy contrasts with the conventional NC-CC dichotomy observed for all other elements (\textit{e.g.,} Burkhardt et al., 2019; Kleine et al., 2020) and is possibly due to the inadequate correction for natural mass dependent fractionation with the exponential law, especially considering the relatively small dispersion of Si anomalies across SS materials (Dauphas et al., 2023). We emphasize that pebbles, given a prolonged accretion of Earth, are not expected to contribute significantly to rocky planet formation (\hl{Fig. 14}).\\
\indent The heliocentric, \textit{s-process} trend observed for NC meteorites in multi-element space (Burkhardt et al., 2021) point to a missing, isotopically Earth-like endmember in the innermost disk that has yet to be isolated. Recently, Yap \& Tissot (2023) suggested that this missing component may represent the bulk composition of the SS parent molecular cloud. Regardless of its true identity, the inclusion of this component to mixing models for the Earth serves to diminish the fraction of CC material required to explain the terrestrial composition. 

\section{Implications of Omitted Details}
\indent Having dedicated much of the last section to rocky planet formation, we now return to the disk-wide picture of the characteristic \textit{St} and its implications, as addressed in Sections 3, 4.1, \& 4.2. Specifically, we discuss here the potential implications of several significant assumptions made in our disk and dust-gas coupling model. 
\begin{figure} 
\scalebox{0.575}{\includegraphics{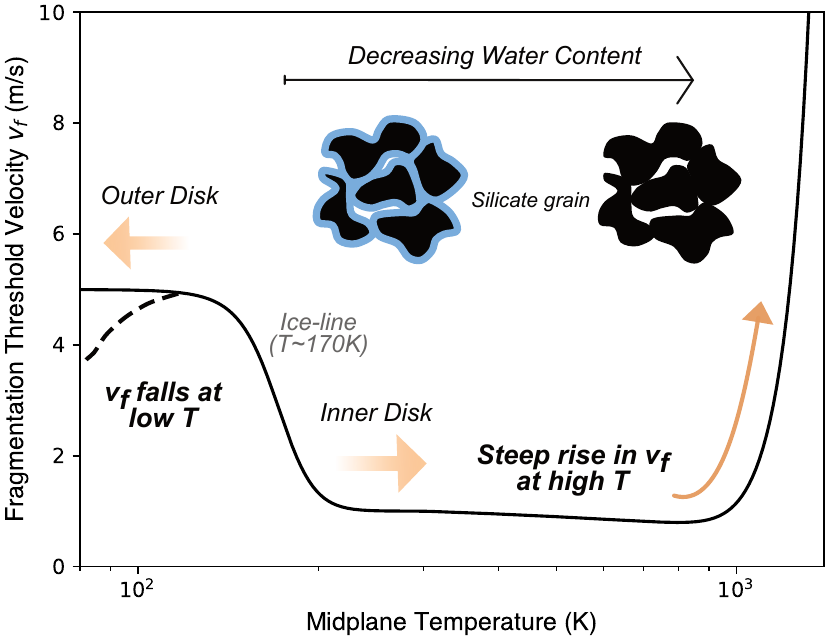}}
\caption{\textbf{\footnotesize Profile of the fragmentation threshold velocity as a function of temperature, having implemented the increase in surface energy at high temperatures due to extensive drying (\textit{i.e.,} the evaporation of the monolayer of water on grain surfaces/interstices.} At $T\simeq 1300K$, $v_{f,rock}$ rises to $\simeq 9$ m/s, close to $v_{f,ice}\simeq 10$ m/s. See Section 5.}
\label{fig:Figure 15}
\end{figure}
\indent As we have shown, fragmentation is the dominant process limiting growth in the absence of the bouncing barrier, resulting in the dramatic rise in $St$ across the \hl{ice-line}. This dichotomy in $St$ translates to that in characteristic particle sizes, radial drift velocities, and degrees of vertical settling, the latter controlling the dominant mode of planetary accretion. This result reflects the fragmentation threshold velocities $v_{f}$ we have prescribed to silicate and icy particles within and without the ice-line, respectively. That said, it is worthwhile if not proper to consider the temperature dependence of $v_{f}$, neglected above for simplicity. In particular,  $v_{f,rock}$ is expected to increase towards the silicate sublimation line at $T\simeq 1500 K$, while $v_{f,ice}$ is expected to decrease at temperatures lower than $T\sim 200K$. Indeed, Pillich et al. (2021) found that at temperatures exceeding $\sim 1200K$, the evaporation of water at the interstices and surfaces of silicate dust grains (coupled with phase changes such as the reduction of Fe in silicates to metallic Fe; Pillich et al., 2023) leads to an increase in surface energies, and thus adhesion, by orders of magnitude. \hl{Sintering may also contribute to higher $v_{f,rock}$ (Poppe, 2003).} Musiolik \& Wurm (2019) found that the surface energy of water ice falls by a factor of $\simeq60$ simply in going from $T>200K$ to $T\simeq 175K$, suggesting that icy particles a few AU or less beyond the ice-line will not have an advantage over silicate particles in sticking.\\
\indent Let us first consider the expected increase in $v_{f,rock}$ at high temperatures, using dimensional analysis as a guide to a quantitative determination of how $v_{f,rock}$ may vary towards the disk inner edge. From Pillich et al. (2021), the effective surface energy $\gamma_{e}$ as a function of temperature is modeled as such: 
\begin{equation}
\gamma_{e} = \gamma_{e_{w}}(f_{w}e^{\frac{300-T}{T_{w}}} + f_{sd}e^{\frac{T-300}{T_{sd}}}),
\end{equation}
where $300K\lesssim T\lesssim1300K$, and the normalizing constants $\gamma_{e_{w}}\simeq 0.07$ J/m$^{2}$, $f_{w}\simeq 8$, $f_{sd}\simeq 2\times10^{-4}$, $T_{w}\simeq 950.06K$, and $T_{sd}\simeq 67.16K$. The functional form of $\gamma_{e}$ sets that of $v_{f} (T>300K)$, and we relate the two quantities by setting the specific energy for catastrophic disruption (Benz \& Asphaug, 1999; Leinhardt \& Stewart, 2009)  $Q = m_{1}v_{f}^{2}/2m_{2}$ equal to $C\gamma_{e}$ where $C$ is a constant of normalization with units of m$^{2}$/kg (such that $C\gamma_{e}$ has units of velocity squared):
\begin{equation}
v_{f}(T>300K) = \sqrt{2C\gamma_{e}} = C'\sqrt{\gamma_{e}},
\end{equation}
where we have assumed that the colliding particles are of equal mass and grouped the constant $\sqrt{2C}$ into another constant $C'$, which will be used to to normalize $v_{f}$ at $T\simeq300K$ to $v_{f,rock}(T=300K) = 1$ m/s. The profile of $v_{f}(T)$ which results from implementing Eq. 35 is shown in \hl{Fig. 15}. At $T\simeq 1300K$, $v_{f,rock}$ has steeply risen to $\simeq 8.5$ m/s, higher than the value of $v_{f,ice}$ assumed. This rise in $v_{f,rock}$ appears to suggest that there are conditions (\textit{i.e.,} low $\tilde{\alpha}$ and high $\psi$) for which the dry silicate particle in the inner disk can reach relatively high $St$, settle considerably, and consequently partake in 2DPA. Nonetheless, the image of a thin sub-disk of dust at the disk inner edge is at odds with intuition, namely the notion the inner region ought to be extremely turbulent due to high ionization fractions for which the MRI operates close to the ideal MHD limit (Armitage, 2020). Here, we call attention to another key assumption we have made in our analysis: that $\tilde{\alpha}$ (including both $\alpha_{\nu}$ and $\alpha_{DW}$ themselves) are constant in space and time.\\
\indent To first-order, $\tilde{\alpha}$ likely peaks close to the inner edge of the disk, and falls off with distance from the star. This is true for disks dominated by MRI-driven turbulence (\textit{e.g.,} Penna et al., 2013) and disk winds so long as the strength of the magnetic field threading the disk decreases outward as well. \hl{Moreover, measurements of IR molecular line broadening suggests the inner edge of disks are characterized by $\alpha_{\nu}\gtrsim 10^{-2}$ (\textit{e.g.,} Carr et al., 2004), far higher than values estimated by other techniques (\textit{e.g.,} sub-mm line broadening, dust scale height) for tens or hundreds of AU (\textit{i.e.,} $\alpha_{\nu}\sim 10^{-4}$ to $10^{-3}$) (Rosotti, 2023).} As $\tilde{\alpha}$ rises for a given $\psi$, the characteristic $St$ as set by turbulent fragmentation and fragmentation from relative drift velocities will decrease through both $\tilde{\alpha}$ explicitly and the isothermal sound speed $c_{s}\sim T^{1/2}$ (Eqs. 17 \& 19). High $\tilde{\alpha} (\gtrsim 10^{-2})$ likely outweighs the increase in $v_{f,rock}$ in the innermost regions of the disk such that 2DPA is only attainable after several units of  $t_{Acc,0}$ have passed ($\tilde{\alpha}$ may also decay with time; see below), at which point the disk has probably dissipated. \\
\indent The counteraction of effects from $\tilde{\alpha}$ and $v_{f}$ may also apply to the outer disk. With $\tilde{\alpha}$ kept constant, the decrease in $v_{f,ice}$ with $T\lesssim200K$ implies that the characteristic $St$ of icy particles (or where $T\lesssim150K$;$\gtrsim 10$ AU for the turbulence-dominated disk, and even closer for the wind-dominated disk) will plunge towards values close to those of their silicate counterparts in the inner disk, dampening 2DPA therein. In the event that $\tilde{\alpha}$ decays with distance, it would serve to increase $St$. If $v_{f,ice}\simeq$$v_{f,rock}$ for most of the disk at low temperatures ($\lesssim 150$K), the dramatic rise in $St$ across the ice-line would be limited in extent (resembling a bump if you will). With the continual decrease of $\tilde{\alpha}$ beyond the ice-line, the outer disk will still host particles larger and more readily settled than the inner disk. \\
\indent Having considered the spatial variation in $\tilde{\alpha}$, let us consider briefly the implications of its temporal variation.  The decay of $\alpha_{\nu}$ (or $\tilde{\alpha}$ for low $\psi$) early in the disk's life is expected as infall onto the disk and gravitational instabilities dwindle (\textit{e.g.,} Morbidelli et al., 2022). Following the short-lived infall stage, the evolution of $\alpha_{\nu}$ will largely be dictated by temperature, which controls the extent of ionization (of elements such as K) and thus the vigor of the MRI (Armitage, 2020). The evolution of $\alpha_{DW}$, similarly, is governed by the strength of magnetic field lines threading the disk, giving rise to MHD winds.  We proceed by assuming $\tilde{\alpha}$ globally decays as time passes, noting that the ensuing points can simply be reversed for a growing $\tilde{\alpha}$. The decrease in $\tilde{\alpha}$ results in an increase in characteristic $St$ across the disk. Graphically, this means that the $St$ profiles in Figs. 5c \& 5d will not only evolve to the left towards the star, but also up towards higher values. Over time, then, particles across the disk are more prone to gravitational settling and higher drift rates due to increased headwind drag. Moreover, the threshold to transition from 3DPA to 2DPA decreases more rapidly at any given location in the disk, \hl{and PA rates increase}. The decay rate of $\tilde{\alpha}$ may scale with $\tilde{\alpha}$ itself, such the effects discussed are most applicable to disks with high ($\gtrsim 10^{-3}$) $\tilde{\alpha}$. \\
\indent Several, second-order, avenues for improving the model at hand include the implementation of a size dependence for $v_{f}$ (see Section 4.1) and ice-lines for other prominent condensates such as CO and CH$_{4}$. We do not foresee these details changing the key takeaways from this work to any significant degree. 

\section{Concluding Points}
\indent In this work, we model the degree of dust-gas coupling across protoplanetary disks evolving under both turbulence and MHD disk winds. The key conclusions of our work are the following:
\begin{itemize}
\item Absent of the bouncing barrier, fragmentation is the dominant mechanism limiting particle growth, trumping radial drift in setting the characteristic Stokes number of particles across the disk. 
\item Assuming icy particles fragment at higher collisional velocities than silicate particles, the water-ice sublimation line constitutes a crucial transition point in particle Stokes number, and consequently, dust settling, drift velocity, and planetary accretion regimes. Namely, higher $St$ translates to a thinner solid sub-disk, faster radial drift velocities, and a lower threshold to enter the 2D pebble accretion regime. 
\item For $\tilde{\alpha}=10^{-3}$, silicate particles interior to the ice-line in turbulence-dominated ($\psi\simeq0.1$) and wind-dominated ($\psi\simeq 10$) disks are characterized by $St\lesssim 10^{-3}$ and $St\lesssim 10^{-2}$, respectively, corresponding to sizes in the mm- to 1-cm-scale. Icy particles beyond the ice-line possess Stokes numbers $\sim$ two orders of magnitude larger than their silicate counterparts, at $St\gtrsim10^{-2}$ and $St\gtrsim 0.1$, corresponding to the cm- to dm-scale. Bouncing-limited particles are smaller than fragmentation-limited particles, and assume sizes from the tens to hundreds of microns, depending on disk parameters. 
\item For most of parameter space, with the exception of a rather small corner at low $\tilde{\alpha}$ (\textit{i.e.,} $\lesssim 10^{-3.5}$) and moderately high $\psi$ (\textit{i.e.,} $\gtrsim 1$), planetary embryos interior to the ice-line (\textit{i.e.,} at $r=1$ AU) are confined to the 3D regime of pebble accretion. Inaccessibility to 2D pebble accretion in the inner disk, however,  does not necessarily imply the dominance of classical growth \hl{by pairwise collisions between embryos.} 
\item       For pebble accretion to dominate rocky planet formation, the main stage of growth must be (i) completed within a few million years, prior to disk dissipation, and be (ii) attributed mostly to pebbles. This can only be realized in disks whose evolution is (i) driven by MHD winds (\textit{i.e.,} high $\psi$) and (ii) relatively slow owing to low torques (\textit{i.e.,} low $\tilde{\alpha}$), and (iii) in the absence of pressure maxima that can facilitate the concentration of solids in the planetary feeding zone, resulting in planetesimal rings that significantly accelerate classical growth. \hl{Such disks are extremely quiescent, with $\alpha_{\nu}\lesssim 10^{-4}$.} Thus, \hl{for almost all of parameter space corresponding to $\alpha_{\nu}$ values observed for protoplanetary disks ($\gtrsim$ few $\times 10^{-4}$)}, pairwise collisional growth is predicted to dominate pebble accretion interior to the water ice-line, serving as the primary mode of rocky planet accretion.
\item For planetary embryos exterior to the ice-line (\textit{i.e.,} at $r= 20$ AU), pebble accretion dominates classical collisional growth for all parameter space, regardless of whether the planetary embryo lies in the 2D or 3D pebble accretion regime. The rapid accretion of giant planet cores suggests the \hl{outer} SS disk was not turbulence-dominated, such that 2D pebble accretion was accessible \hl{therein}. \hl{Rapid growth can be facilitated even in the 3D regime with the aid of sublimation fronts that elevate local solid densities.}
\item Super-Earth formation in planetesimal rings is supported by the rapidity of rocky planet accretion facilitated by collisional growth within such rings, allowing for disk-driven migration to take place before disk dissipation. The planetesimal ring framework is also favored from an observational standpoint, and for its ability to explain the galactic abundance and intra-system uniformity of super-Earths. 
\item The dominance of classical growth argues against a significant ($\gtrsim 10\%$) contribution of outer disk, carbonaceous material to the Earth in the form of pebbles, in accordance with chemical and isotopic investigations of Earth's accretion history. \\
\end{itemize}
\indent We conclude by emphasizing that growth rates associated with different accretionary processes depend on both the overall (solid and gas) disk structure as well as the Stokes number, and quantitative measures must be undertaken to delineate the details of their competition. Furthermore, insofar as our results on rocky planet accretion are concerned, MHD winds can be replaced by any mechanism aside from turbulent viscosity that exerts a torque on the disk, driving stellar accretion. Seeing as fragmentation is key in setting the characteristic Stokes number throughout the disk, future models should feature a more refined prescription of fragmentation threshold velocities as constrained by laboratory experiments. A prescription of spatially and/or temporally variable $\alpha$ parameters also represents a worthwhile pursuit.

\section*{Declaration of Competing Interest}
\indent The authors declare that they have no known competing financial interests or personal relationships that could have appeared to influence the work reported in this paper. 

\section*{Acknowledgements}
\indent \hl{We thank three anonymous reviewers for their constructive reviews on the manuscript, and editor Sean Raymond for prompt and careful editorial handling. We also thank François L. H. Tissot for advice on improving clarity in communicating our results. This work was supported by a Packard fellowship and National Science Foundation grant (AST 2109276) to KB, as well as the Caltech Center for Comparative Planetary Evolution (3CPE).}

\section*{References}
Ansdell, M., Williams, J. P., Manara, C. F., Miotello, A., Facchini, S., van der Marel, N., ... \& van Dishoeck, E. F. (2017). An ALMA survey of protoplanetary disks in the $\sigma$ Orionis cluster. \textit{The Astronomical Journal}, \textit{153}(5), 240.\\

Andrews, S. M., Wilner, D. J., Hughes, A. M., Qi, C., \& Dullemond, C. P. (2009). Protoplanetary disk structures in Ophiuchus. The Astrophysical Journal, 700(2), 1502.\\

Andrews, S. M., Wilner, D. J., Zhu, Z., Birnstiel, T., Carpenter, J. M., Pérez, L. M., ... \& Ricci, L. (2016). Ringed substructure and a gap at 1 au in the nearest protoplanetary disk. \textit{The Astrophysical Journal Letters}, \textit{820}(2), L40.\\

Armitage, P. J. (2020). Astrophysics of planet formation. Cambridge University Press.\\

Balbus, S. A., \& Hawley, J. F. (1998). Instability, turbulence, and enhanced transport in accretion disks. Reviews of modern physics, 70(1), 1.\\

Barboni, M., Boehnke, P., Keller, B., Kohl, I. E., Schoene, B., Young, E. D., \& McKeegan, K. D. (2017). Early formation of the Moon 4.51 billion years ago. \textit{Science advances}, \textit{3}(1), e1602365.\\

Batygin, K., \& Morbidelli, A. (2020). Formation of giant planet satellites. \textit{The Astrophysical Journal}, \textit{894}(2), 143.\\
 
Batygin, K., \& Morbidelli, A. (2022). Self-consistent model for dust-gas coupling in protoplanetary disks. Astronomy \& Astrophysics, 666, A19.\\

Batygin, K., \& Morbidelli, A. (2023). Formation of rocky super-earths from a narrow ring of planetesimals. Nature Astronomy, 7(3), 330-338.\\

Bean, J. L., Raymond, S. N., \& Owen, J. E. (2021). The Nature and Origins of Sub‐Neptune Size Planets. Journal of Geophysical Research: Planets, 126(1), e2020JE006639.\\

Beitz, E., Güttler, C., Blum, J., Meisner, T., Teiser, J., \& Wurm, G. (2011). Low-velocity collisions of centimeter-sized dust aggregates. The Astrophysical Journal, 736(1), 34.\\

Benz, W., \& Asphaug, E. (1999). Catastrophic disruptions revisited. Icarus, 142(1), 5-20.\\

Birnstiel, T., Ormel, C. W., \& Dullemond, C. P. (2011). Dust size distributions in coagulation/fragmentation equilibrium: numerical solutions and analytical fits. Astronomy \& Astrophysics, 525, A11.\\

Birnstiel, T., Klahr, H., \& Ercolano, B. (2012). A simple model for the evolution of the dust population in protoplanetary disks. Astronomy \& Astrophysics, 539, A148.\\

Bitsch, B., Morbidelli, A., Lega, E., \& Crida, A. (2014). Stellar irradiated discs and implications on migration of embedded planets-II. Accreting-discs. Astronomy \& Astrophysics, 564, A135.\\

Bitsch, B., Izidoro, A., Johansen, A., Raymond, S. N., Morbidelli, A., Lambrechts, M., \& Jacobson, S. A. (2019). Formation of planetary systems by pebble accretion and migration: growth of gas giants. \textit{Astronomy \& Astrophysics}, \textit{623}, A88.\\

Bitsch, B., Raymond, S. N., \& Izidoro, A. (2019). Rocky super-Earths or waterworlds: The interplay of planet migration, pebble accretion, and disc evolution. Astronomy \& Astrophysics, 624, A109.\\

Borlina, C. S., Weiss, B. P., Bryson, J. F., \& Armitage, P. J. (2022). Lifetime of the outer solar system nebula from carbonaceous chondrites. \textit{Journal of Geophysical Research: Planets}, \textit{127}(7), e2021JE007139.\\

Boss, A. P., \& Ciesla, F. J. (2014). The solar nebula. Planets, Asteroids, Comets and The Solar System, 2, 37-53.\\

Blandford, R. D., \& Payne, D. G. (1982). Hydromagnetic flows from accretion discs and the production of radio jets. Monthly Notices of the Royal Astronomical Society, 199(4), 883-903.\\

Blum, J., \& Münch, M. (1993). Experimental investigations on aggregate-aggregate collisions in the early solar nebula. Icarus, 106(1), 151-167.\\

Blum, J., \& Wurm, G. (2008). The growth mechanisms of macroscopic bodies in protoplanetary disks. \textit{Annu. Rev. Astron. Astrophys.}, \textit{46}, 21-56.\\

Briceno, C., Vivas, A. K., Calvet, N., Hartmann, L., Pacheco, R., Herrera, D., ... \& Andrews, P. (2001). The CIDA-QUEST large-scale survey of Orion OB1: Evidence for rapid disk dissipation in a dispersed stellar population. Science, 291(5501), 93-96.\\

Burkhardt, C., Spitzer, F., Morbidelli, A., Budde, G., Render, J. H., Kruijer, T. S., \& Kleine, T. (2021). Terrestrial planet formation from lost inner solar system material. Science advances, 7(52), eabj7601.\\

Cai, M. X., Tan, J. C., \& Portegies Zwart, S. (2022). Inside–out planet formation: VI. oligarchic coagulation of planetesimals from a pebble ring?. \textit{Monthly Notices of the Royal Astronomical Society}, \textit{510}(4), 5486-5499.\\

Carr, J. S., Tokunaga, A. T., \& Najita, J. (2004). Hot H2O emission and evidence for turbulence in the disk of a young star. \textit{The Astrophysical Journal}, \textit{603}(1), 213.\\

Chambers, J. E., \& Wetherill, G. W. (1998). Making the terrestrial planets: N-body integrations of planetary embryos in three dimensions. Icarus, 136(2), 304-327.\\

Chambers, J. E. (2009). An analytic model for the evolution of a viscous, irradiated disk. \textit{The Astrophysical Journal}, \textit{705}(2), 1206.\\

Chatterjee, S., \& Tan, J. C. (2013). Inside-out planet formation. The Astrophysical Journal, 780(1), 53.\\

Chiang, E. I., \& Goldreich, P. (1997). Spectral energy distributions of T Tauri stars with passive circumstellar disks. The Astrophysical Journal, 490(1), 368.\\

Concha-Ramírez, F., Wilhelm, M. J., Portegies Zwart, S., \& Haworth, T. J. (2019). External photoevaporation of circumstellar discs constrains the time-scale for planet formation. \textit{Monthly Notices of the Royal Astronomical Society}, \textit{490}(4), 5678-5690.\\

Cuzzi, J. N., Hogan, R. C., Paque, J. M., \& Dobrovolskis, A. R. (2001). Size-selective concentration of chondrules and other small particles in protoplanetary nebula turbulence. The Astrophysical Journal, 546(1), 496.\\

Dauphas, N., \& Chaussidon, M. (2011). A perspective from extinct radionuclides on a young stellar object: the Sun and its accretion disk. \textit{Annual Review of Earth and Planetary Sciences}, \textit{39}, 351-386.\\

Dauphas, N. (2017). The isotopic nature of the Earth’s accreting material through time. Nature, 541(7638), 521-524.\\
 
Dauphas, N., Hopp, T., \& Nesvorný, D. (2024). Bayesian inference on the isotopic building blocks of Mars and Earth. \textit{Icarus}, \textit{408}, 115805.\\
 
Dubrulle, B., Morfill, G., \& Sterzik, M. (1995). The dust subdisk in the protoplanetary nebula. icarus, 114(2), 237-246.\\

Dullemond, C. P., Birnstiel, T., Huang, J., Kurtovic, N. T., Andrews, S. M., Guzmán, V. V., ... \& Ricci, L. (2018). The disk substructures at high angular resolution project (DSHARP). VI. Dust trapping in thin-ringed protoplanetary disks. \textit{The Astrophysical Journal Letters}, \textit{869}(2), L46.\\

Ercolano, B., \& Pascucci, I. (2017). The dispersal of planet-forming discs: theory confronts observations. \textit{Royal Society Open Science}, \textit{4}(4), 170114.\\

Flock, M., Turner, N. J., Mulders, G. D., Hasegawa, Y., Nelson, R. P., \& Bitsch, B. (2019). Planet formation and migration near the silicate sublimation front in protoplanetary disks. Astronomy \& Astrophysics, 630, A147.\\

Friedrich, J. M., Weisberg, M. K., Ebel, D. S., Biltz, A. E., Corbett, B. M., Iotzov, I. V., ... \& Wolman, M. D. (2015). Chondrule size and related physical properties: A compilation and evaluation of current data across all meteorite groups. \textit{Geochemistry}, \textit{75}(4), 419-443.\\

Fulton, B. J., Petigura, E. A., Howard, A. W., Isaacson, H., Marcy, G. W., Cargile, P. A., ... \& Hirsch, L. A. (2017). The California-Kepler survey. III. A gap in the radius distribution of small planets. The Astronomical Journal, 154(3), 109.\\

Gerbig, K., \& Li, R. (2023). Planetesimal Initial Mass Functions Following Diffusion-regulated Gravitational Collapse. \textit{The Astrophysical Journal}, \textit{949}(2), 81.\\

Gundlach, B., Kilias, S., Beitz, E., \& Blum, J. (2011). Micrometer-sized ice particles for planetary-science experiments–I. Preparation, critical rolling friction force, and specific surface energy. Icarus, 214(2), 717-723.\\
 
Gundlach, B., \& Blum, J. (2014). The stickiness of micrometer-sized water-ice particles. \textit{The Astrophysical Journal}, \textit{798}(1), 34.\\

Gundlach, B., Schmidt, K. P., Kreuzig, C., Bischoff, D., Rezaei, F., Kothe, S., ... \& Stoll, E. (2018). The tensile strength of ice and dust aggregates and its dependence on particle properties. \textit{Monthly Notices of the Royal Astronomical Society}, \textit{479}(1), 1273-1277.\\

Gupta, A., \& Schlichting, H. E. (2019). Sculpting the valley in the radius distribution of small exoplanets as a by-product of planet formation: the core-powered mass-loss mechanism. Monthly Notices of the Royal Astronomical Society, 487(1), 24-33.\\

Güttler, C., Blum, J., Zsom, A., Ormel, C. W., \& Dullemond, C. P. (2010). The outcome of protoplanetary dust growth: pebbles, boulders, or planetesimals?-I. Mapping the zoo of laboratory collision experiments. Astronomy \& Astrophysics, 513, A56.\\

Hartmann, L., Calvet, N., Gullbring, E., \& D'Alessio, P. (1998). Accretion and the evolution of T Tauri disks. The Astrophysical Journal, 495(1), 385.\\

Hayashi, C. (1981). Structure of the solar nebula, growth and decay of magnetic fields and effects of magnetic and turbulent viscosities on the nebula. \textit{Progress of Theoretical Physics Supplement}, \textit{70}, 35-53.\\

Hunt, A. C., Cook, D. L., Lichtenberg, T., Reger, P. M., Ek, M., Golabek, G. J., \& Schönbächler, M. (2018). Late metal–silicate separation on the IAB parent asteroid: constraints from combined W and Pt isotopes and thermal modelling. \textit{Earth and Planetary Science Letters}, \textit{482}, 490-500.\\\\

Ida, S., Guillot, T., \& Morbidelli, A. (2016). The radial dependence of pebble accretion rates: A source of diversity in planetary systems-I. Analytical formulation. Astronomy \& Astrophysics, 591, A72.\\

Ida, S., \& Lin, D. N. C. (2010). Toward a deterministic model of planetary formation. VI. Dynamical interaction and coagulation of multiple rocky embryos and super-Earth systems around solar-type stars. \textit{The Astrophysical Journal}, \textit{719}(1), 810.\\

Inamdar, N. K., \& Schlichting, H. E. (2015). The formation of super-Earths and mini-Neptunes with giant impacts. \textit{Monthly Notices of the Royal Astronomical Society}, \textit{448}(2), 1751-1760.\\

Izidoro, A., Bitsch, B., \& Dasgupta, R. (2021a). The effect of a strong pressure bump in the Sun’s natal disk: terrestrial planet formation via planetesimal accretion rather than pebble accretion. \textit{The Astrophysical Journal}, \textit{915}(1), 62.\\

Izidoro, A., Bitsch, B., Raymond, S. N., Johansen, A., Morbidelli, A., Lambrechts, M., \& Jacobson, S. A. (2021b). Formation of planetary systems by pebble accretion and migration-Hot super-Earth systems from breaking compact resonant chains. Astronomy \& Astrophysics, 650, A152.\\

Izidoro, A., Dasgupta, R., Raymond, S. N., Deienno, R., Bitsch, B., \& Isella, A. (2022). Planetesimal rings as the cause of the Solar System’s planetary architecture. \textit{Nature Astronomy}, \textit{6}(3), 357-366.\\

Johansen, A., Low, M. M. M., Lacerda, P., \& Bizzarro, M. (2015). Growth of asteroids, planetary embryos, and Kuiper belt objects by chondrule accretion. Science Advances, 1(3), e1500109.\\

Kenyon, S. J., \& Hartmann, L. W. (1987). Spectral energy distributions of T Tauri stars-Disk flaring and limits on accretion. \textit{Astrophysical Journal, Part 1 (ISSN 0004-637X), vol. 323, Dec. 15, 1987, p. 714-733.}, \textit{323}, 714-733.\\

Klahr, H., \& Schreiber, A. (2020). Turbulence sets the length scale for planetesimal formation: Local 2D simulations of streaming instability and planetesimal formation. \textit{The Astrophysical Journal}, \textit{901}(1), 54.\\

 Kleine, T., Budde, G., Burkhardt, C., Kruijer, T. S., Worsham, E. A., Morbidelli, A., \& Nimmo, F. (2020). The non-carbonaceous–carbonaceous meteorite dichotomy. Space Science Reviews, 216, 1-27.\\

Kokubo, E., \& Ida, S. (2000). Formation of protoplanets from planetesimals in the solar nebula. \textit{Icarus}, \textit{143}(1), 15-27.\\

Kruijer, T. S., Burkhardt, C., Budde, G., \& Kleine, T. (2017a). Age of Jupiter inferred from the distinct genetics and formation times of meteorites. \textit{Proceedings of the National Academy of Sciences}, \textit{114}(26), 6712-6716.\\

Kruijer, T. S., \& Kleine, T. (2017b). Tungsten isotopes and the origin of the Moon. \textit{Earth and Planetary Science Letters}, \textit{475}, 15-24.\\

Kruijer, T. S., Touboul, M., Fischer-Gödde, M., Bermingham, K. R., Walker, R. J., \& Kleine, T. (2014). Protracted core formation and rapid accretion of protoplanets. \textit{Science}, \textit{344}(6188), 1150-1154.\\\\

Kurosaki, K., Ikoma, M., \& Hori, Y. (2014). Impact of photo-evaporative mass loss on masses and radii of water-rich sub/super-Earths. \textit{Astronomy \& Astrophysics}, \textit{562}, A80.\\
 
Lambrechts, M., \& Johansen, A. (2012). Rapid growth of gas-giant cores by pebble accretion. Astronomy \& Astrophysics, 544, A32.\\

Lambrechts, M., Johansen, A., \& Morbidelli, A. (2014). Separating gas-giant and ice-giant planets by halting pebble accretion. Astronomy \& Astrophysics, 572, A35.\\

Leinhardt, Z. M., \& Stewart, S. T. (2009). Full numerical simulations of catastrophic small body collisions. Icarus, 199(2), 542-559.\\

Lenz, C. T., Klahr, H., \& Birnstiel, T. (2019). Planetesimal population synthesis: Pebble flux-regulated planetesimal formation. \textit{The Astrophysical Journal}, \textit{874}(1), 36.\\

Liu, W., Zhang, Y., Tissot, F. L., Avice, G., Ye, Z., \& Yin, Q. Z. (2023). I/Pu reveals Earth mainly accreted from volatile-poor differentiated planetesimals. Science Advances, 9(27), eadg9213.\\

Lissauer, J. J. (1993). Planet formation. \textit{Annual review of astronomy and astrophysics}, \textit{31}(1), 129-172.\\

Lynden-Bell, D., \& Pringle, J. E. (1974). The evolution of viscous discs and the origin of the nebular variables. Monthly Notices of the Royal Astronomical Society, 168(3), 603-637.\\

Miotello, A., Kamp, I., Birnstiel, T., Cleeves, L. I., \& Kataoka, A. (2022). Setting the stage for planet formation: measurements and implications of the fundamental disk properties. \textit{arXiv preprint arXiv:2203.09818}.\\
 
Morbidelli, A., Baillie, K., Batygin, K., Charnoz, S., Guillot, T., Rubie, D. C., \& Kleine, T. (2022). Contemporary formation of early Solar System planetesimals at two distinct radial locations. Nature Astronomy, 6(1), 72-79.\\

Mulders, G. D., Pascucci, I., Apai, D., \& Ciesla, F. J. (2018). The exoplanet population observation simulator. I. the inner edges of planetary systems. \textit{The Astronomical Journal}, \textit{156}(1), 24.\\

Musiolik, G., \& Wurm, G. (2019). Contacts of water ice in protoplanetary disks—laboratory experiments. The Astrophysical Journal, 873(1), 58.\\

Nakagawa, Y., Sekiya, M., \& Hayashi, C. (1986). Settling and growth of dust particles in a laminar phase of a low-mass solar nebula. Icarus, 67(3), 375-390.\\

Nakamoto, T., \& Nakagawa, Y. (1994). Formation, early evolution, and gravitational stability of protoplanetary disks. Astrophysical Journal, Part 1 (ISSN 0004-637X), vol. 421, no. 2, p. 640-650, 421, 640-650.\\

Nelson, R. P., Gressel, O., \& Umurhan, O. M. (2013). Linear and non-linear evolution of the vertical shear instability in accretion discs. Monthly Notices of the Royal Astronomical Society, 435(3), 2610-2632.\\

Ogihara, M., Morbidelli, A., \& Guillot, T. (2015). A reassessment of the in situ formation of close-in super-Earths. Astronomy \& Astrophysics, 578, A36.\\

Olson, P. L., \& Sharp, Z. D. (2023). Hafnium-tungsten evolution with pebble accretion during Earth formation. \textit{Earth and Planetary Science Letters}, \textit{622}, 118418.\\

Onyett, I. J., Schiller, M., Makhatadze, G. V., Deng, Z., Johansen, A., \& Bizzarro, M. (2023). Silicon isotope constraints on terrestrial planet accretion. Nature, 1-6.\\

Ormel, C. W., \& Cuzzi, J. N. (2007). Closed-form expressions for particle relative velocities induced by turbulence. Astronomy \& Astrophysics, 466(2), 413-420.\\

Ormel, C. W., \& Klahr, H. H. (2010). The effect of gas drag on the growth of protoplanets-analytical expressions for the accretion of small bodies in laminar disks. Astronomy \& Astrophysics, 520, A43.\\

Ormel, C. W. (2017). The emerging paradigm of pebble accretion. Formation, Evolution, and Dynamics of Young Solar Systems, 197-228.\\

Owen, J. E., \& Wu, Y. (2013). Kepler planets: a tale of evaporation. The Astrophysical Journal, 775(2), 105.\\

Owen, J. E., \& Wu, Y. (2017). The evaporation valley in the Kepler planets. The Astrophysical Journal, 847(1), 29.\\

Penna, R. F., Sadowski, A., Kulkarni, A. K., \& Narayan, R. (2013). The Shakura-Sunyaev viscosity prescription with variable $\alpha$(r). Monthly Notices of the Royal Astronomical Society, 428(3), 2255-2274.\\

Petigura, E. A., Howard, A. W., \& Marcy, G. W. (2013). Prevalence of Earth-size planets orbiting Sun-like stars. Proceedings of the National Academy of Sciences, 110(48), 19273-19278.\\

Pillich, C., Bogdan, T., Landers, J., Wurm, G., \& Wende, H. (2021). Drifting inwards in protoplanetary discs-II. The effect of water on sticking properties at increasing temperatures. Astronomy \& Astrophysics, 652, A106.\\

Pillich, C., Bogdan, T., Tasto, J., Landers, J., Wurm, G., \& Wende, H. (2023). Composition and Sticking of Hot Chondritic Dust in a Protoplanetary Hydrogen Atmosphere. \textit{The Planetary Science Journal}, \textit{4}(10), 195.\\

Pinilla, P., Lenz, C. T., \& Stammler, S. M. (2021). Growing and trapping pebbles with fragile collisions of particles in protoplanetary disks. \textit{Astronomy \& Astrophysics}, \textit{645}, A70.\\

Podosek, F. A., \& Cassen, P. (1994). Theoretical, observational, and isotopic estimates of the lifetime of the solar nebula. Meteoritics, 29(1), 6-25.\\

Pollack, J. B., McKay, C. P., \& Christofferson, B. M. (1985). A calculation of the Rosseland mean opacity of dust grains in primordial solar system nebulae. \textit{Icarus}, \textit{64}(3), 471-492.\\

Poppe, T. (2003). Sintering of highly porous silica-particle samples: analogues of early Solar-System aggregates. \textit{Icarus}, \textit{164}(1), 139-148.\\

Rafikov, R. R. (2011). Constraint on the giant planet production by core accretion. The Astrophysical Journal, 727(2), 86.\\

Ricci, L., Testi, L., Natta, A., Neri, R., Cabrit, S., \& Herczeg, G. J. (2010). Dust properties of protoplanetary disks in the Taurus-Auriga star forming region from millimeter wavelengths. \textit{Astronomy \& Astrophysics}, \textit{512}, A15.\\

Rosenthal, L. J., Knutson, H. A., Chachan, Y., Dai, F., Howard, A. W., Fulton, B. J., ... \& Wright, J. T. (2022). The California legacy survey. III. On the shoulders of (some) giants: the relationship between inner small planets and outer massive planets. The Astrophysical Journal Supplement Series, 262(1), 1.\\

Rosotti, G. P. (2023). Empirical constraints on turbulence in proto-planetary discs. New Astronomy Reviews, 101674.\\

Rozitis, B., Ryan, A. J., Emery, J. P., Christensen, P. R., Hamilton, V. E., Simon, A. A., ... \& Lauretta, D. S. (2020). Asteroid (101955) Bennu’s weak boulders and thermally anomalous equator. \textit{Science Advances}, \textit{6}(41), eabc3699.\\

Safronov, V. S. (1972). Evolution of the Protoplanetary Cloud and Formation of the Earth and the Planets.\\

Schäfer, U., Yang, C. C., \& Johansen, A. (2017). Initial mass function of planetesimals formed by the streaming instability. \textit{Astronomy \& Astrophysics}, \textit{597}, A69.\\

Schiller, M., Bizzarro, M., \& Siebert, J. (2020). Iron isotope evidence for very rapid accretion and differentiation of the proto-Earth. Science advances, 6(7), eaay7604.\\

Sears, D. W. (1998). The case for rarity of chondrules and calcium-aluminum-rich inclusions in the early solar system and some implications for astrophysical models. \textit{The Astrophysical Journal}, \textit{498}(2), 773.\\
 
Shakura, N. I., \& Sunyaev, R. A. (1973). Black holes in binary systems. Observational appearance. Astronomy and Astrophysics, Vol. 24, p. 337-355, 24, 337-355.\\

Sossi, P. A., Stotz, I. L., Jacobson, S. A., Morbidelli, A., \& O’Neill, H. S. C. (2022). Stochastic accretion of the Earth. \textit{Nature astronomy}, \textit{6}(8), 951-960.\\

Spitzer, F., Burkhardt, C., Nimmo, F., \& Kleine, T. (2021). Nucleosynthetic Pt isotope anomalies and the Hf-W chronology of core formation in inner and outer solar system planetesimals. \textit{Earth and Planetary Science Letters}, \textit{576}, 117211.\\

Squire, J., \& Hopkins, P. F. (2018). Resonant drag instabilities in protoplanetary discs: the streaming instability and new, faster growing instabilities. \textit{Monthly Notices of the Royal Astronomical Society}, \textit{477}(4), 5011-5040.\\

Steinpilz, T., Joeris, K., Jungmann, F., Wolf, D., Brendel, L., Teiser, J., ... \& Wurm, G. (2020). Electrical charging overcomes the bouncing barrier in planet formation. Nature Physics, 16(2), 225-229.\\

Stepinski, T. F. (1998). The solar nebula as a process–An analytic model. \textit{Icarus}, \textit{132}(1), 100-112.\\

Störzer, H., \& Hollenbach, D. (1999). Photodissociation region models of photoevaporating circumstellar disks and application to the proplyds in Orion. \textit{The Astrophysical Journal}, \textit{515}(2), 669.\\
 
Tabone, B., Rosotti, G. P., Cridland, A. J., Armitage, P. J., \& Lodato, G. (2022). Secular evolution of MHD wind-driven discs: analytical solutions in the expanded $\alpha$-framework. Monthly Notices of the Royal Astronomical Society, 512(2), 2290-2309.\\

Teske, J. K., Ciardi, D. R., Howell, S. B., Hirsch, L. A., \& Johnson, R. A. (2018). The effects of stellar companions on the observed transiting exoplanet radius distribution. \textit{The Astronomical Journal}, \textit{156}(6), 292.\\

Testi, L., Natta, A., Shepherd, D. S., \& Wilner, D. J. (2003). Large grains in the disk of CQ Tau. \textit{Astronomy \& Astrophysics}, \textit{403}(1), 323-328.\\

Thiemens, M. M., Tusch, J., OC Fonseca, R., Leitzke, F., Fischer-Gödde, M., Debaille, V., ... \& Münker, C. (2021). Reply to: No 182W evidence for early Moon formation. \textit{Nature Geoscience}, \textit{14}(10), 716-718.\\

Tissot, F. L., Collinet, M., Namur, O., \& Grove, T. L. (2022). The case for the angrite parent body as the archetypal first-generation planetesimal: Large, reduced and Mg-enriched. Geochimica et Cosmochimica Acta, 338, 278-301.\\

Trapman, L., Tabone, B., Rosotti, G., \& Zhang, K. (2022). Effect of MHD wind-driven disk evolution on the observed sizes of protoplanetary disks. The Astrophysical Journal, 926(1), 61.\\

Wada, K., Tanaka, H., Suyama, T., Kimura, H., \& Yamamoto, T. (2011). The rebound condition of dust aggregates revealed by numerical simulation of their collisions. The Astrophysical Journal, 737(1), 36.\\

Wang, H., Weiss, B. P., Bai, X. N., Downey, B. G., Wang, J., Wang, J., ... \& Zucolotto, M. E. (2017). Lifetime of the solar nebula constrained by meteorite paleomagnetism. \textit{Science}, \textit{355}(6325), 623-627.\\
 
Weidenschilling, S. J. (1977). Aerodynamics of solid bodies in the solar nebula. Monthly Notices of the Royal Astronomical Society, 180(2), 57-70.\\

Weidenschilling, S. J., \& Cuzzi, J. N. (1993). Formation of planetesimals in the solar nebula. In \textit{Protostars and planets III} (pp. 1031-1060).\\

Weidling, R., Güttler, C., \& Blum, J. (2012). Free collisions in a microgravity many-particle experiment. I. Dust aggregate sticking at low velocities. Icarus, 218(1), 688-700.\\

Whipple, F. L. (1972). On certain aerodynamic processes for asteroids and comets. In From plasma to planet (p. 211).\\

Williams, J. P., \& Cieza, L. A. (2011). Protoplanetary disks and their evolution. Annual Review of Astronomy and Astrophysics, 49, 67-117.\\

Windmark, F., Birnstiel, T., Güttler, C., Blum, J., Dullemond, C. P., \& Henning, T. (2012). Planetesimal formation by sweep-up: how the bouncing barrier can be beneficial to growth. Astronomy \& Astrophysics, 540, A73.\\

Wu, Y. (2019). Mass and mass scalings of super-Earths. The Astrophysical Journal, 874(1), 91.\\

Yang, C. C., Johansen, A., \& Carrera, D. (2017). Concentrating small particles in protoplanetary disks through the streaming instability. Astronomy \& Astrophysics, 606, A80.\\

Yap, T. E., \& Tissot, F. L. (2023). The NC-CC dichotomy explained by significant addition of CAI-like dust to the Bulk Molecular Cloud (BMC) composition. Icarus, 405, 115680.\\

Yokoyama, T., Nagashima, K., Nakai, I., Young, E. D., Abe, Y., Aléon, J., ... \& Yurimoto, H. (2022). Samples returned from the asteroid Ryugu are similar to Ivuna-type carbonaceous meteorites. \textit{Science}, \textit{379}(6634), eabn7850.\\

Youdin, A. N., \& Goodman, J. (2005). Streaming instabilities in protoplanetary disks. The Astrophysical Journal, 620(1), 459.\\

Zsom, A., Ormel, C. W., Güttler, C., Blum, J., \& Dullemond, C. P. (2010). The outcome of protoplanetary dust growth: pebbles, boulders, or planetesimals?-II. Introducing the bouncing barrier. Astronomy \& Astrophysics, 513, A57.\\

\end{document}